\documentclass[twocolumn]{aastex631}
\usepackage{amsmath}
\usepackage{mathrsfs}
\usepackage{CJK}
\usepackage{multirow}

\shorttitle{Origins of GW231123}
\shortauthors{Li et al.}

\graphicspath{{./}{figures/}}

\begin{document}
\begin{CJK*}{UTF8}{gbsn}

\title{GW231123: Likely a product of successive mergers from $\sim 10 $ stellar-mass black holes}

\author[0000-0001-5087-9613]{Yin-Jie Li (李银杰)}
\affiliation{Key Laboratory of Dark Matter and Space Astronomy, Purple Mountain Observatory, Chinese Academy of Sciences, Nanjing 210023, People's Republic of China}

\author[0000-0001-9120-7733]{Shao-Peng Tang (唐少鹏)}
\affiliation{Key Laboratory of Dark Matter and Space Astronomy, Purple Mountain Observatory, Chinese Academy of Sciences, Nanjing 210023, People's Republic of China}

\author[0009-0001-6242-0823]{Ling-Qin Xue (薛灵钦)}
\affiliation{Department of Physics, University of Florida, PO Box 118440, Gainesville, FL 32611-8440, USA}

\author[0000-0002-8966-6911]{Yi-Zhong Fan (范一中)}
\affiliation{Key Laboratory of Dark Matter and Space Astronomy, Purple Mountain Observatory, Chinese Academy of Sciences, Nanjing 210023, People's Republic of China}
\affiliation{School of Astronomy and Space Science, University of Science and Technology of China, Hefei, Anhui 230026, People's Republic of China}
\email{The corresponding author: yzfan@pmo.ac.cn (Y.Z.F)}

\begin{abstract}
GW231123 is an exceptionally massive binary black hole (BBH) merger with unusually high component spins. Such extreme properties challenge conventional stellar evolution models predicting a black hole mass gap due to pair-instability supernovae. 
{We test possible formation scenarios for GW231123 using population-informed priors on BH spin distributions, in light of population properties built on the previous (GWTC-3) data.
Our analysis shows that GW231123 belongs to the high-spin subpopulation that is naturally interpreted as hierarchical BBH mergers. By comparing the spin magnitudes and component masses of GW231123 to those of the remnants of previous mergers, we show that both components of GW231123 are multi-generation ($>$2G) merger remnants,} and plausibly originated from the successive mergers of $\sim 6$ and $\sim 4$ first-generation BHs, respectively. This suggests that repeated mergers can be frequent and even more massive intermediate-mass black holes may be produced. Thus mechanisms that can efficiently harden the BBHs' orbits are required, e.g., gas dynamical friction in the disks of active galactic nuclei.
\end{abstract}

\keywords{Binary Black Holes; Gravitational Waves; Stellar Evolution; Active Galactic Nuclei}

\section{Introduction}
Very recently, the LIGO-Virgo-KAGRA (LVK) collaboration reported the detection of GW231123 \citep{2025ApJ...993L..25A}, a binary black hole (BBH) merger with a remarkably large total source-frame mass of $\sim190$-$265\,M_\odot$ and component dimensionless spins of $\chi_1\approx0.9$ and $\chi_2\approx0.8$. Such extreme masses and spins are unprecedented in previous LVK observations \citep[e.g.,][]{2023PhRvX..13d1039A}, and they provide a unique opportunity to probe the formation and evolution of intermediate-mass black holes (IMBHs) \citep{2004IJMPD..13....1M, 2017arXiv170501881L, 2022A&A...659A..84A}.

Theories of stellar evolution suggest that stars with helium core masses up to $\sim135\,M_\odot$ can only leave behind black holes (BHs) lighter than $\sim65\,M_{\odot}$, due to (pulsational) pair-instability supernovae \citep[(P)PISN;][]{1964ApJS....9..201F, 1967PhRvL..18..379B, 2017ApJ...836..244W, 2021ApJ...912L..31W}. This leads to a so-called pair-instability mass gap (PIMG) in the BH mass spectrum, although the exact edges of the gap remain uncertain currently \citep{2017ApJ...836..244W, 2018MNRAS.474.2959G, 2019ApJ...878...49W, 2019ApJ...887...53F, 2020ApJ...902L..36F, 2020ApJ...890..113B}. However, a star with a helium core mass $\gtrsim 135\,M_\odot$ is expected to directly collapse into an IMBH, providing a possible origin for the components of GW231123.

Alternatively, in dynamical formation channels BHs can undergo repeated mergers and substantially grow in mass, as long as the merger remnants are efficiently retained in their host environments \citep[see][]{2021NatAs...5..749G, 2022ApJ...935L..20Z, 2025ApJ...984...63L}. GW231123 may thus represent a hierarchical merger. However, how massive BHs can grow via hierarchical assembly remains unclear and likely depends on the host environment. For example, in active galactic nucleus (AGN) disk channels the maximum BH mass may be limited by the finite lifetime of the disk \citep{2025arXiv250419570X}. Additionally, primordial BHs (PBHs) formed in the early Universe \citep{1974MNRAS.168..399C,2010RAA....10..495K} could also populate mass ranges that are hard to reach via stellar collapse \citep{2018PhRvL.121h1306C, 2024bheg.book..261E}.

Previously, the BBH merger GW190521 \citep{2020PhRvL.125j1102A} attracted significant attention, as it contained at least one component with a mass apparently above the lower edge of the PIMG  \cite{2020ApJ...904L..26F}. \citet{2021ApJ...907L...9N} proposed that GW190521 might be a \textit{straddling binary}, meaning one BH lay below and the other above the PIMG. However,  population analyses inferred that at least one component of GW190521 was likely a higher-generation BH based on its spin and mass properties \citep{2021ApJ...913...42W, 2021ApJ...915L..35K, 2022ApJ...941L..39W, 2024PhRvL.133e1401L, 2024arXiv240601679P, 2025PhRvL.134a1401A, 2025ApJ...987...65L, 2025PhRvD.112f3040A}.

The formation history of a GW event is encoded in the observed source parameters. By performing parameter estimation (inferring the component masses, spins, etc.), one can test various astrophysical origin hypotheses. However, the marginal distributions of these parameters often have large uncertainties \citep{2019PhRvX...9c1040A, 2020ApJ...900L..13A, 2021PhRvX..11b1053A, 2023PhRvX..13d1039A}, making it challenging to determine the origins of these events from standard analyses alone. Crucially, some source parameters are correlated or degenerate with each other \citep{2016PhRvD..93h4042P, 2018ApJ...868..140T}. For example, adopting a particular prior for the spin distribution can influence the inferred mass distribution for the same event. We therefore employ \textit{population-informed priors} -- priors { in light of population analysis built on the previous BBH catalog -- to better understand the origin and evolution history of GW231123.}

The rest of the paper is organized as follows: In Section~\ref{case}, we describe our population-informed prior choices, and present the results of applying these priors to GW231123 and compare the evidence for each scenario. Besides, we also estimate the number of progenitors needed to generate GW231123 in the hierarchical merger scenario.  In Section~\ref{sec:diss}, we discuss the implications and draw conclusions.

\section{Testing the Formation Hypotheses with Population Information}\label{case}

The spins of BHs are expected to differ based on their formation mechanism. 
Stellar-collapse (first-generation) BHs are generally predicted to be born with low spins \citep[$\chi \lesssim 0.1$;][]{2019ApJ...881L...1F}, although binary interactions can spin them up modestly \citep{2018A&A...616A..28Q,2020A&A...635A..97B,2022ApJ...930...26S}. 
Primordial BHs would also have effectively negligible spins \citep{2017PTEP.2017h3E01C,2019JCAP...05..018D, 2021JPhG...48d3001G}, neglecting any significant accretions that PBHs may undergo \citep{2020JCAP...04..052D} 
\footnote{ There is no evidence for an exactly non-spinning subpopulation \citep{2022ApJ...937L..13C, 2022PhRvD.106j3019T,2024arXiv241102252H} within current GW data. Additionally, it would be challenging to identify mergers of PBHs from astrophysical within era of second-generation interferometers \citep{2022PhRvD.105h3526F}.}.
In contrast, BHs formed by previous BBH mergers tend to have significantly larger spins, typically peaking around $\chi \sim 0.7$ \citep{2017PhRvD..95l4046G, 2017ApJ...840L..24F, 2021NatAs...5..749G}. { Particularly, spin magnitudes of higher-generation BHs assembled in the AGN disks may shift to even larger values due to gas accretion \citep{2024A&A...685A..51V,2025arXiv250608801F,2025arXiv250808558B}.
}

Indeed, recent population studies of GWTC-3 have found evidence for two distinct subpopulations in the plane of spin magnitude versus component mass \citep{2024PhRvL.133e1401L, 2024A&A...692A..80P}. The first subpopulation (with smaller BH masses and $\chi \lesssim  0.3$) is consistent with BHs born from stellar core-collapse, whereas the second subpopulation (with more massive BHs and $\chi \sim 0.75$) is indicative of BHs that have undergone at least one merger already.
{Similar subpopulations are also identified {with different approaches \citep{2024arXiv241102252H,2025PhRvL.134a1401A,2025ApJ...987...65L},} although some population analyses that concentrate on the marginal distributions of spins (or that do not account for the mass-spin correlations) did not find evident subpopulations \citep[e.g.][]{2023PhRvX..13a1048A,2023ApJ...946...16E,2022MNRAS.517.2738M,2018PhRvD..97d3014W,2023PhRvD.108j3009G,2024ApJ...964L...6A}. With the newly released data of GWTC-4, the subpopulations of {low-spin} and {high-spin} BHs 
are more evident \citep[e.g.,][]{2025arXiv250904151T,2025arXiv250904637A,2025arXiv250915646B,2025arXiv250923897L,2025arXiv251022698W}.
}

\subsection{Both components are likely higher-generation BHs}
\begin{figure*}
	\centering  
\includegraphics[width=0.8\linewidth]{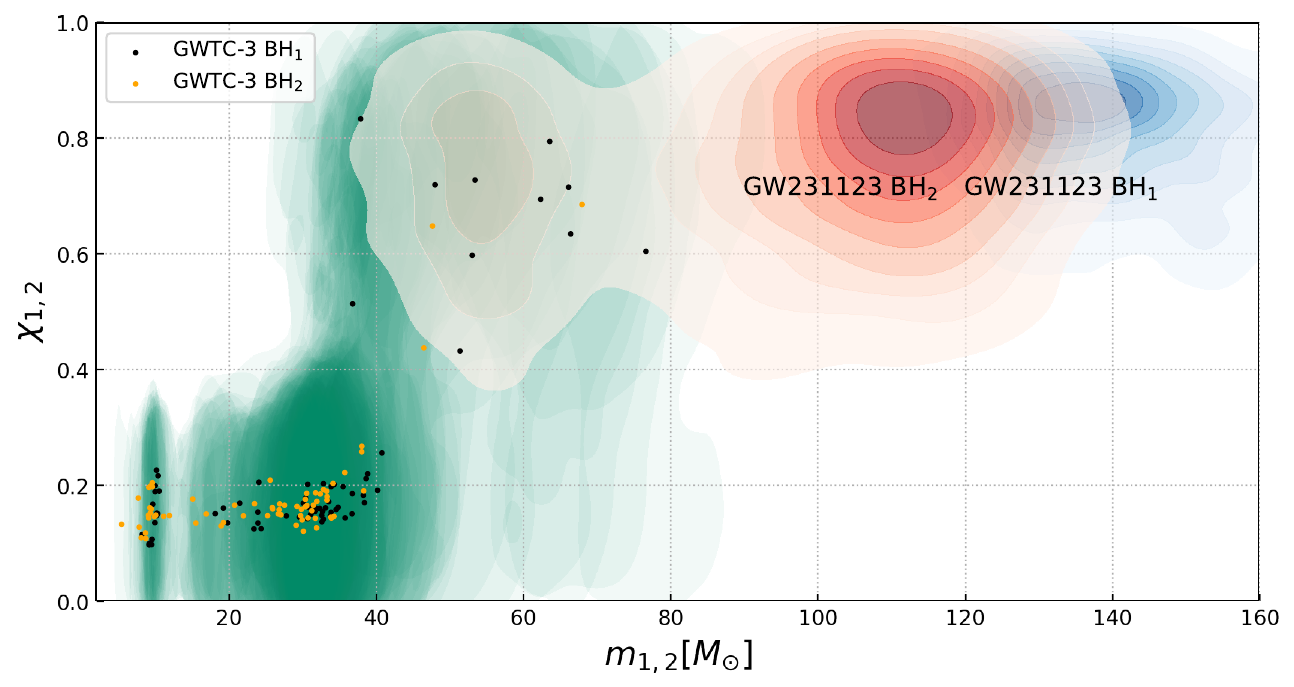}
\caption{
Spin magnitude v.s. component mass distribution of GW231123 ({ reweighed by HS-HS prior}) compared to those of BBH events in GWTC-3 informed by population model of \citet{2024PhRvL.133e1401L}. The red and blue contours denote the primary and secondary components of GW231123. The green shaded regions show the $90\%$ credible regions for component BHs in the GWTC-3 population, and the black (orange) points mark the mean component masses for the primary (secondary) BHs in those events.}
\label{fig:m-a-dist}
\end{figure*}

Motivated by the subpopulations that are identified with previous data \citep{2024PhRvL.133e1401L}
and the potential links \footnote{Some work also suggested the existence of a high spin population \citep{2021PhRvD.104h3010R,2022ApJ...928...75H,2021ApJ...921L..15G,2024arXiv241102252H}, though some are unlikely hierarchical mergers, due to divergent model assumptions. The identification and interpretation of first-generation and higher-generation BHs is being common \citep[e.g.][]{2022ApJ...941L..39W,2024ApJ...975...54G,2025PhRvD.112f3040A}, especially with GWTC-4 \citep[e.g.,][]{2025arXiv250904151T,2025arXiv250904637A,2025arXiv250909123A,2025arXiv250915646B,2025arXiv250923897L,2025arXiv251022698W}.} to the origins of BHs \citep{2021NatAs...5..749G}, we apply the population-informed priors based on the subpopulations mentioned above to test the origin of GW231123.
Specifically, we use the priors informed by the population of BHs inferred from GWTC-3 \citep{2024PhRvL.133e1401L}, see Appendix~\ref{app:BF} for detailed illustration.

Comparing the Bayes factors between different priors (see Appendix~\ref{app:BF}), we find the most favored scenario is that both BHs of GW231123 belong to the high-spin subpopulation. 
{ Considering the origin of GW231123 in light of the subpopulation properties inferred with previous data \citep{2021NatAs...5..749G,2024PhRvL.133e1401L}, the interpretation that GW231123 is a hierarchical merger is very natural. Such a conclusion is clearly illustrated in}
the (population-informed) component-mass versus spin-magnitude distributions of GW231123\footnote{The posterior samples are obtained from https://zenodo.org/records/16004263  \dataset  []{ligo_scientific_collaboration_2025_16004263}, { and the `Mixed sample' are adopted.}}, comparing to that of the previous events reweighed\footnote{The data and codes can be download from \href{https://github.com/JackLee0214/Resolving-the-stellar-collapse- and-hierarchical-merger-origins-of-the-coalescing-black-holes}{GitHub: Stellar-formed V.S. Merger-formed} \dataset[]{jacklee0214_2025_17572948}, { which built on the `Mixed samples' of GWTC-3.}} by the population model in \citet{2024PhRvL.133e1401L}, see Figure~\ref{fig:m-a-dist}. 

{ Under the population-informed priors that both components belong to the high-spin subpopulation (i.e., HS-HS), we obtain $\chi_1=0.82^{+0.13}_{-0.22}$ and $\chi_2= 0.77^{+0.17}_{-0.31}$, see Appendix~\ref{app:reweight} for the details of reweighting. 
We find the component masses are also changed after reweighting, since the source parameters are degenerated in the parameter estimation of individual event. Under HS-HS priors, we find the secondary mass has fewer samples at low-mass range, providing further support for the higher-generation scenario. It is worth mentioning that under the HS-LS prior (assuming the secondary BH is slowly spinning), the secondary mass shows increasing support at $\sim 60M_\odot$. However, the HS-LS prior is less favored compared to the HS-HS prior by $\log_{10}\mathcal{B}=-0.7$.
}

\subsection{Likely higher than second-generation BHs}
{ The final spin of the merger remnant also depends on the spin of BH progenitors \citep{2017PhRvD..95l4046G,2017ApJ...840L..24F,2024ApJ...975..117M}. Mergers of low-spin equal-mass progenitors have final spin narrowly peaking at $0.7$, while mergers of high-spin progenitors have a broad distribution of final spins. Particularly, if the highly spinning progenitor BBHs tend to align their spins with the orbits, the final spins will shift to larger values. 
We generate the final spins of the BBHs in three cases based on the subpopulations of \citet{2024PhRvL.133e1401L}: one is that both BHs come from the LS subpopulation (1G+1G), the second is that the two BHs respectively come from LS and HS subpopulations (nG+1G), the third is that both BHs come from the HS subpopulation (nG+nG). 
We find that both spin magnitudes of GW231123 are more consistent with the remnants of nG+nG or nG+1G rather than the remnants of 1G+1G. 
}

\begin{figure*}
	\centering  
\includegraphics[width=0.49\linewidth]{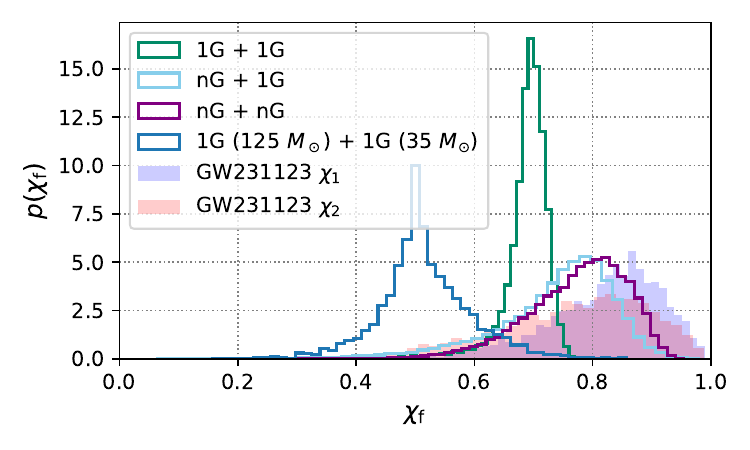}
\includegraphics[width=0.49\linewidth]{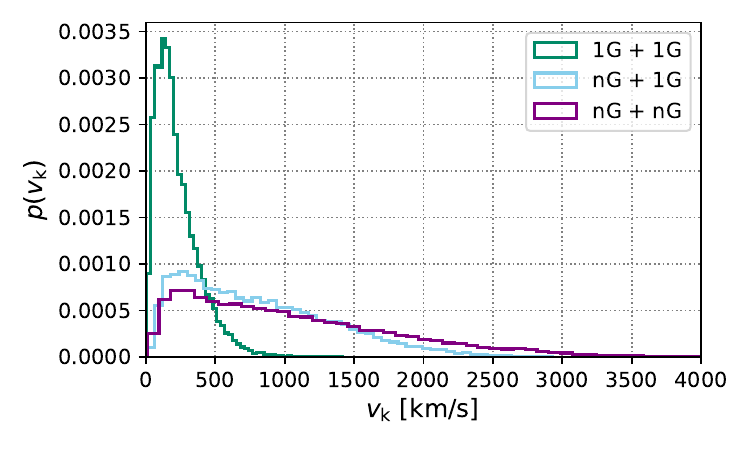}
\caption{Comparison of final remnant spins (left panel) and GW recoil kick velocities(right panel) for mergers with BHs of different generations. The left panel shows the corresponding distributions of the remnants' final spin magnitudes, with shaded areas comparing the component spin magnitudes of GW231123, { reweighed with HS-HS priors}. The right panel shows the kick velocity distributions, indicating that hierarchical mergers involving higher-generation component BHs produce significantly larger kicks than 1G+1G mergers. The 1G+1G, nG+1G, and nG+nG BBHs samples are drawn from the posterior population distribution in \citet{2024PhRvL.133e1401L}.}
\label{fig:vk}
\end{figure*}

\begin{figure*}
	\centering  
\includegraphics[width=0.8\linewidth]{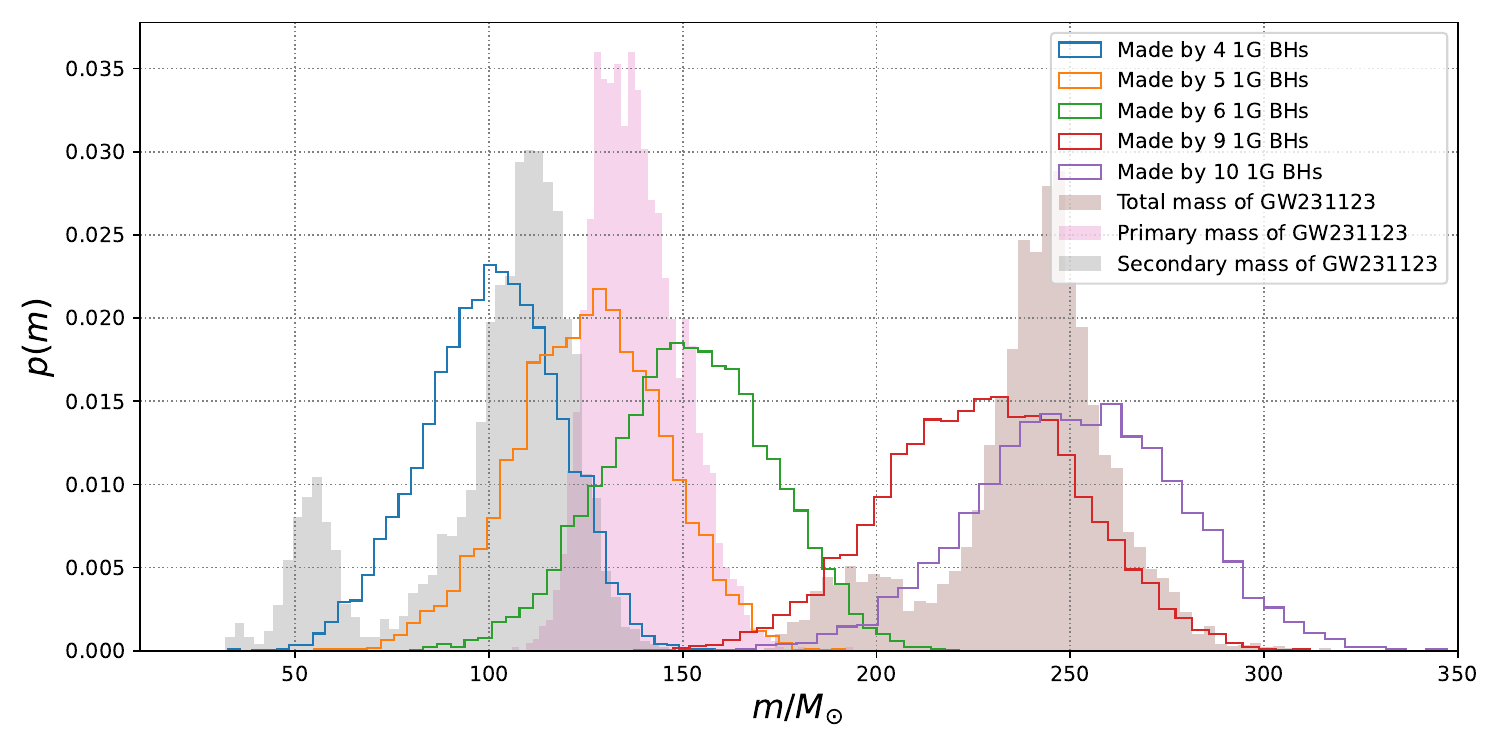}
\caption{
Mass distributions of merger remnants formed from first-generation BHs in dynamical formation channels, compared to the component masses of GW231123, { reweighed with HS-HS prior}. The first-generation BH mass distribution is taken from \citet{2024ApJ...977...67L}. The pink and grey shaded regions indicate GW231123's primary and secondary mass distributions, which align with mass distribution of remnants assembled from $\sim6$ (green) and $\sim4$ (blue) first-generation BHs, respectively. A self-consistency check comparing these model predictions to the observed higher-generation BH population is provided in Appendix~\ref{app:check} (see Figure~\ref{fig:check-dist}).}
\label{fig:generations}
\end{figure*}

{ As we have found that both components of GW231123 are more likely to be BHs beyond the second generation, it is worth investigating how many stellar-formed BHs constitute them. For this purpose, we have to specify the underlying first-generation BHs that contribute to the successive mergers.} Theoretical predictions and population analysis suggest that isolated binary evolution channels contribute to the LVK's BBH mergers \citep[see e.g.][]{2021ApJ...910..152Z, 2022ApJ...941L..39W, 2023arXiv230401288G, 2024ApJ...977...67L}, which are not expected to participate in subsequent hierarchical mergers \citep{2021NatAs...5..749G}. Only the BHs in dynamical formation channels may undergo hierarchical mergers, and potentially contribute to the GW231123-like events.
{ Some} population analyses revealed a subpopulation of BHs, which dominates the mass range $\sim[15,40]~M_{\odot}$ \citep[see e.g.][]{2022ApJ...941L..39W, 2024arXiv240403166R, 2024ApJ...977...67L}\footnote{See also \citet{2025arXiv250915646B} for analysis on GWTC-4, though their interpretation is different from us.}. { This subpopulation is consistent with dynamical formation channels predicted by simulations \citep[see e.g.,][]{ 2020PhRvD.102l3016A,2022MNRAS.517.2953T,2023MNRAS.522..466A,2025arXiv250608801F}. We note that the results are slightly different in \citet{2023arXiv230401288G}, which may be caused by the fact that the models in \citet{2023arXiv230401288G} have only two subpopulations when fitting the data that includes high-spin events, whereas models in \citet{2024ApJ...977...67L} consider three potential subpopulations. In general, there will be {at least} three subpopulations of BBHs: the first is for the field channels, the second is for the first-generation BBHs from dynamical channels, and the third is for the hierarchical mergers. We note that there are two modes in the posterior distribution of analysis from \citet{2023arXiv230401288G} (see their Figure 6), one mode (taking 76\% of samples) may have mixed BBHs from field channels and first-generation BBHs from dynamical channels (Dyn-1G) together, the other mode (taking 24\% of samples) may have mixed Dyn-1G BBHs with hierarchical mergers. }
Therefore, in estimating how many stellar-mass BHs are required to build GW231123's components, we adopt the { possible} mass distribution of the Dyn-1G BHs identified by \citet{2024ApJ...977...67L}\footnote{Data and codes are available at \href{https://github.com/JackLee0214/Exploring-field-evolution-and-dynamical-capture-coalescing-binary-black-holes-in-GWTC-3/tree/main}{GitHub: Field V.S. Dyanmical}\citep{jacklee0214_2025_17572869}}. { Though the inferred mass function remains largely uncertain, it is broadly consistent with the prediction from some simulations \citep[e.g.,][]{2020PhRvD.102l3016A,2022MNRAS.517.2953T,2023MNRAS.522..466A,2025arXiv250608801F}.} { Please see Appendix~\ref{app:generation} for the details about generating the mass functions of higher-generation BHs.}

{ To validate our choice of mass function for the Dyn-1G BHs, we have also made a self-consistency check, see Appendix~\ref{app:check}.
}
We compare the mass functions of remnants (made by $n$ Dyn-1G BHs) with the subpopulation of { potentially}\footnote{Such a subpopulation is quite solid now, as was revealed with many different approaches in GWTC-3\citep[e.g.][]{2022ApJ...941L..39W,2024A&A...692A..80P,2025PhRvL.134a1401A}, and recently confirmed with GWTC-4 \citep[e.g.][]{2025arXiv250904637A,2025arXiv250915646B,2025arXiv250923897L,2025arXiv251022698W}.} higher-generation BHs observed in GWTC-3 \citep{2024PhRvL.133e1401L, 2024ApJ...977...67L}. We find that the distribution of higher-generation BH masses is consistent with remnants produced by the mergers with $\gtrsim 2$ first-generation BHs, supporting the idea that many of the heavy BHs in GWTC-3 (and GW231123 itself) are products of hierarchical growth.

Figure~\ref{fig:generations} shows the mass distribution of hierarchical merger remnants produced by BHs drawn from this subpopulation, compared against the observed masses of GW231123. From these distributions, we infer that producing a BH of GW231123's primary mass would require merging roughly 6 first-generation BHs, while the secondary could be built from about 4 first-generation BHs (implying a total equivalent of $\sim10$ stellar-mass BHs to assemble the system's $\sim250~M_{\odot}$ total mass).

{ If only considering the masses of GW231123, }the primary component could also arise from a direct merger between a $\sim 125 M_{\odot}$ stellar-collapse IMBH { above the PIMG} and a $\sim 35 M_{\odot}$ BH, which would shorten the merger chain to generate GW231123. However, the predicted final spin in such a scenario is significantly lower than the measured primary spin of GW231123, see right panel of Figure~\ref{fig:vk}. { Note that we assume the first-generation BHs are slowly spinning as was found in \citet{2024PhRvL.133e1401L} \citep[though see also] []{2025arXiv250900154P,2025arXiv250810088C}.} 
Consequently, it is more plausible that both components of GW231123 were assembled through successive mergers of first-generation BHs below the PIMG.

\section{Conclusions and Discussion}\label{sec:diss}
GW231123 is the most massive BBH merger detected by LVK to date \citep{2023PhRvX..13d1039A}, making it one of the most intriguing events for tests of astrophysical and fundamental physics \citep{2025ApJ...993L..25A,2025arXiv250902047W,2025arXiv250903480T,2025arXiv250721788C}. 
{ We establish an association between GW231123 and the populations inferred from previous observations, exploring the formation and evolution of GW231123 in light of BBH subpopulation studies that were carried out with GWTC-3 \citep{2024PhRvL.133e1401L} \citep[see also e.g.,][for similar results]{2022ApJ...941L..39W, 2024arXiv240601679P, 2025PhRvL.134a1401A, 2025ApJ...987...65L, 2025PhRvD.112f3040A}, with further confirmation from GWTC-4 analyses \citep[e.g.,][]{2025arXiv250904151T,2025arXiv250904637A,2025arXiv250915646B,2025arXiv250923897L,2025arXiv251022698W}}.
We show that both components of GW231123 are best described by the high-spin subpopulation, which are natural for the higher-generation BHs \citep{2021NatAs...5..749G}. 
Besides, the spin magnitudes of the two components tend to be more consistent with the remnants of mergers with higher-generation BHs, as shown in Figure~\ref{fig:vk} (right panel). 
Therefore, it is more likely that both components of GW231123 have undergone successive mergers with several first-generation (stellar-collapse) BHs.

Using the distribution of first-generation BH masses from dynamical formation channels identified in GWTC-3 by \citet{2024ApJ...977...67L} (see also \citealt{2022ApJ...941L..39W, 2024arXiv240403166R}), we estimate that the primary (secondary) BH in GW231123 could be built up by the merger of $\sim6$ ($\sim4$) first-generation BHs. The assembly of such a massive system via hierarchical mergers requires that two key conditions be met. First, earlier merger remnants must be retained in the same environment to undergo subsequent mergers: the host's escape velocity must exceed the gravitational recoil (kick) velocity of each merger remnant. Hierarchical BH mergers impart substantially larger kicks than first-generation mergers (see the left panel of Figure~\ref{fig:vk}), so only environments with sufficiently deep potential wells--such as nuclear star clusters or AGN disks--could keep the GW231123 progenitors bound after each merger. Second, the binary orbits must be { hardened} efficiently so that multiple mergers can occur within a reasonable timescale. In dense stellar systems, dynamical interactions (e.g., binary-single encounters) help harden binaries, but an even more efficient mechanism is likely needed for rapid, repeated mergers. Gas dynamical friction in AGN disks can rapidly shrink binary orbits and shorten merger timescales \citep{2019PhRvL.123r1101Y}, and simulations suggest that this gas-assisted hardening may dominate the upper limit on BBH merger masses \citep{2024A&A...685A..51V, 2025arXiv250419570X}. Therefore, among known environments, AGN disks are the more plausible sites for producing BBH mergers as massive as GW231123.

After our initial manuscript appeared online, we noted that several alternative scenarios for GW231123 have also been discussed. \citet{2025arXiv250808558B} showed that accretion is also possible to produce mergers with near-equal, high-masses, and strongly aligned high spin mergers like GW231123, particularly in binaries embedded in AGN disks and Population~\uppercase\expandafter{\romannumeral 3} remnants. Meanwhile, \citet{2025arXiv250810088C} found that stellar evolution alone is sufficient to explain GW231123 without accretion (see also \citep{2025arXiv250904574B}).
The direct formation of GW231123 via Population~\uppercase\expandafter{\romannumeral 3} was also proposed by \citet{2025ApJ...993L..30L,2025arXiv250900154P}, and the possible primordial black hole origins were investigated by \citet{2025arXiv250809965D,2025PhRvD.112h1306Y}. 
{Additionally, some other explanations for the signal of GW231123 were also suggested \citep[see e.g.,][]{2025arXiv251105144H,2025arXiv251113820L,2025arXiv251007228R,2025arXiv251219077C,2025arXiv251220890Y,2025arXiv251217550H,2026arXiv260109678B}.
}

If AGN-disk channels indeed dominate the formation of GW231123-like events, then BBH merger masses up to $\sim\mathcal{O}(10^4)M_{\odot}$ might be achievable \citep{2024A&A...685A..51V}, which are well beyond the detection range of the current generation of ground-based detectors \citep{2015CQGra..32g4001L, 2015CQGra..32b4001A, 2018PTEP.2018a3F01A}. The next generation of gravitational-wave observatories, including planned space-based detectors \citep{2016CQGra..33c5010L,2017NSRev...4..685H,2017arXiv170200786A,2017CQGra..34d4001A,2010CQGra..27s4002P}, will be required to observe such extremely massive mergers, offering a more complete view of the BBH mass spectrum and evolutionary pathways. 
Beyond standard astrophysical channels, a few exotic formation scenarios have been proposed. For example, ``dark stars" (hypothetical early-universe stars powered by dark matter annihilation) could form IMBHs; however, the expected masses of such remnants are far larger than those of GW231123 \citep{2025ApJ...980..249L}. Another proposal involves a cosmological coupling mechanism that gradually increases BH masses, which in principle might produce BHs as massive as GW231123 \citep{2021ApJ...921L..22C}. This cosmological coupling hypothesis, however, has been ruled out by very recent JWST observations
\citep{2025arXiv250619589L}. Ultimately, a comprehensive population analysis of BBH mergers will be necessary to fully elucidate the origins and formation channels of GW231123 and similar exceptional events \citep[e.g.,][]{2021ApJ...917...33L, 2022ApJ...933L..14L, 2022PhRvD.105l3024F, 2023PhRvX..13a1048A, 2024PhRvL.133e1401L, 2024ApJ...975...54G, 2025arXiv250620731A}.


\begin{acknowledgments}
This work is supported by the National Natural Science Foundation of China (No. 12233011, No. 12503059, and No. 12303056), the General Fund (No. 2024M753495) of the China Postdoctoral Science Foundation, and the Priority Research Program of the Chinese Academy of Sciences (No. XDB0550400). This research has made use of data and software obtained from the Gravitational Wave Open Science Center (https://www.gw-openscience.org), a service of LIGO Laboratory, the LIGO Scientific Collaboration and the Virgo Collaboration. LIGO is funded by the U.S. National Science Foundation. Virgo is funded by the French Centre National de Recherche Scientifique (CNRS), the Italian Istituto Nazionale della Fisica Nucleare (INFN) and the Dutch Nikhef, with contributions by Polish and Hungarian institutes.
\end{acknowledgments}

\vspace{5mm}

\software{Bilby \citep[version 1.1.4, ascl:1901.011,      \url{https://git.ligo.org/lscsoft/bilby/}]{2019ascl.soft01011A},
          Dynesty \citep[version 1.0.1, \url{https://github.com/joshspeagle/dynesty}]{2020MNRAS.493.3132S},
          PyMultiNest \citep[version 2.11, ascl:1606.005, \url{https://github.com/JohannesBuchner/PyMultiNest}]{2016ascl.soft06005B},
          Precession \citep[version 1.0.3, \url{https://github.com/dgerosa/precession} ]{2016PhRvD..93l4066G}
          }


\appendix

\section{Reweighted by Population-informed Priors}\label{app:reweight}

{In hierarchical Bayesian inference, combining multiple events within a population allows the prior for individual event parameters to be informed by the learned population distribution \citep{2020ApJ...891L..31F,2020ApJ...895..128M}. This approach enables the self-consistent joint inference of both the population shape, such as mass and spin distributions, and the parameters of individual events \citep{2010PhRvD..81h4029M}. A comprehensive and rigorous approach, requires a population analysis that fits for the parameters of GW231123 together with full BBH observations \citep[see][for a technical guidance]{Callister:2021rew}. However, at the time of the initial submission of this paper, the data from GWTC-4 had not been fully released \footnote{Please refer \citet{2025arXiv250923897L,2025arXiv251022698W} for our consequent analysis when the full GWTC-4 data was released. The distribution of the populations are broadly consistent with previous for GWTC-3\citep{2024PhRvL.133e1401L}, except for the mass range of high-spin subpopulation. The difference is likely caused by the prior range of maximum mass, which was assumed to be $<100M_\odot$ in GWTC-3.}. Therefore, in this work we employ a simpler and less rigorous approach \citep[see also][for analysis of GW190521]{2020ApJ...904L..26F}, and adopt the population-informed prior that is given by the posterior population distribution inferred with GWTC-3 \citep{2024PhRvL.133e1401L},
}
\begin{equation}
P_{\mathcal{H}}(\lambda| d_{\rm GWTC-3}) \propto \int{p_{\mathcal{H}}(\lambda | \Lambda) p(\Lambda | d_{\rm GWTC-3}) {\rm d}\Lambda}, 
 \end{equation} 
To obtain the parameters of GW231123 ($\lambda_0$) under these priors, we reweight the posterior samples of LVK \citep{2025ApJ...993L..25A} by
\begin{equation}
 \frac{P_{\mathcal{H}}(\lambda_0| d_{\rm GWTC-3}) }{p_{\rm default}(\lambda_0)},
 \end{equation} 
 where ${p_{\rm default}(\lambda_0)}$ is the default prior for parameter estimation of GW231123 \citep{2025ApJ...993L..25A}.
We consider two different scenarios to construct population-informed priors. One ($P_{\rm HS-HS}(\chi_1,\chi_2| d_{\rm GWTC-3})$) is that both components of GW231123 belong to the high-spin (HS) subpopulation, the other ($P_{\rm HS-LS}(\chi_1,\chi_2| d_{\rm GWTC-3})$) is that primary / secondary BH belong to HS / low-spin (LS) subpopulation. The scenarios that the primary BH belong to the LS subpopulation are not considered, because the $\chi_1$ posterior samples of GW231123 are scarcely supported in distribution of HS subpopulation, which is also reflected by the Bayes factors between different priors (see Table~\ref{tab:reweighed_BF}).
{In principle, GW231123 could have been directly reweighted using the predictive distribution of the two-component population (i.e., the mixture of low-spin (LS) and high-spin (HS) components) rather than by manually assigning each black hole to a specific component. We did not adopt this approach because the two-component model in \citet{2024PhRvL.133e1401L} was constructed in the $m-\chi$ plane and its applicable mass range, as inferred from GWTC-3, is limited to $100M_\odot$. Instead, for GW231123 we adopted a uniform prior distribution for $m_1$ and $m_2$, similar to the default prior used in parameter estimation \citep{2025ApJ...993L..25A}.}
In both scenarios, the priors for spin tilt angles $P(\cos\theta_1,\cos\theta_2| d_{\rm GWTC-3})$ are the same as that inferred in \citet{2024PhRvL.133e1401L}. 
 
\begin{figure}
	\centering  
\includegraphics[width=0.5\linewidth]{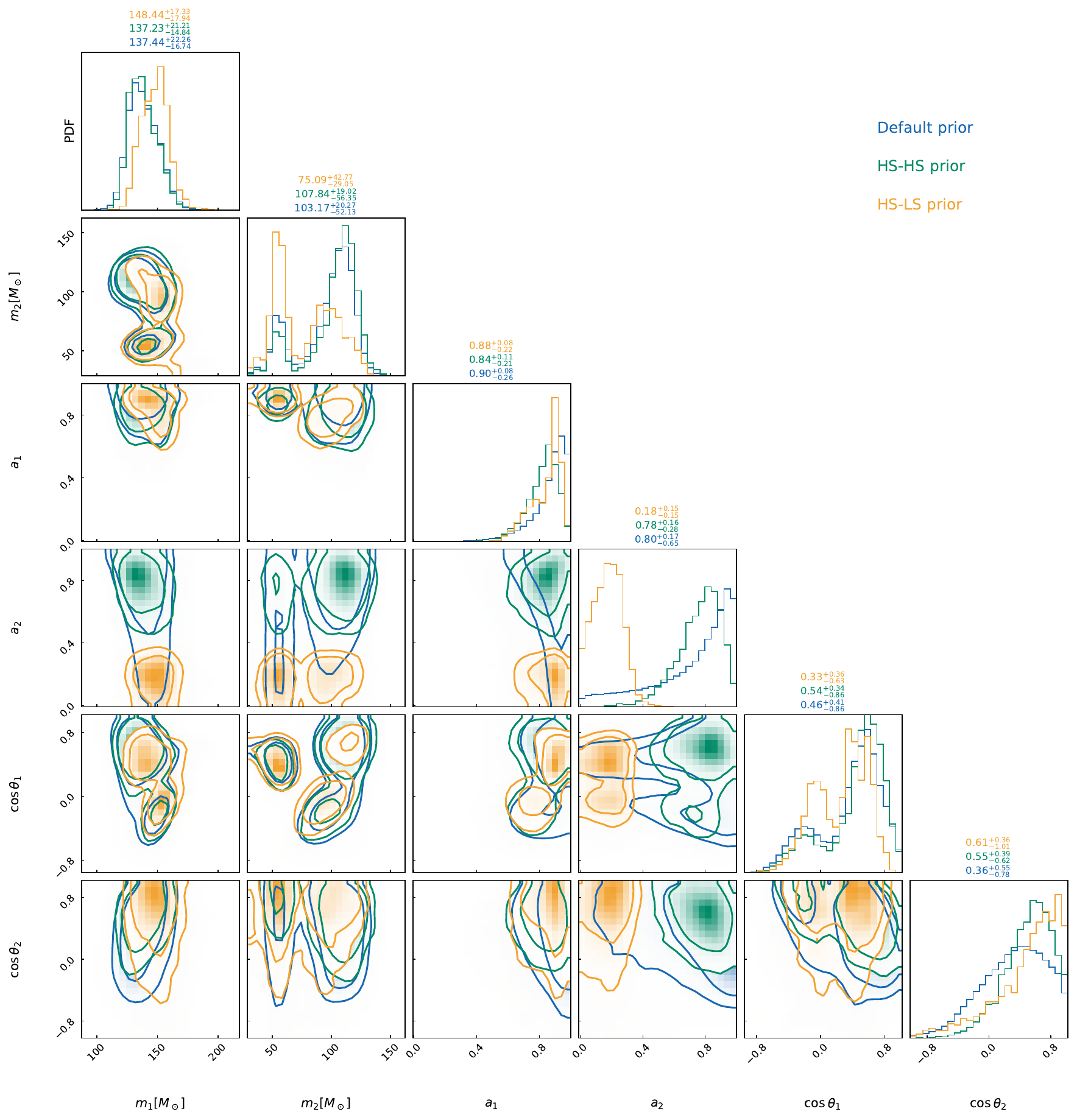}
\caption{Posterior distributions of component masses, spin magnitudes, and spin orientations of GW231123 under different priors.}
\label{fig:corner}
\end{figure}

Figure~\ref{fig:corner} shows the reweighed posterior samples compared to the samples under default prior. 
We find that under HS-HS prior, the $m_2$ shows more / less support for the high-mass / low-mass peak. Whereas $m_2$ shows more support for the peak at $\sim 60M_{\odot}$ under the HS-LS prior. If the secondary of GW231123 was a $\sim 60M_{\odot}$ and slowly spinning BH, it would support the existence of a group of low-spin $\sim 50 - 70 M_{\odot}$ BHs \citep{2025arXiv251022698W}. However the HS-LS prior is less favored compared to the HS-HS prior, as show in Table~\ref{tab:reweighed_BF}.

\section{Bayes Factors between population-informed Priors}\label{app:BF}

In this section, we perform a systematic comparison of the different spin priors discussed in the main text via the computation of their Bayes factors using GW231123 data.
The aim of this section is not to identify the parameter distribution that best fits the data -- such an endeavor would require a joint analysis of multiple BBH events -- but rather to evaluate the plausibility of a given prior assumption $\mathcal{H}$: GW231123 belong to one of the inferred subpopulation from GWTC-3, $p_{\mathcal{H}}(\lambda |d_{\rm GWTC-3})$, in the context of the single-event likelihood $p(d_{\rm GW231123} | \lambda)$ \citep[see also][for GW190521]{2020ApJ...904L..26F}. 
 Then the Bayesian evidence for $\mathcal{H}$ given data $d_{\rm GW231123}$ is 
\begin{equation}\label{eq:BF}
p(d_{\rm GW231123} | \mathcal{H}, d_{\rm GWTC-3} ) = \int{ p(d_{\rm GW231123} | \lambda) p_{\mathcal{H}}(\lambda| \Lambda){\rm d}\lambda p(\Lambda | d_{\rm GWTC-3}) {\rm d}\Lambda },
\end{equation} 
We calculate Eq.(\ref{eq:BF}) by Monte Carlo integration (or importance sampling):
\begin{equation}
p(d_{\rm GW231123} | \mathcal{H}, d_{\rm GWTC-3} ) = 
\left\langle \left\langle \frac{p_{\mathcal{H}}(\lambda | \Lambda)}{p_{\rm default}(\lambda)} \right\rangle_{\{\lambda\}}\right\rangle_{\{\Lambda\} | d_{\rm GWTC-3}}.
\end{equation}
The Bayes factor of $\mathcal{H}_{A}$ over $\mathcal{H}_{B}$ is $\mathcal{B}^{A}_{B}=\frac{p(d_{\rm GW231123} | \mathcal{H}_{A}, d_{\rm GWTC-3} )}{p(d_{\rm GW231123} | \mathcal{H}_{B}, d_{\rm GWTC-3} )}$. The Bayes factors between hypotheses that the two components of GW231123 belong to LS or HS subpopulation are summarized in Table~\ref{tab:reweighed_BF}. 
We find that both BHs of GW231123 belong to the HS subpopulation is most favored, and the hypothesis that the primary BH belongs to the LS subpopulation is ruled out.

\begin{table*}[htpb]
\centering
\caption{Bayes factors  of hypothesis $\mathcal{H}_x$ relative to the hypothesis that both components belong to the high-spin subpopulation $\mathcal{H}_{\rm HS-HS}$.}\label{tab:reweighed_BF}
\begin{tabular}{c|cc}
\hline
\hline
 $\log_{10}\mathcal{B}_{\rm HS-HS}^x$ & \multicolumn{2}{c}{BH$_1$ Spin magnitude} \\
\cline{2-3}
BH$_2$ Spin magnitude & $\chi_1 \sim P_{\rm LS}(\chi | d_{\rm GWTC-3} )$ & $\chi_1 \sim P_{\rm HS}(\chi  | d_{\rm GWTC-3} )$ \\
\hline
$\chi_2 \sim P_{\rm LS}(\chi  | d_{\rm GWTC-3} )$ & $-2.7$ & ${-0.8}$ \\
$\chi_2 \sim P_{\rm HS}(\chi  | d_{\rm GWTC-3} )$ & $-2.7$ & $0.0$ \\
\hline
\hline
\end{tabular}
\end{table*}

\section{Mass function of higher-generation BHs}
\subsection{Details about mass function of different generations of BHs}\label{app:generation}
{To generate the mass functions of the BHs made by $n$ stellar-mass BHs, we randomly draw $n$ BHs from the potential first-generation subpopulation of dynamical formation channels \citep{2024ApJ...977...67L}. Since the mass function of the subpopulation, where we draw BHs, has a large uncertainty, especially in the lower range as shown in Figure~\ref{fig:check-dist}. We randomly choose 500 independent mass functions from the posterior samples obtained in \citet{2024ApJ...977...67L}, and then randomly draw 100 BHs for each mass function, so that totally 50000 BHs are generated, which are served as the ingredient BHs. For the remnant BH made by $n$ stellar-mass BHs, we sum the masses of $n$ BHs that are randomly drawn from ingredient sample. 
} 
In practice, we neglect the mass loss due to GW radiation because it only reduces the remnant mass by a few percent, an effect that is small compared to current measurement uncertainties. We also neglect mass accretion in plausible gas‐rich environments (e.g., AGN disks), which could increase BH masses by $\sim10\%$-$20\%$ \citep{2020ApJ...898...25T,2025arXiv250419570X,2024A&A...685A..51V}. Including these effects would not alter our qualitative conclusions but would slightly change the number (at most one) of first‐generation progenitors required to assemble GW231123.  

\subsection{Self-consistency check for BH generations}\label{app:check}
{It is the purpose of \citet{2024PhRvL.133e1401L} to resolve the first-generation and higher-generation subpopulations of BHs, whereas the purpose of \citet{2024ApJ...977...67L} is to resolve the isolated channels and dynamical channels (which contribute to hierarchical mergers) for the first-generation mergers. Therefore, the first-generation subpopulation obtained in \citet{2024PhRvL.133e1401L} are not expected to stand for the ingredient BHs to generate the higher-generation BHs, since the mergers of isolated formation channels are hard to contribute to hierarchical mergers. In \citet{2024ApJ...977...67L}, the BBHs from potential isolated channels (referred to the aligned assembly) are separated from the potential dynamical channels (referred to the isotropic assembly). Therefore the potential subpopulation of the first-generation dynamically formed BHs, are reasonably more appropriate to stand for the ingredient BHs to generate the higher-generation BHs. 

Note that the mass function of the dynamical first-generation subpopulation adopted by \citet{2024ApJ...977...67L} is consistent with the prediction from many simulations \citep[e.g.][]{2020PhRvD.102l3016A,2022MNRAS.517.2953T,2023MNRAS.522..466A,2025arXiv250608801F}. This also supports us that the generated functions for the higher-generation BHs are reliable. 
}

\begin{figure*}
	\centering  
\includegraphics[width=0.8\linewidth]{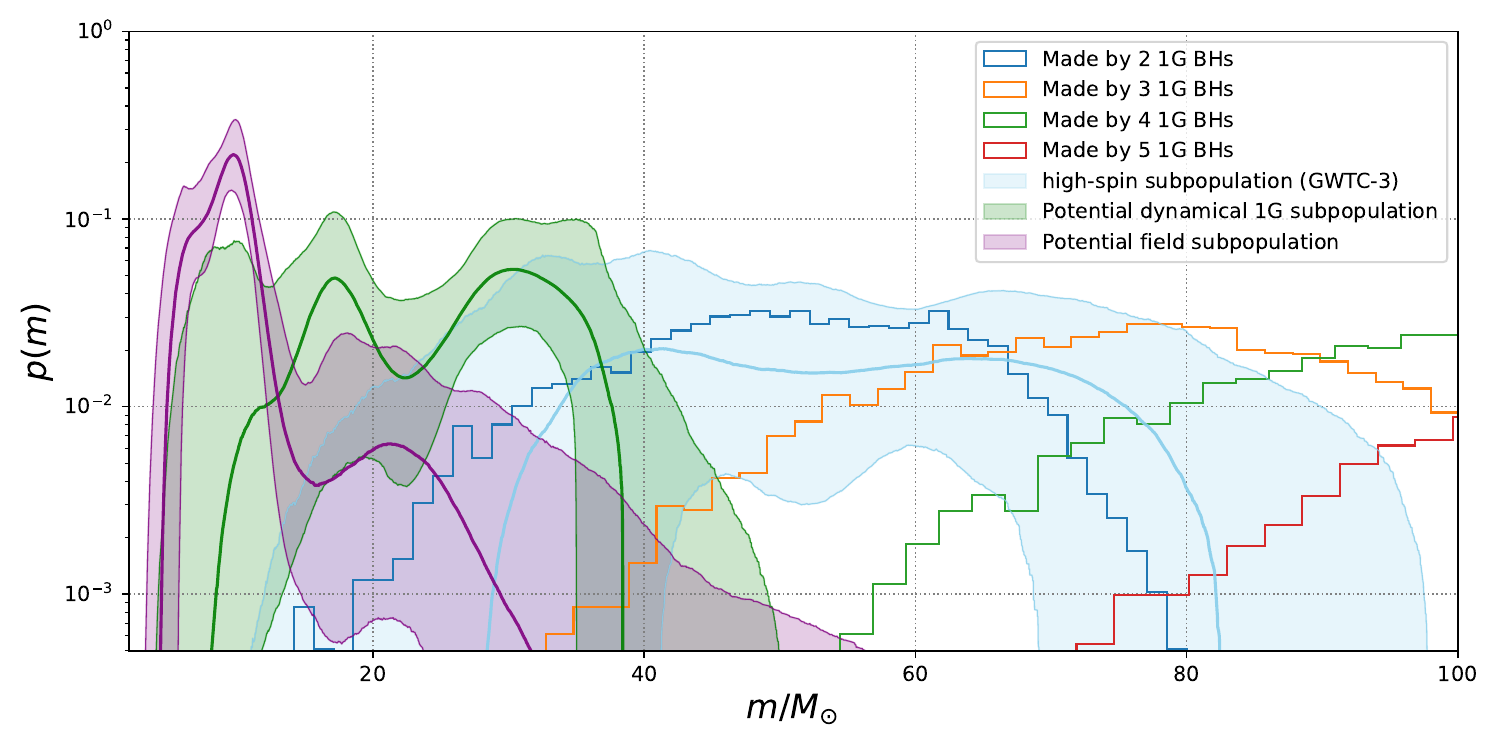}
\includegraphics[width=0.8\linewidth]{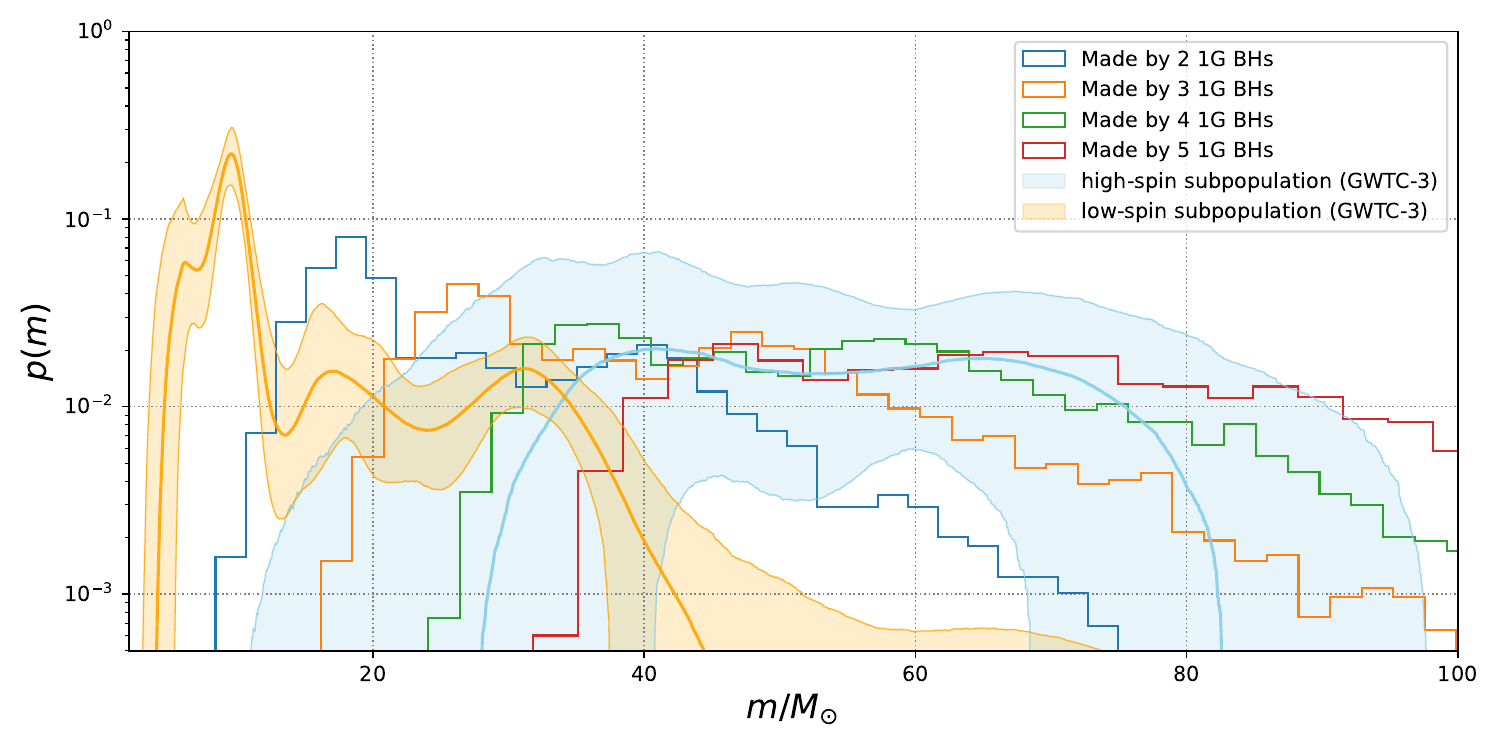}
\caption{Mass distributions of the subpopulations comparing to the remnants of mergers drawn from the potential first-generation subpopulation. Top: the first-generation BHs are drawn from the potential {\it dynamical} first-generation subpopulation (green region), the subpopulation in purple are potentially associated with {\it field evolution channels}. Bottom: the first-generation BHs are drawn from the low-spin subpopulation (orange region), which is likely the mixture of dynamical and field channels. Note that each distribution of subpopulation is normalized, and the shaded regions are for $90\%$ credible levels.}
\label{fig:check-dist}
\end{figure*}

To {further} validate our choice of the first-generation mass distribution in dynamical formation channels, we compare the mass distribution of remnants made by multiple ($n$) first-generation BHs with those of  subpopulations revealed in \citet{2024ApJ...977...67L} and \citet{2024PhRvL.133e1401L}. As shown in the top panel of Figure~\ref{fig:check-dist}, if we assume the 1G BHs are drawn from the potential dynamical 1G subpopulation (green shaded area) revealed in \citet{2024ApJ...977...67L}, remnants from $\gtrsim2$ progenitors can well reproduce the higher-generation subpopulation (blue shaded area). Additionally, the higher-generation BHs are dominated by the remnants made by 2 1G BHs, which is natural for the scenario in hierarchical mergers.
However, if we assume the first-generation BHs are drawn from the low-spin subpopulation (orange shaded area) as revealed in \citet{2024PhRvL.133e1401L}, then the mass distribution of remnants made by 2 1G BHs would peak at $\sim 20 M_{\odot}$, which is inconsistent with the distribution of the higher-generation subpopulation (blue shaded area), see the bottom panel of Figure~\ref{fig:check-dist}. This discrepancy further supports our adoption of the dynamical 1G subpopulation for modeling hierarchical growth.

\bibliography{export-bibtex}{}

\begin{thebibliography}{}
\expandafter\ifx\csname natexlab\endcsname\relax\def\natexlab#1{#1}\fi
\providecommand{\url}[1]{\href{#1}{#1}}
\providecommand{\dodoi}[1]{doi:~\href{http://doi.org/#1}{\nolinkurl{#1}}}
\providecommand{\doeprint}[1]{\href{http://ascl.net/#1}{\nolinkurl{http://ascl.net/#1}}}
\providecommand{\doarXiv}[1]{\href{https://arxiv.org/abs/#1}{\nolinkurl{https://arxiv.org/abs/#1}}}

\bibitem[{{Abac} {et~al.}(2025){Abac}, {Abouelfettouh}, {Acernese}, {Ackley},
  {Adamcewicz}, {Adhicary}, {Adhikari}, {Adhikari}, {Adhikari}, {Adkins},
  {Afroz}, {Agapito}, {Agarwal}, {Agathos}, {Aggarwal}, {Aggarwal}, {Aguiar},
  {Ahrend}, {Aiello}, {Ain}, {Ajith}, {Akutsu}, {Albanesi}, {Ali}, {Al-Kershi},
  {All{\'e}n{\'e}}, {Allocca}, {Al-Shammari}, {Altin}, {Alvarez-Lopez}, {Amar},
  {Amarasinghe}, {Amato}, {Amicucci}, {Amra}, {Ananyeva}, {Anderson},
  {Anderson}, {Andia}, {Ando}, {Andr{\'e}s-Carcasona}, {Andri{\'c}}, {Anglin},
  {Ansoldi}, {Antelis}, {Antier}, {Aoumi}, {Appavuravther}, {Appert}, {Apple},
  {Arai}, {Alvarez}, {Araya}, {Araya}, {Arca Sedda}, {Areeda}, {Aritomi},
  {Armato}, {Armstrong}, {Arnaud}, {Arogeti}, {Aronson}, {Arun}, {Ashton},
  {Aso}, {Asprea}, {Assiduo}, {Assis de Souza Melo}, {Aston}, {Astone},
  {Attadio}, {Aubin}, {AultONeal}, {Avallone}, {Avila}, {Babak}, {Badger},
  {Bae}, {Bagnasco}, {Baiotti}, {Bajpai}, {Baka}, {Baker}, {Baker}, {Baker},
  {Baldi}, {Baldicchi}, {Ball}, {Ballardin}, {Ballmer}, {Banagiri}, {Banerjee},
  {Bankar}, {Baptiste}, {Baral}, {Baratti}, {Barayoga}, {Barish}, {Barker},
  {Barman}, {Barneo}, {Barone}, {Barr}, {Barsotti}, {Barsuglia}, {Barta},
  {Bartoletti}, {Barton}, {Bartos}, {Basalaev}, {Bassiri}, {Basti}, {Bawaj},
  {Baxi}, {Bayley}, {Baylor}, {Baynard}, {Bazzan}, {Bedakihale}, {Beirnaert},
  {Bejger}, {Belardinelli}, {Bell}, {Bellie}, {Bellizzi}, {Benoit}, {Bentara},
  {Bentley}, {Ben Yaala}, {Bera}, {Bergamin}, {Berger}, {Bernuzzi}, {Beroiz},
  {Berry}, {Bersanetti}, {Bertheas}, {Bertolini}, {Betzwieser}, {Beveridge},
  {Bevilacqua}, {Bevins}, {Bhandare}, {Bhatt}, {Bhattacharjee},
  {Bhattacharyya}, {Bhaumik}, {Bhagwat}, {Biancalana}, {Bianchi}, {Bilenko},
  {Billingsley}, {Binetti}, {Bini}, {Binu}, {Biot}, {Birnholtz}, {Biscoveanu},
  {Bisht}, {Bitossi}, {Bizouard}, {Blaber}, {Blackburn}, {Blagg}, {Blair},
  {Blair}, {Bode}, {Boettner}, {Boileau}, {Boldrini}, {Bolingbroke},
  {Bolliand}, {Bonavena}, {Bondarescu}, {Bondu}, {Bonilla}, {Bonilla},
  {Bonino}, {Bonnand}, {Borchers}, {Borhanian}, {Boschi}, {Bose}, {Bossilkov},
  {Bothra}, {Boudon}, {Bourg}, {Boyle}, {Bozzi}, {Bradaschia}, {Brady},
  {Branch}, {Branchesi}, {Braun}, {Briant}, {Brillet}, {Brinkmann}, {Brockill},
  {Brockmueller}, \& {Brooks}}]{2025ApJ...993L..25A}
{Abac}, A.~G., {Abouelfettouh}, I., {Acernese}, F., {et~al.} 2025, \apjl, 993,
  L25, \dodoi{10.3847/2041-8213/ae0c9c}

\bibitem[{{Abbott} {et~al.}(2017){Abbott}, {Abbott}, {Abbott}, {Abernathy},
  {Ackley}, {Adams}, {Addesso}, {Adhikari}, {Adya}, {Affeldt}, {Aggarwal},
  {Aguiar}, {Ain}, {Ajith}, {Allen}, {Altin}, {Anderson}, {Anderson}, {Arai},
  {Araya}, {Arceneaux}, {Areeda}, {Arun}, {Ashton}, {Ast}, {Aston}, {Aufmuth},
  {Aulbert}, {Babak}, {Baker}, {Ballmer}, {Barayoga}, {Barclay}, {Barish},
  {Barker}, {Barr}, {Barsotti}, {Bartlett}, {Bartos}, {Bassiri}, {Batch},
  {Baune}, {Bell}, {Berger}, {Bergmann}, {Berry}, {Betzwieser}, {Bhagwat},
  {Bhandare}, {Bilenko}, {Billingsley}, {Birch}, {Birney}, {Biscans}, {Bisht},
  {Biwer}, {Blackburn}, {Blair}, {Blair}, {Blair}, {Bock}, {Bogan}, {Bohe},
  {Bond}, {Bork}, {Bose}, {Brady}, {Braginsky}, {Brau}, {Brinkmann},
  {Brockill}, {Broida}, {Brooks}, {Brown}, {Brown}, {Brown}, {Brunett},
  {Buchanan}, {Buikema}, {Buonanno}, {Byer}, {Cabero}, {Cadonati}, {Cahillane},
  {Calder{\'o}n Bustillo}, {Callister}, {Camp}, {Cannon}, {Cao}, {Capano},
  {Caride}, {Caudill}, {Cavagli{\`a}}, {Cepeda}, {Chamberlin}, {Chan}, {Chao},
  {Charlton}, {Cheeseboro}, {Chen}, {Chen}, {Cheng}, {Cho}, {Cho}, {Chow},
  {Christensen}, {Chu}, {Chung}, {Ciani}, {Clara}, {Clark}, {Collette},
  {Cominsky}, {Constancio}, {Cook}, {Corbitt}, {Cornish}, {Corsi}, {Costa},
  {Coughlin}, {Coughlin}, {Countryman}, {Couvares}, {Cowan}, {Coward},
  {Cowart}, {Coyne}, {Coyne}, {Craig}, {Creighton}, {Cripe}, {Crowder},
  {Cumming}, {Cunningham}, {Dal Canton}, {Danilishin}, {Danzmann}, {Darman},
  {Dasgupta}, {Da Silva Costa}, {Dave}, {Davies}, {Daw}, {De}, {DeBra}, {Del
  Pozzo}, {Denker}, {Dent}, {Dergachev}, {DeRosa}, {DeSalvo}, {Devine},
  {Dhurandhar}, {D{\'\i}az}, {Di Palma}, {Donovan}, {Dooley}, {Doravari},
  {Douglas}, {Downes}, {Drago}, {Drever}, {Driggers}, {Dwyer}, {Edo},
  {Edwards}, {Effler}, {Eggenstein}, {Ehrens}, {Eichholz}, {Eikenberry},
  {Engels}, {Essick}, {Etzel}, {Evans}, {Evans}, {Everett}, {Factourovich},
  {Fair}, {Fairhurst}, {Fan}, {Fang}, {Farr}, {Farr}, {Favata}, {Fays},
  {Fehrmann}, {Fejer}, {Fenyvesi}, {Ferreira}, {Fisher}, {Fletcher}, {Frei},
  {Freise}, {Frey}, {Fritschel}, {Frolov}, {Fulda}, {Fyffe}, {Gabbard}, {Gair},
  {Gaonkar}, {Gaur}, {Gehrels}, {Geng}, {George}, {Gergely}, {Ghosh}, {Ghosh},
  {Giaime}, {Giardina}, {Gill}, {Glaefke}, {Goetz}, {Goetz}, {Gondan},
  {Gonz{\'a}lez}, {Gopakumar}, {Gordon}, {Gorodetsky}, {Gossan}, {Graef},
  {Graff}, {Grant}, {Gras}, {Gray}, {Green}, {Grote}, {Grunewald}, {Guo},
  {Gupta}, {Gupta}, {Gushwa}, {Gustafson}, {Gustafson}, {Hacker}, {Hall},
  {Hall}, {Hammond}, {Haney}, {Hanke}, {Hanks}, {Hanna}, {Hannam}, {Hanson},
  {Hardwick}, {Harry}, {Harry}, {Hart}, {Hartman}, {Haster}, {Haughian},
  {Heintze}, {Hendry}, {Heng}, {Hennig}, {Henry}, {Heptonstall}, {Heurs},
  {Hild}, {Hoak}, {Holt}, {Holz}, {Hopkins}, {Hough}, {Houston}, {Howell},
  {Hu}, {Huang}, {Huerta}, {Hughey}, {Husa}, {Huttner}, {Huynh-Dinh}, {Indik},
  {Ingram}, {Inta}, {Isa}, {Isi}, {Isogai}, {Iyer}, {Izumi}, {Jang}, {Jani},
  {Jawahar}, {Jian}, {Jim{\'e}nez-Forteza}, {Johnson}, {Jones}, {Jones}, {Ju},
  {Haris}, {Kalaghatgi}, {Kalogera}, {Kandhasamy}, {Kang}, {Kanner}, {Kapadia},
  {Karki}, {Karvinen}, {Kasprzack}, {Katsavounidis}, {Katzman}, {Kaufer},
  {Kaur}, {Kawabe}, {Kehl}, {Keitel}, {Kelley}, {Kells}, {Kennedy}, {Key},
  {Khalili}, {Khan}, {Khan}, {Khazanov}, {Kijbunchoo}, {Kim}, {Kim}, {Kim},
  {Kim}, {Kim}, {Kim}, {Kim}, {Kimbrell}, {King}, {King}, {Kissel}, {Klein},
  {Kleybolte}, {Klimenko}, {Koehlenbeck}, {Kondrashov}, {Kontos}, {Korobko},
  {Korth}, {Kozak}, {Kringel}, {Krueger}, {Kuehn}, {Kumar}, {Kumar}, {Kuo},
  {Lackey}, {Landry}, {Lange}, {Lantz}, {Lasky}, {Laxen}, {Lazzarini},
  {Leavey}, {Lebigot}, {Lee}, {Lee}, {Lee}, {Lee}, {Lenon}, {Leong}, {Levin},
  {Lewis}, {Li}, {Libson}, {Littenberg}, {Lockerbie}, {Lombardi}, {London},
  {Lord}, {Lormand}, {Lough}, {L{\"u}ck}, {Lundgren}, {Lynch}, {Ma},
  {Machenschalk}, {MacInnis}, {Macleod}, {Maga{\~n}a-Sandoval}, {Maga{\~n}a
  Zertuche}, {Magee}, {Mandic}, {Mangano}, {Mansell}, {Manske}, {M{\'a}rka},
  {M{\'a}rka}, {Markosyan}, {Maros}, {Martin}, {Martynov}, {Mason},
  {Massinger}, {Masso-Reid}, {Matichard}, {Matone}, {Mavalvala}, {Mazumder},
  {McCarthy}, {McClelland}, {McCormick}, {McGuire}, {McIntyre}, {McIver},
  {McManus}, {McRae}, {McWilliams}, {Meacher}, {Meadors}, {Melatos}, {Mendell},
  {Mercer}, {Merilh}, {Meshkov}, {Messenger}, {Messick}, {Meyers}, {Miao},
  {Middleton}, {Mikhailov}, {Miller}, {Miller}, {Miller}, {Miller},
  {Millhouse}, {Ming}, {Mirshekari}, {Mishra}, {Mitra}, {Mitrofanov},
  {Mitselmakher}, {Mittleman}, {Mohapatra}, {Moore}, {Moore}, {Moraru},
  {Moreno}, {Morriss}, {Mossavi}, {Mow-Lowry}, {Mueller}, {Muir}, {Mukherjee},
  {Mukherjee}, {Mukherjee}, {Mukund}, {Mullavey}, {Munch}, {Murphy}, {Murray},
  {Mytidis}, {Nayak}, {Nedkova}, {Nelson}, {Neunzert}, {Newton}, {Nguyen},
  {Nielsen}, {Nitz}, {Nolting}, {Normandin}, {Nuttall}, {Oberling}, {Ochsner},
  {O'Dell}, {Oelker}, {Ogin}, {Oh}, {Oh}, {Ohme}, {Oliver}, {Oppermann},
  {Oram}, {O'Reilly}, {O'Shaughnessy}, {Ottaway}, {Overmier}, {Owen}, {Pai},
  {Pai}, {Palamos}, {Palashov}, {Pal-Singh}, {Pan}, {Pankow}, {Pannarale},
  {Pant}, {Papa}, {Paris}, {Parker}, {Pascucci}, {Patrick}, {Pearlstone},
  {Pedraza}, {Pekowsky}, {Pele}, {Penn}, {Perreca}, {Perri}, {Phelps},
  {Pierro}, {Pinto}, {Pitkin}, {Poe}, {Post}, {Powell}, {Prasad}, {Predoi},
  {Prestegard}, {Price}, {Prijatelj}, {Principe}, {Privitera}, {Prokhorov},
  {Puncken}, {P{\"u}rrer}, {Qi}, {Qin}, {Qiu}, {Quetschke}, {Quintero},
  {Quitzow-James}, {Raab}, {Rabeling}, {Radkins}, {Raffai}, {Raja}, {Rajan},
  {Rakhmanov}, {Raymond}, {Read}, {Reed}, {Reid}, {Reitze}, {Rew}, {Reyes},
  {Riles}, {Rizzo}, {Robertson}, {Robie}, {Rollins}, {Roma}, {Romanov},
  {Romie}, {Rowan}, {R{\"u}diger}, {Ryan}, {Sachdev}, {Sadecki}, {Sadeghian},
  {Sakellariadou}, {Saleem}, {Salemi}, {Samajdar}, {Sammut}, {Sanchez},
  {Sandberg}, {Sandeen}, {Sanders}, {Sathyaprakash}, {Saulson}, {Sauter},
  {Savage}, {Sawadsky}, {Schale}, {Schilling}, {Schmidt}, {Schmidt},
  {Schnabel}, {Schofield}, {Sch{\"o}nbeck}, {Schreiber}, {Schuette}, {Schutz},
  {Scott}, {Scott}, {Sellers}, {Sengupta}, {Sergeev}, {Shaddock}, {Shaffer},
  {Shahriar}, {Shaltev}, {Shapiro}, {Shawhan}, {Sheperd}, {Shoemaker},
  {Shoemaker}, {Siellez}, {Siemens}, {Sigg}, {Silva}, {Singer}, {Singer},
  {Singh}, {Singh}, {Sintes}, {Slagmolen}, {Smith}, {Smith}, {Smith}, {Son},
  {Sorazu}, {Souradeep}, {Srivastava}, {Staley}, {Steinke}, {Steinlechner},
  {Steinlechner}, {Steinmeyer}, {Stephens}, {Stone}, {Strain}, {Strauss},
  {Strigin}, {Sturani}, {Stuver}, {Summerscales}, {Sun}, {Sunil}, {Sutton},
  {Szczepa{\'n}czyk}, {Talukder}, {Tanner}, {T{\'a}pai}, {Tarabrin},
  {Taracchini}, {Taylor}, {Theeg}, {Thirugnanasambandam}, {Thomas}, {Thomas},
  {Thomas}, {Thorne}, {Thrane}, {Tiwari}, {Tokmakov}, {Toland}, {Tomlinson},
  {Tornasi}, {Torres}, {Torrie}, {T{\"o}yr{\"a}}, {Traylor}, {Trifir{\`o}},
  {Tse}, {Tuyenbayev}, {Ugolini}, {Unnikrishnan}, {Urban}, {Usman},
  {Vahlbruch}, {Vajente}, {Valdes}, {Vander-Hyde}, {van Veggel}, {Vass},
  {Vaulin}, {Vecchio}, {Veitch}, {Veitch}, {Venkateswara}, {Vinciguerra},
  {Vine}, {Vitale}, {Vo}, {Vorvick}, {Voss}, {Vousden}, {Vyatchanin}, {Wade},
  {Wade}, {Wade}, {Walker}, {Wallace}, {Walsh}, {Wang}, {Wang}, {Wang}, {Wang},
  {Ward}, {Warner}, {Weaver}, {Weinert}, {Weinstein}, {Weiss}, {Wen},
  {We{\ss}els}, {Westphal}, {Wette}, {Whelan}, {Whiting}, {Williams},
  {Williamson}, {Willis}, {Willke}, {Wimmer}, {Winkler}, {Wipf}, {Wittel},
  {Woan}, {Woehler}, {Worden}, {Wright}, {Wu}, {Wu}, {Yablon}, {Yam},
  {Yamamoto}, {Yancey}, {Yu}, {Zanolin}, {Zevin}, {Zhang}, {Zhang}, {Zhang},
  {Zhao}, {Zhou}, {Zhou}, {Zhu}, {Zucker}, {Zuraw}, {Zweizig}, {(LIGO
  Scientific Collaboration}, \& {Harms}}]{2017CQGra..34d4001A}
{Abbott}, B.~P., {Abbott}, R., {Abbott}, T.~D., {et~al.} 2017, Classical and
  Quantum Gravity, 34, 044001, \dodoi{10.1088/1361-6382/aa51f4}

\bibitem[{{Abbott} {et~al.}(2019){Abbott}, {Abbott}, {Abbott}, {Abraham},
  {Acernese}, {Ackley}, {Adams}, {Adhikari}, {Adya}, {Affeldt}, {Agathos},
  {Agatsuma}, {Aggarwal}, {Aguiar}, {Aiello}, {Ain}, {Ajith}, {Allen},
  {Allocca}, {Aloy}, {Altin}, {Amato}, {Ananyeva}, {Anderson}, {Anderson},
  {Angelova}, {Antier}, {Appert}, {Arai}, {Araya}, {Areeda}, {Ar{\`e}ne},
  {Arnaud}, {Arun}, {Ascenzi}, {Ashton}, {Aston}, {Astone}, {Aubin}, {Aufmuth},
  {AultONeal}, {Austin}, {Avendano}, {Avila-Alvarez}, {Babak}, {Bacon},
  {Badaracco}, {Bader}, {Bae}, {Baker}, {Baldaccini}, {Ballardin}, {Ballmer},
  {Banagiri}, {Barayoga}, {Barclay}, {Barish}, {Barker}, {Barkett}, {Barnum},
  {Barone}, {Barr}, {Barsotti}, {Barsuglia}, {Barta}, {Bartlett}, {Bartos},
  {Bassiri}, {Basti}, {Bawaj}, {Bayley}, {Bazzan}, {B{\'e}csy}, {Bejger},
  {Belahcene}, {Bell}, {Beniwal}, {Berger}, {Bergmann}, {Bernuzzi}, {Bero},
  {Berry}, {Bersanetti}, {Bertolini}, {Betzwieser}, {Bhandare}, {Bidler},
  {Bilenko}, {Bilgili}, {Billingsley}, {Birch}, {Birney}, {Birnholtz},
  {Biscans}, {Biscoveanu}, {Bisht}, {Bitossi}, {Bizouard}, {Blackburn},
  {Blackman}, {Blair}, {Blair}, {Blair}, {Bloemen}, {Bode}, {Boer}, {Boetzel},
  {Bogaert}, {Bondu}, {Bonilla}, {Bonnand}, {Booker}, {Boom}, {Booth}, {Bork},
  {Boschi}, {Bose}, {Bossie}, {Bossilkov}, {Bosveld}, {Bouffanais}, {Bozzi},
  {Bradaschia}, {Brady}, {Bramley}, {Branchesi}, {Brau}, {Briant}, {Briggs},
  {Brighenti}, {Brillet}, {Brinkmann}, {Brisson}, {Brockill}, {Brooks},
  {Brown}, {Brunett}, {Buikema}, {Bulik}, {Bulten}, {Buonanno}, {Buskulic},
  {Bustamante Rosell}, {Buy}, {Byer}, {Cabero}, {Cadonati}, {Cagnoli},
  {Cahillane}, {Calder{\'o}n Bustillo}, {Callister}, {Calloni}, {Camp},
  {Campbell}, {Canepa}, {Cannon}, {Cao}, {Cao}, {Capocasa}, {Carbognani},
  {Caride}, {Carney}, {Carullo}, {Casanueva Diaz}, {Casentini}, {Caudill},
  {Cavagli{\`a}}, {Cavalier}, {Cavalieri}, {Cella}, {Cerd{\'a}-Dur{\'a}n},
  {Cerretani}, {Cesarini}, {Chaibi}, {Chakravarti}, {Chamberlin}, {Chan},
  {Chao}, {Charlton}, {Chase}, {Chassande-Mottin}, {Chatterjee}, {Chaturvedi},
  {Chatziioannou}, {Cheeseboro}, {Chen}, {Chen}, {Chen}, {Cheng}, {Cheong},
  {Chia}, {Chincarini}, {Chiummo}, {Cho}, {Cho}, {Cho}, {Christensen}, {Chu},
  {Chua}, {Chung}, {Chung}, {Ciani}, {Ciobanu}, {Ciolfi}, {Cipriano}, {Cirone},
  {Clara}, {Clark}, {Clearwater}, {Cleva}, {Cocchieri}, {Coccia}, {Cohadon},
  {Cohen}, {Colgan}, {Colleoni}, {Collette}, {Collins}, {Cominsky},
  {Constancio}, {Conti}, {Cooper}, {Corban}, {Corbitt}, {Cordero-Carri{\'o}n},
  {Corley}, {Cornish}, {Corsi}, {Cortese}, {Costa}, {Cotesta}, {Coughlin},
  {Coughlin}, {Coulon}, {Countryman}, {Couvares}, {Covas}, {Cowan}, {Coward},
  {Cowart}, {Coyne}, {Coyne}, {Creighton}, {Creighton}, {Cripe}, {Croquette},
  {Crowder}, {Cullen}, {Cumming}, {Cunningham}, {Cuoco}, {Canton}, {D{\'a}lya},
  {Danilishin}, {D'Antonio}, {Danzmann}, {Dasgupta}, {Da Silva Costa},
  {Datrier}, {Dattilo}, {Dave}, {Davier}, {Davis}, {Daw}, {DeBra},
  {Deenadayalan}, {Degallaix}, {De Laurentis}, {Del{\'e}glise}, {Del Pozzo},
  {DeMarchi}, {Demos}, {Dent}, {De Pietri}, {Derby}, {De Rosa}, {De Rossi},
  {DeSalvo}, {de Varona}, {Dhurandhar}, {D{\'\i}az}, {Dietrich}, {Di Fiore},
  {Di Giovanni}, {Di Girolamo}, {Di Lieto}, {Ding}, {Di Pace}, {Di Palma}, {Di
  Renzo}, {Dmitriev}, {Doctor}, {Donovan}, {Dooley}, {Doravari}, {Dorrington},
  {Downes}, {Drago}, {Driggers}, {Du}, {Ducoin}, {Dupej}, {Dwyer}, {Easter},
  {Edo}, {Edwards}, {Effler}, {Ehrens}, {Eichholz}, {Eikenberry}, {Eisenmann},
  {Eisenstein}, {Essick}, {Estelles}, {Estevez}, {Etienne}, {Etzel}, {Evans},
  {Evans}, {Fafone}, {Fair}, {Fairhurst}, {Fan}, {Farinon}, {Farr}, {Farr},
  {Fauchon-Jones}, {Favata}, {Fays}, {Fazio}, {Fee}, {Feicht}, {Fejer}, {Feng},
  {Fernandez-Galiana}, {Ferrante}, {Ferreira}, {Ferreira}, {Ferrini},
  {Fidecaro}, {Fiori}, {Fiorucci}, {Fishbach}, {Fisher}, {Fishner},
  {Fitz-Axen}, {Flaminio}, {Fletcher}, {Flynn}, {Fong}, {Font}, {Forsyth},
  {Fournier}, {Frasca}, {Frasconi}, {Frei}, {Freise}, {Frey}, {Frey},
  {Fritschel}, {Frolov}, {Fulda}, {Fyffe}, {Gabbard}, {Gadre}, {Gaebel},
  {Gair}, {Gammaitoni}, {Ganija}, {Gaonkar}, {Garcia},
  {Garc{\'\i}a-Quir{\'o}s}, {Garufi}, {Gateley}, {Gaudio}, {Gaur}, {Gayathri},
  {Gemme}, {Genin}, {Gennai}, {George}, {George}, {Gergely}, {Germain},
  {Ghonge}, {Ghosh}, {Ghosh}, {Ghosh}, {Giacomazzo}, {Giaime}, {Giardina},
  {Giazotto}, {Gill}, {Giordano}, {Glover}, {Godwin}, {Goetz}, {Goetz},
  {Goncharov}, {Gonz{\'a}lez}, {Gonzalez Castro}, {Gopakumar}, {Gorodetsky},
  {Gossan}, {Gosselin}, {Gouaty}, {Grado}, {Graef}, {Granata}, {Grant}, {Gras},
  {Grassia}, {Gray}, {Gray}, {Greco}, {Green}, {Green}, {Gretarsson}, {Groot},
  {Grote}, {Grunewald}, {Gruning}, {Guidi}, {Gulati}, {Guo}, {Gupta}, {Gupta},
  {Gustafson}, {Gustafson}, {Haegel}, {Halim}, {Hall}, {Hall}, {Hamilton},
  {Hammond}, {Haney}, {Hanke}, {Hanks}, {Hanna}, {Hannam}, {Hannuksela},
  {Hanson}, {Hardwick}, {Haris}, {Harms}, {Harry}, {Harry}, {Haster},
  {Haughian}, {Hayes}, {Healy}, {Heidmann}, {Heintze}, {Heitmann}, {Hello},
  {Hemming}, {Hendry}, {Heng}, {Hennig}, {Heptonstall}, {Hernandez Vivanco},
  {Heurs}, {Hild}, {Hinderer}, {Hoak}, {Hochheim}, {Hofman}, {Holgado},
  {Holland}, {Holt}, {Holz}, {Hopkins}, {Horst}, {Hough}, {Howell}, {Hoy},
  {Hreibi}, {Huang}, {Huerta}, {Huet}, {Hughey}, {Hulko}, {Husa}, {Huttner},
  {Huynh-Dinh}, {Idzkowski}, {Iess}, {Ingram}, {Inta}, {Intini}, {Irwin},
  {Isa}, {Isac}, {Isi}, {Iyer}, {Izumi}, {Jacqmin}, {Jadhav}, {Jani},
  {Janthalur}, {Jaranowski}, {Jenkins}, {Jiang}, {Johnson}, {Johnson-McDaniel},
  {Jones}, {Jones}, {Jones}, {Jonker}, {Ju}, {Junker}, {Kalaghatgi},
  {Kalogera}, {Kamai}, {Kandhasamy}, {Kang}, {Kanner}, {Kapadia}, {Karki},
  {Karvinen}, {Kashyap}, {Kasprzack}, {Katsanevas}, {Katsavounidis}, {Katzman},
  {Kaufer}, {Kawabe}, {Keerthana}, {K{\'e}f{\'e}lian}, {Keitel}, {Kennedy},
  {Key}, {Khalili}, {Khan}, {Khan}, {Khan}, {Khan}, {Khazanov}, {Khursheed},
  {Kijbunchoo}, {Kim}, {Kim}, {Kim}, {Kim}, {Kim}, {Kim}, {Kimball}, {King},
  {King}, {Kinley-Hanlon}, {Kirchhoff}, {Kissel}, {Kleybolte}, {Klika},
  {Klimenko}, {Knowles}, {Koch}, {Koehlenbeck}, {Koekoek}, {Koley},
  {Kondrashov}, {Kontos}, {Koper}, {Korobko}, {Korth}, {Kowalska}, {Kozak},
  {Kringel}, {Krishnendu}, {Kr{\'o}lak}, {Kuehn}, {Kumar}, {Kumar}, {Kumar},
  {Kumar}, {Kuo}, {Kutynia}, {Kwang}, {Lackey}, {Lai}, {Lam}, {Landry}, {Lane},
  {Lang}, {Lange}, {Lantz}, {Lanza}, {Lartaux-Vollard}, {Lasky}, {Laxen},
  {Lazzarini}, {Lazzaro}, {Leaci}, {Leavey}, {Lecoeuche}, {Lee}, {Lee}, {Lee},
  {Lee}, {Lee}, {Lee}, {Lehmann}, {Lenon}, {Leroy}, {Letendre}, {Levin}, {Li},
  {Li}, {Li}, {Li}, {Lin}, {Linde}, {Linker}, {Littenberg}, {Liu}, {Liu}, {Lo},
  {Lockerbie}, {London}, {Longo}, {Lorenzini}, {Loriette}, {Lormand},
  {Losurdo}, {Lough}, {Lousto}, {Lovelace}, {Lower}, {L{\"u}ck}, {Lumaca},
  {Lundgren}, {Lynch}, {Ma}, {Macas}, {Macfoy}, {MacInnis}, {Macleod},
  {Macquet}, {Maga{\~n}a-Sandoval}, {Maga{\~n}a Zertuche}, {Magee}, {Majorana},
  {Maksimovic}, {Malik}, {Man}, {Mandic}, {Mangano}, {Mansell}, {Manske},
  {Mantovani}, {Marchesoni}, {Marion}, {M{\'a}rka}, {M{\'a}rka}, {Markakis},
  {Markosyan}, {Markowitz}, {Maros}, {Marquina}, {Marsat}, {Martelli},
  {Martin}, {Martin}, {Martynov}, {Mason}, {Massera}, {Masserot}, {Massinger},
  {Masso-Reid}, {Mastrogiovanni}, {Matas}, {Matichard}, {Matone}, {Mavalvala},
  {Mazumder}, {McCann}, {McCarthy}, {McClelland}, {McCormick}, {McCuller},
  {McGuire}, {McIver}, {McManus}, {McRae}, {McWilliams}, {Meacher}, {Meadors},
  {Mehmet}, {Mehta}, {Meidam}, {Melatos}, {Mendell}, {Mercer}, {Mereni},
  {Merilh}, {Merzougui}, {Meshkov}, {Messenger}, {Messick}, {Metzdorff},
  {Meyers}, {Miao}, {Michel}, {Middleton}, {Mikhailov}, {Milano}, {Miller},
  {Miller}, {Millhouse}, {Mills}, {Milovich-Goff}, {Minazzoli}, {Minenkov},
  {Mishkin}, {Mishra}, {Mistry}, {Mitra}, {Mitrofanov}, {Mitselmakher},
  {Mittleman}, {Mo}, {Moffa}, {Mogushi}, {Mohapatra}, {Montani}, {Moore},
  {Moraru}, {Moreno}, {Morisaki}, {Mours}, {Mow-Lowry}, {Mukherjee},
  {Mukherjee}, {Mukherjee}, {Mukund}, {Mullavey}, {Munch}, {Mu{\~n}iz},
  {Muratore}, {Murray}, {Nagar}, {Nardecchia}, {Naticchioni}, {Nayak},
  {Neilson}, {Nelemans}, {Nelson}, {Nery}, {Neunzert}, {Ng}, {Ng}, {Nguyen},
  {Nichols}, {Nielsen}, {Nissanke}, {Nitz}, {Nocera}, {North}, {Nuttall},
  {Obergaulinger}, {Oberling}, {O'Brien}, {O'Dea}, {Ogin}, {Oh}, {Oh}, {Ohme},
  {Ohta}, {Okada}, {Oliver}, {Oppermann}, {Oram}, {O'Reilly}, {Ormiston},
  {Ortega}, {O'Shaughnessy}, {Ossokine}, {Ottaway}, {Overmier}, {Owen}, {Pace},
  {Pagano}, {Page}, {Pai}, {Pai}, {Palamos}, {Palashov}, {Palomba},
  {Pal-Singh}, {Pan}, {Pang}, {Pang}, {Pankow}, {Pannarale}, {Pant},
  {Paoletti}, {Paoli}, {Papa}, {Parida}, {Parker}, {Pascucci}, {Pasqualetti},
  {Passaquieti}, {Passuello}, {Patil}, {Patricelli}, {Pearlstone}, {Pedersen},
  {Pedraza}, {Pedurand}, {Pele}, {Penn}, {Perego}, {Perez}, {Perreca},
  {Pfeiffer}, {Phelps}, {Phukon}, {Piccinni}, {Pichot}, {Piergiovanni},
  {Pillant}, {Pinard}, {Pirello}, {Pitkin}, {Poggiani}, {Pong}, {Ponrathnam},
  {Popolizio}, {Porter}, {Powell}, {Prajapati}, {Prasad}, {Prasai}, {Prasanna},
  {Pratten}, {Prestegard}, {Privitera}, {Prodi}, {Prokhorov}, {Puncken},
  {Punturo}, {Puppo}, {P{\"u}rrer}, {Qi}, {Quetschke}, {Quinonez}, {Quintero},
  {Quitzow-James}, {Raab}, {Radkins}, {Radulescu}, {Raffai}, {Raja}, {Rajan},
  {Rajbhandari}, {Rakhmanov}, {Ramirez}, {Ramos-Buades}, {Rana}, {Rao},
  {Rapagnani}, {Raymond}, {Razzano}, {Read}, {Regimbau}, {Rei}, {Reid},
  {Reitze}, {Ren}, {Ricci}, {Richardson}, {Richardson}, {Ricker},
  {Riemenschneider}, {Riles}, {Rizzo}, {Robertson}, {Robie}, {Robinet},
  {Rocchi}, {Rolland}, {Rollins}, {Roma}, {Romanelli}, {Romano}, {Romel},
  {Romie}, {Rose}, {Rosi{\'n}ska}, {Rosofsky}, {Ross}, {Rowan}, {R{\"u}diger},
  {Ruggi}, {Rutins}, {Ryan}, {Sachdev}, {Sadecki}, {Sakellariadou}, {Salafia},
  {Salconi}, {Saleem}, {Salemi}, {Samajdar}, {Sammut}, {Sanchez}, {Sanchez},
  {Sanchis-Gual}, {Sandberg}, {Sanders}, {Santiago}, {Sarin}, {Sassolas},
  {Sathyaprakash}, {Saulson}, {Sauter}, {Savage}, {Schale}, {Scheel},
  {Scheuer}, {Schmidt}, {Schnabel}, {Schofield}, {Sch{\"o}nbeck}, {Schreiber},
  {Schulte}, {Schutz}, {Schwalbe}, {Scott}, {Scott}, {Seidel}, {Sellers},
  {Sengupta}, {Sennett}, {Sentenac}, {Sequino}, {Sergeev}, {Setyawati},
  {Shaddock}, {Shaffer}, {Shahriar}, {Shaner}, {Shao}, {Sharma}, {Shawhan},
  {Shen}, {Shink}, {Shoemaker}, {Shoemaker}, {ShyamSundar}, {Siellez},
  {Sieniawska}, {Sigg}, {Silva}, {Singer}, {Singh}, {Singhal}, {Sintes},
  {Sitmukhambetov}, {Skliris}, {Slagmolen}, {Slaven-Blair}, {Smith}, {Smith},
  {Somala}, {Son}, {Sorazu}, {Sorrentino}, {Souradeep}, {Sowell}, {Spencer},
  {Srivastava}, {Srivastava}, {Staats}, {Stachie}, {Standke}, {Steer},
  {Steinke}, {Steinlechner}, {Steinlechner}, {Steinmeyer}, {Stevenson},
  {Stocks}, {Stone}, {Stops}, {Strain}, {Stratta}, {Strigin}, {Strunk},
  {Sturani}, {Stuver}, {Sudhir}, {Summerscales}, {Sun}, {Sunil}, {Suresh},
  {Sutton}, {Swinkels}, {Szczepa{\'n}czyk}, {Tacca}, {Tait}, {Talbot},
  {Talukder}, {Tanner}, {T{\'a}pai}, {Taracchini}, {Tasson}, {Taylor}, {Thies},
  {Thomas}, {Thomas}, {Thondapu}, {Thorne}, {Thrane}, {Tiwari}, {Tiwari},
  {Tiwari}, {Toland}, {Tonelli}, {Tornasi}, {Torres-Forn{\'e}}, {Torrie},
  {T{\"o}yr{\"a}}, {Travasso}, {Traylor}, {Tringali}, {Trovato}, {Trozzo},
  {Trudeau}, {Tsang}, {Tse}, {Tso}, {Tsukada}, {Tsuna}, {Tuyenbayev}, {Ueno},
  {Ugolini}, {Unnikrishnan}, {Urban}, {Usman}, {Vahlbruch}, {Vajente},
  {Valdes}, {van Bakel}, {van Beuzekom}, {van den Brand}, {Van Den Broeck},
  {Vander-Hyde}, {van Heijningen}, {van der Schaaf}, {van Veggel}, {Vardaro},
  {Varma}, {Vass}, {Vas{\'u}th}, {Vecchio}, {Vedovato}, {Veitch}, {Veitch},
  {Venkateswara}, {Venugopalan}, {Verkindt}, {Vetrano}, {Vicer{\'e}}, {Viets},
  {Vine}, {Vinet}, {Vitale}, {Vo}, {Vocca}, {Vorvick}, {Vyatchanin}, {Wade},
  {Wade}, {Wade}, {Walet}, {Walker}, {Wallace}, {Walsh}, {Wang}, {Wang},
  {Wang}, {Wang}, {Wang}, {Ward}, {Warden}, {Warner}, {Was}, {Watchi},
  {Weaver}, {Wei}, {Weinert}, {Weinstein}, {Weiss}, {Wellmann}, {Wen},
  {Wessel}, {We{\ss}els}, {Westhouse}, {Wette}, {Whelan}, {White}, {Whiting},
  {Whittle}, {Wilken}, {Williams}, {Williamson}, {Willis}, {Willke}, {Wimmer},
  {Winkler}, {Wipf}, {Wittel}, {Woan}, {Woehler}, {Wofford}, {Worden},
  {Wright}, {Wu}, {Wysocki}, {Xiao}, {Yamamoto}, {Yancey}, {Yang}, {Yap},
  {Yazback}, {Yeeles}, {Yu}, {Yu}, {Yuen}, {Yvert}, {Zadro{\.Z}ny}, {Zanolin},
  {Zappa}, {Zelenova}, {Zendri}, {Zevin}, {Zhang}, {Zhang}, {Zhang}, {Zhao},
  {Zhou}, {Zhou}, {Zhu}, {Zimmerman}, {Zlochower}, {Zucker}, {Zweizig}, {LIGO
  Scientific Collaboration}, \& {Virgo Collaboration}}]{2019PhRvX...9c1040A}
---. 2019, Physical Review X, 9, 031040, \dodoi{10.1103/PhysRevX.9.031040}

\bibitem[{{Abbott} {et~al.}(2020{\natexlab{a}}){Abbott}, {Abbott}, {Abraham},
  {Acernese}, {Ackley}, {Adams}, {Adhikari}, {Adya}, {Affeldt}, {Agathos},
  {Agatsuma}, {Aggarwal}, {Aguiar}, {Aich}, {Aiello}, {Ain}, {Ajith}, {Akcay},
  {Allen}, {Allocca}, {Altin}, {Amato}, {Anand}, {Ananyeva}, {Anderson},
  {Anderson}, {Angelova}, {Ansoldi}, {Antier}, {Appert}, {Arai}, {Araya},
  {Areeda}, {Ar{\`e}ne}, {Arnaud}, {Aronson}, {Arun}, {Asali}, {Ascenzi},
  {Ashton}, {Aston}, {Astone}, {Aubin}, {Aufmuth}, {AultONeal}, {Austin},
  {Avendano}, {Babak}, {Bacon}, {Badaracco}, {Bader}, {Bae}, {Baer}, {Baird},
  {Baldaccini}, {Ballardin}, {Ballmer}, {Bals}, {Balsamo}, {Baltus},
  {Banagiri}, {Bankar}, {Bankar}, {Barayoga}, {Barbieri}, {Barish}, {Barker},
  {Barkett}, {Barneo}, {Barone}, {Barr}, {Barsotti}, {Barsuglia}, {Barta},
  {Bartlett}, {Bartos}, {Bassiri}, {Basti}, {Bawaj}, {Bayley}, {Bazzan},
  {B{\'e}csy}, {Bejger}, {Belahcene}, {Bell}, {Beniwal}, {Benjamin}, {Bentley},
  {Bergamin}, {Berger}, {Bergmann}, {Bernuzzi}, {Berry}, {Bersanetti},
  {Bertolini}, {Betzwieser}, {Bhandare}, {Bhandari}, {Bidler}, {Biggs},
  {Bilenko}, {Billingsley}, {Birney}, {Birnholtz}, {Biscans}, {Bischi},
  {Biscoveanu}, {Bisht}, {Bissenbayeva}, {Bitossi}, {Bizouard}, {Blackburn},
  {Blackman}, {Blair}, {Blair}, {Blair}, {Bobba}, {Bode}, {Boer}, {Boetzel},
  {Bogaert}, {Bondu}, {Bonilla}, {Bonnand}, {Booker}, {Boom}, {Bork}, {Boschi},
  {Bose}, {Bossilkov}, {Bosveld}, {Bouffanais}, {Bozzi}, {Bradaschia}, {Brady},
  {Bramley}, {Branchesi}, {Brau}, {Breschi}, {Briant}, {Briggs}, {Brighenti},
  {Brillet}, {Brinkmann}, {Brockill}, {Brooks}, {Brooks}, {Brown}, {Brunett},
  {Bruno}, {Bruntz}, {Buikema}, {Bulik}, {Bulten}, {Buonanno}, {Buscicchio},
  {Buskulic}, {Byer}, {Cabero}, {Cadonati}, {Cagnoli}, {Cahillane},
  {Calder{\'o}n Bustillo}, {Callaghan}, {Callister}, {Calloni}, {Camp},
  {Canepa}, {Cannon}, {Cao}, {Cao}, {Carapella}, {Carbognani}, {Caride},
  {Carney}, {Carullo}, {Casanueva Diaz}, {Casentini}, {Casta{\~n}eda},
  {Caudill}, {Cavagli{\`a}}, {Cavalier}, {Cavalieri}, {Cella},
  {Cerd{\'a}-Dur{\'a}n}, {Cesarini}, {Chaibi}, {Chakravarti}, {Chan}, {Chan},
  {Chandra}, {Chao}, {Charlton}, {Chase}, {Chassande-Mottin}, {Chatterjee},
  {Chaturvedi}, {Chatziioannou}, {Chen}, {Chen}, {Chen}, {Cheng}, {Cheong},
  {Chia}, {Chiadini}, {Chierici}, {Chincarini}, {Chiummo}, {Cho}, {Cho}, {Cho},
  {Christensen}, {Chu}, {Chua}, {Chung}, {Chung}, {Ciani}, {Ciecielag},
  {Cie{\'s}lar}, {Ciobanu}, {Ciolfi}, {Cipriano}, {Cirone}, {Clara}, {Clark},
  {Clearwater}, {Clesse}, {Cleva}, {Coccia}, {Cohadon}, {Cohen}, {Colleoni},
  {Collette}, {Collins}, {Colpi}, {Constancio}, {Conti}, {Cooper}, {Corban},
  {Corbitt}, {Cordero-Carri{\'o}n}, {Corezzi}, {Corley}, {Cornish}, {Corre},
  {Corsi}, {Cortese}, {Costa}, {Cotesta}, {Coughlin}, {Coughlin}, {Coulon},
  {Countryman}, {Couvares}, {Covas}, {Coward}, {Cowart}, {Coyne}, {Coyne},
  {Creighton}, {Creighton}, {Cripe}, {Croquette}, {Crowder}, {Cudell},
  {Cullen}, {Cumming}, {Cummings}, {Cunningham}, {Cuoco}, {Curylo}, {Canton},
  {D{\'a}lya}, {Dana}, {Daneshgaran-Bajastani}, {D'Angelo}, {Danilishin},
  {D'Antonio}, {Danzmann}, {Darsow-Fromm}, {Dasgupta}, {Datrier}, {Dattilo},
  {Dave}, {Davier}, {Davies}, {Davis}, {Daw}, {DeBra}, {Deenadayalan},
  {Degallaix}, {De Laurentis}, {Del{\'e}glise}, {Delfavero}, {De Lillo}, {Del
  Pozzo}, {DeMarchi}, {D'Emilio}, {Demos}, {Dent}, {De Pietri}, {De Rosa}, {De
  Rossi}, {DeSalvo}, {de Varona}, {Dhurandhar}, {D{\'\i}az}, {Diaz-Ortiz},
  {Dietrich}, {Di Fiore}, {Di Fronzo}, {Di Giorgio}, {Di Giovanni}, {Di
  Giovanni}, {Di Girolamo}, {Di Lieto}, {Ding}, {Di Pace}, {Di Palma}, {Di
  Renzo}, {Divakarla}, {Dmitriev}, {Doctor}, {Donovan}, {Dooley}, {Doravari},
  {Dorrington}, {Downes}, {Drago}, {Driggers}, {Du}, {Ducoin}, {Dupej},
  {Durante}, {D'Urso}, {Dwyer}, {Easter}, {Eddolls}, {Edelman}, {Edo}, {Edy},
  {Effler}, {Ehrens}, {Eichholz}, {Eikenberry}, {Eisenmann}, {Eisenstein},
  {Ejlli}, {Errico}, {Essick}, {Estelles}, {Estevez}, {Etienne}, {Etzel},
  {Evans}, {Evans}, {Ewing}, {Fafone}, {Fairhurst}, {Fan}, {Farinon}, {Farr},
  {Farr}, {Fauchon-Jones}, {Favata}, {Fays}, {Fazio}, {Feicht}, {Fejer},
  {Feng}, {Fenyvesi}, {Ferguson}, {Fernandez-Galiana}, {Ferrante}, {Ferreira},
  {Ferreira}, {Fidecaro}, {Fiori}, {Fiorucci}, {Fishbach}, {Fisher},
  {Fittipaldi}, {Fitz-Axen}, {Fiumara}, {Flaminio}, {Floden}, {Flynn}, {Fong},
  {Font}, {Forsyth}, {Fournier}, {Frasca}, {Frasconi}, {Frei}, {Freise},
  {Frey}, {Frey}, {Fritschel}, {Frolov}, {Fronz{\`e}}, {Fulda}, {Fyffe},
  {Gabbard}, {Gadre}, {Gaebel}, {Gair}, {Galaudage}, {Ganapathy}, {Ganguly},
  {Gaonkar}, {Garc{\'\i}a-Quir{\'o}s}, {Garufi}, {Gateley}, {Gaudio},
  {Gayathri}, {Gemme}, {Genin}, {Gennai}, {George}, {George}, {Gergely},
  {Ghonge}, {Ghosh}, {Ghosh}, {Ghosh}, {Giacomazzo}, {Giaime}, {Giardina},
  {Gibson}, {Gier}, {Gill}, {Glanzer}, {Gniesmer}, {Godwin}, {Goetz}, {Goetz},
  {Gohlke}, {Goncharov}, {Gonz{\'a}lez}, {Gopakumar}, {Gossan}, {Gosselin},
  {Gouaty}, {Grace}, {Grado}, {Granata}, {Grant}, {Gras}, {Grassia}, {Gray},
  {Gray}, {Greco}, {Green}, {Green}, {Gretarsson}, {Griggs}, {Grignani},
  {Grimaldi}, {Grimm}, {Grote}, {Grunewald}, {Gruning}, {Guidi}, {Guimaraes},
  {Guix{\'e}}, {Gulati}, {Guo}, {Gupta}, {Gupta}, {Gupta}, {Gustafson},
  {Gustafson}, {Haegel}, {Halim}, {Hall}, {Hamilton}, {Hammond}, {Haney},
  {Hanke}, {Hanks}, {Hanna}, {Hannam}, {Hannuksela}, {Hansen}, {Hanson},
  {Harder}, {Hardwick}, {Haris}, {Harms}, {Harry}, {Harry}, {Hasskew},
  {Haster}, {Haughian}, {Hayes}, {Healy}, {Heidmann}, {Heintze}, {Heinze},
  {Heitmann}, {Hellman}, {Hello}, {Hemming}, {Hendry}, {Heng}, {Hennes},
  {Hennig}, {Heurs}, {Hild}, {Hinderer}, {Hoback}, {Hochheim}, {Hofgard},
  {Hofman}, {Holgado}, {Holland}, {Holt}, {Holz}, {Hopkins}, {Horst}, {Hough},
  {Howell}, {Hoy}, {Huang}, {H{\"u}bner}, {Huerta}, {Huet}, {Hughey}, {Hui},
  {Husa}, {Huttner}, {Huxford}, {Huynh-Dinh}, {Idzkowski}, {Iess}, {Inchauspe},
  {Ingram}, {Intini}, {Isac}, {Isi}, {Iyer}, {Jacqmin}, {Jadhav}, {Jadhav},
  {James}, {Jani}, {Janthalur}, {Jaranowski}, {Jariwala}, {Jaume}, {Jenkins},
  {Jiang}, {Johns}, {Johnson-McDaniel}, {Jones}, {Jones}, {Jones}, {Jones},
  {Jones}, {Jonker}, {Ju}, {Junker}, {Kalaghatgi}, {Kalogera}, {Kamai},
  {Kandhasamy}, {Kang}, {Kanner}, {Kapadia}, {Karki}, {Kashyap}, {Kasprzack},
  {Kastaun}, {Katsanevas}, {Katsavounidis}, {Katzman}, {Kaufer}, {Kawabe},
  {K{\'e}f{\'e}lian}, {Keitel}, {Keivani}, {Kennedy}, {Key}, {Khadka},
  {Khalili}, {Khan}, {Khan}, {Khan}, {Khazanov}, {Khetan}, {Khursheed},
  {Kijbunchoo}, {Kim}, {Kim}, {Kim}, {Kim}, {Kim}, {Kim}, {Kim}, {Kimball},
  {King}, {Kinley-Hanlon}, {Kirchhoff}, {Kissel}, {Kleybolte}, {Klimenko},
  {Knowles}, {Knyazev}, {Koch}, {Koehlenbeck}, {Koekoek}, {Koley},
  {Kondrashov}, {Kontos}, {Koper}, {Korobko}, {Korth}, {Kovalam}, {Kozak},
  {Kringel}, {Krishnendu}, {Kr{\'o}lak}, {Krupinski}, {Kuehn}, {Kumar},
  {Kumar}, {Kumar}, {Kumar}, {Kumar}, {Kuo}, {Kutynia}, {Lackey}, {Laghi},
  {Lalande}, {Lam}, {Lamberts}, {Landry}, {Lane}, {Lang}, {Lange}, {Lantz},
  {Lanza}, {La Rosa}, {Lartaux-Vollard}, {Lasky}, {Laxen}, {Lazzarini},
  {Lazzaro}, {Leaci}, {Leavey}, {Lecoeuche}, {Lee}, {Lee}, {Lee}, {Lee}, {Lee},
  {Lehmann}, {Leroy}, {Letendre}, {Levin}, {Li}, {Li}, {li}, {Li}, {Li},
  {Linde}, {Linker}, {Linley}, {Littenberg}, {Liu}, {Liu},
  {Llorens-Monteagudo}, {Lo}, {Lockwood}, {London}, {Longo}, {Lorenzini},
  {Loriette}, {Lormand}, {Losurdo}, {Lough}, {Lousto}, {Lovelace}, {L{\"u}ck},
  {Lumaca}, {Lundgren}, {Ma}, {Macas}, {Macfoy}, {MacInnis}, {Macleod},
  {MacMillan}, {Macquet}, {Maga{\~n}a Hernandez}, {Maga{\~n}a-Sandoval},
  {Magee}, {Majorana}, {Maksimovic}, {Malik}, {Man}, {Mandic}, {Mangano},
  {Mansell}, {Manske}, {Mantovani}, {Mapelli}, {Marchesoni}, {Marion},
  {M{\'a}rka}, {M{\'a}rka}, {Markakis}, {Markosyan}, {Markowitz}, {Maros},
  {Marquina}, {Marsat}, {Martelli}, {Martin}, {Martin}, {Martinez}, {Martynov},
  {Masalehdan}, {Mason}, {Massera}, {Masserot}, {Massinger}, {Masso-Reid},
  {Mastrogiovanni}, {Matas}, {Matichard}, {Mavalvala}, {Maynard}, {McCann},
  {McCarthy}, {McClelland}, {McCormick}, {McCuller}, {McGuire}, {McIsaac},
  {McIver}, {McManus}, {McRae}, {McWilliams}, {Meacher}, {Meadors}, {Mehmet},
  {Mehta}, {Mejuto Villa}, {Melatos}, {Mendell}, {Mercer}, {Mereni}, {Merfeld},
  {Merilh}, {Merritt}, {Merzougui}, {Meshkov}, {Messenger}, {Messick},
  {Metzdorff}, {Meyers}, {Meylahn}, {Mhaske}, {Miani}, {Miao}, {Michaloliakos},
  {Michel}, {Middleton}, {Milano}, {Miller}, {Millhouse}, {Mills}, {Milotti},
  {Milovich-Goff}, {Minazzoli}, {Minenkov}, {Mishkin}, {Mishra}, {Mistry},
  {Mitra}, {Mitrofanov}, {Mitselmakher}, {Mittleman}, {Mo}, {Mogushi},
  {Mohapatra}, {Mohite}, {Molina-Ruiz}, {Mondin}, {Montani}, {Moore}, {Moraru},
  {Morawski}, {Moreno}, {Morisaki}, {Mours}, {Mow-Lowry}, {Mozzon},
  {Muciaccia}, {Mukherjee}, {Mukherjee}, {Mukherjee}, {Mukherjee}, {Mukund},
  {Mullavey}, {Munch}, {Mu{\~n}iz}, {Murray}, {Nagar}, {Nardecchia},
  {Naticchioni}, {Nayak}, {Neil}, {Neilson}, {Nelemans}, {Nelson}, {Nery},
  {Neunzert}, {Ng}, {Ng}, {Nguyen}, {Nguyen}, {Nichols}, {Nichols}, {Nissanke},
  {Nitz}, {Nocera}, {Noh}, {North}, {Nothard}, {Nuttall}, {Oberling},
  {O'Brien}, {Oganesyan}, {Ogin}, {Oh}, {Oh}, {Ohme}, {Ohta}, {Okada},
  {Oliver}, {Olivetto}, {Oppermann}, {Oram}, {O'Reilly}, {Ormiston}, {Ortega},
  {O'Shaughnessy}, {Ossokine}, {Osthelder}, {Ottaway}, {Overmier}, {Owen},
  {Pace}, {Pagano}, {Page}, {Pagliaroli}, {Pai}, {Pai}, {Palamos}, {Palashov},
  {Palomba}, {Pan}, {Panda}, {Pang}, {Pankow}, {Pannarale}, {Pant}, {Paoletti},
  {Paoli}, {Parida}, {Parker}, {Pascucci}, {Pasqualetti}, {Passaquieti},
  {Passuello}, {Patricelli}, {Payne}, {Pearlstone}, {Pechsiri}, {Pedersen},
  {Pedraza}, {Pele}, {Penn}, {Perego}, {Perez}, {P{\'e}rigois}, {Perreca},
  {Perri{\`e}s}, {Petermann}, {Pfeiffer}, {Phelps}, {Phukon}, {Piccinni},
  {Pichot}, {Piendibene}, {Piergiovanni}, {Pierro}, {Pillant}, {Pinard},
  {Pinto}, {Piotrzkowski}, {Pirello}, {Pitkin}, {Plastino}, {Poggiani}, {Pong},
  {Ponrathnam}, {Popolizio}, {Porter}, {Powell}, {Prajapati}, {Prasai},
  {Prasanna}, {Pratten}, {Prestegard}, {Principe}, {Prodi}, {Prokhorov},
  {Punturo}, {Puppo}, {P{\"u}rrer}, {Qi}, {Quetschke}, {Quinonez}, {Raab},
  {Raaijmakers}, {Radkins}, {Radulesco}, {Raffai}, {Rafferty}, {Raja}, {Rajan},
  {Rajbhandari}, {Rakhmanov}, {Ramirez}, {Ramos-Buades}, {Rana}, {Rao},
  {Rapagnani}, {Raymond}, {Razzano}, {Read}, {Regimbau}, {Rei}, {Reid},
  {Reitze}, {Rettegno}, {Ricci}, {Richardson}, {Richardson}, {Ricker},
  {Riemenschneider}, {Riles}, {Rizzo}, {Robertson}, {Robinet}, {Rocchi},
  {Rodriguez-Soto}, {Rolland}, {Rollins}, {Roma}, {Romanelli}, {Romano},
  {Romel}, {Romero-Shaw}, {Romie}, {Rose}, {Rose}, {Rose}, {Rosi{\'n}ska},
  {Rosofsky}, {Ross}, {Rowan}, {Rowlinson}, {Roy}, {Roy}, {Roy}, {Ruggi},
  {Rutins}, {Ryan}, {Sachdev}, {Sadecki}, {Sakellariadou}, {Salafia},
  {Salconi}, {Saleem}, {Salemi}, {Samajdar}, {Sanchez}, {Sanchez},
  {Sanchis-Gual}, {Sanders}, {Santiago}, {Santos}, {Sarin}, {Sassolas},
  {Sathyaprakash}, {Sauter}, {Savage}, {Savant}, {Sawant}, {Sayah}, {Schaetzl},
  {Schale}, {Scheel}, {Scheuer}, {Schmidt}, {Schnabel}, {Schofield},
  {Sch{\"o}nbeck}, {Schreiber}, {Schulte}, {Schutz}, {Schwarm}, {Schwartz},
  {Scott}, {Scott}, {Seidel}, {Sellers}, {Sengupta}, {Sennett}, {Sentenac},
  {Sequino}, {Sergeev}, {Setyawati}, {Shaddock}, {Shaffer}, {Sharifi},
  {Shahriar}, {Sharma}, {Sharma}, {Shawhan}, {Shen}, {Shikauchi}, {Shink},
  {Shoemaker}, {Shoemaker}, {Shukla}, {ShyamSundar}, {Siellez}, {Sieniawska},
  {Sigg}, {Singer}, {Singh}, {Singh}, {Singha}, {Singhal}, {Sintes}, {Sipala},
  {Skliris}, {Slagmolen}, {Slaven-Blair}, {Smetana}, {Smith}, {Smith},
  {Somala}, {Son}, {Soni}, {Sorazu}, {Sordini}, {Sorrentino}, {Souradeep},
  {Sowell}, {Spencer}, {Spera}, {Srivastava}, {Srivastava}, {Staats},
  {Stachie}, {Standke}, {Steer}, {Steinke}, {Steinlechner}, {Steinlechner},
  {Steinmeyer}, {Stevenson}, {Stocks}, {Stops}, {Stover}, {Strain}, {Stratta},
  {Strunk}, {Sturani}, {Stuver}, {Sudhagar}, {Sudhir}, {Summerscales}, {Sun},
  {Sunil}, {Sur}, {Suresh}, {Sutton}, {Swinkels}, {Szczepa{\'n}czyk}, {Tacca},
  {Tait}, {Talbot}, {Tanasijczuk}, {Tanner}, {Tao}, {T{\'a}pai}, {Tapia},
  {Tapia San Martin}, {Tasson}, {Taylor}, {Tenorio}, {Terkowski},
  {Thirugnanasambandam}, {Thomas}, {Thomas}, {Thompson}, {Thondapu}, {Thorne},
  {Thrane}, {Tinsman}, {Saravanan}, {Tiwari}, {Tiwari}, {Tiwari}, {Toland},
  {Tonelli}, {Tornasi}, {Torres-Forn{\'e}}, {Torrie}, {Tosta e Melo},
  {T{\"o}yr{\"a}}, {Travasso}, {Traylor}, {Tringali}, {Tripathee}, {Trovato},
  {Trudeau}, {Tsang}, {Tse}, {Tso}, {Tsukada}, {Tsuna}, {Tsutsui}, {Turconi},
  {Ubhi}, {Udall}, {Ueno}, {Ugolini}, {Unnikrishnan}, {Urban}, {Usman},
  {Utina}, {Vahlbruch}, {Vajente}, {Valdes}, {Valentini}, {van Bakel}, {van
  Beuzekom}, {van den Brand}, {Van Den Broeck}, {Vander-Hyde}, {van der
  Schaaf}, {Van Heijningen}, {van Veggel}, {Vardaro}, {Varma}, {Vass},
  {Vas{\'u}th}, {Vecchio}, {Vedovato}, {Veitch}, {Veitch}, {Venkateswara},
  {Venugopalan}, {Verkindt}, {Veske}, {Vetrano}, {Vicer{\'e}}, {Viets},
  {Vinciguerra}, {Vine}, {Vinet}, {Vitale}, {Vivanco}, {Vo}, {Vocca},
  {Vorvick}, {Vyatchanin}, {Wade}, {Wade}, {Wade}, {Walet}, {Walker},
  {Wallace}, {Wallace}, {Walsh}, {Wang}, {Wang}, {Wang}, {Ward}, {Warden},
  {Warner}, {Was}, {Watchi}, {Weaver}, {Wei}, {Weinert}, {Weinstein}, {Weiss},
  {Wellmann}, {Wen}, {We{\ss}els}, {Westhouse}, {Wette}, {Whelan}, {Whiting},
  {Whittle}, {Wilken}, {Williams}, {Willis}, {Willke}, {Winkler}, {Wipf},
  {Wittel}, {Woan}, {Woehler}, {Wofford}, {Wong}, {Wright}, {Wu}, {Wysocki},
  {Xiao}, {Yamamoto}, {Yang}, {Yang}, {Yang}, {Yap}, {Yazback}, {Yeeles}, {Yu},
  {Yu}, {Yuen}, {Zadro{\.Z}ny}, {Zadro{\.Z}ny}, {Zanolin}, {Zelenova},
  {Zendri}, {Zevin}, {Zhang}, {Zhang}, {Zhang}, {Zhao}, {Zhao}, {Zhou}, {Zhou},
  {Zhu}, {Zimmerman}, {Zucker}, {Zweizig}, {LIGO Scientific Collaboration}, \&
  {Virgo Collaboration}}]{2020PhRvL.125j1102A}
{Abbott}, R., {Abbott}, T.~D., {Abraham}, S., {et~al.} 2020{\natexlab{a}},
  \prl, 125, 101102, \dodoi{10.1103/PhysRevLett.125.101102}

\bibitem[{{Abbott} {et~al.}(2020{\natexlab{b}}){Abbott}, {Abbott}, {Abraham},
  {Acernese}, {Ackley}, {Adams}, {Adhikari}, {Adya}, {Affeldt}, {Agathos},
  {Agatsuma}, {Aggarwal}, {Aguiar}, {Aich}, {Aiello}, {Ain}, {Ajith}, {Akcay},
  {Allen}, {Allocca}, {Altin}, {Amato}, {Anand}, {Ananyeva}, {Anderson},
  {Anderson}, {Angelova}, {Ansoldi}, {Antier}, {Appert}, {Arai}, {Araya},
  {Areeda}, {Ar{\`e}ne}, {Arnaud}, {Aronson}, {Arun}, {Asali}, {Ascenzi},
  {Ashton}, {Aston}, {Astone}, {Aubin}, {Aufmuth}, {AultONeal}, {Austin},
  {Avendano}, {Babak}, {Bacon}, {Badaracco}, {Bader}, {Bae}, {Baer}, {Baird},
  {Baldaccini}, {Ballardin}, {Ballmer}, {Bals}, {Balsamo}, {Baltus},
  {Banagiri}, {Bankar}, {Bankar}, {Barayoga}, {Barbieri}, {Barish}, {Barker},
  {Barkett}, {Barneo}, {Barone}, {Barr}, {Barsotti}, {Barsuglia}, {Barta},
  {Bartlett}, {Bartos}, {Bassiri}, {Basti}, {Bawaj}, {Bayley}, {Bazzan},
  {B{\'e}csy}, {Bejger}, {Belahcene}, {Bell}, {Beniwal}, {Benjamin}, {Bentley},
  {Bergamin}, {Berger}, {Bergmann}, {Bernuzzi}, {Berry}, {Bersanetti},
  {Bertolini}, {Betzwieser}, {Bhandare}, {Bhandari}, {Bidler}, {Biggs},
  {Bilenko}, {Billingsley}, {Birney}, {Birnholtz}, {Biscans}, {Bischi},
  {Biscoveanu}, {Bisht}, {Bissenbayeva}, {Bitossi}, {Bizouard}, {Blackburn},
  {Blackman}, {Blair}, {Blair}, {Blair}, {Bobba}, {Bode}, {Boer}, {Boetzel},
  {Bogaert}, {Bondu}, {Bonilla}, {Bonnand}, {Booker}, {Boom}, {Bork}, {Boschi},
  {Bose}, {Bossilkov}, {Bosveld}, {Bouffanais}, {Bozzi}, {Bradaschia}, {Brady},
  {Bramley}, {Branchesi}, {Brau}, {Breschi}, {Briant}, {Briggs}, {Brighenti},
  {Brillet}, {Brinkmann}, {Brockill}, {Brooks}, {Brooks}, {Brown}, {Brunett},
  {Bruno}, {Bruntz}, {Buikema}, {Bulik}, {Bulten}, {Buonanno}, {Buscicchio},
  {Buskulic}, {Byer}, {Cabero}, {Cadonati}, {Cagnoli}, {Cahillane}, {Bustillo},
  {Callaghan}, {Callister}, {Calloni}, {Camp}, {Canepa}, {Cannon}, {Cao},
  {Cao}, {Carapella}, {Carbognani}, {Caride}, {Carney}, {Carullo}, {Diaz},
  {Casentini}, {Casta{\~n}eda}, {Caudill}, {Cavagli{\`a}}, {Cavalier},
  {Cavalieri}, {Cella}, {Cerd{\'a}-Dur{\'a}n}, {Cesarini}, {Chaibi},
  {Chakravarti}, {Chan}, {Chan}, {Chao}, {Charlton}, {Chase},
  {Chassande-Mottin}, {Chatterjee}, {Chaturvedi}, {Chatziioannou}, {Chen},
  {Chen}, {Chen}, {Cheng}, {Cheong}, {Chia}, {Chiadini}, {Chierici},
  {Chincarini}, {Chiummo}, {Cho}, {Cho}, {Cho}, {Christensen}, {Chu}, {Chua},
  {Chung}, {Chung}, {Ciani}, {Ciecielag}, {Cie{\'s}lar}, {Ciobanu}, {Ciolfi},
  {Cipriano}, {Cirone}, {Clara}, {Clark}, {Clearwater}, {Clesse}, {Cleva},
  {Coccia}, {Cohadon}, {Cohen}, {Colleoni}, {Collette}, {Collins}, {Colpi},
  {Constancio}, {Conti}, {Cooper}, {Corban}, {Corbitt}, {Cordero-Carri{\'o}n},
  {Corezzi}, {Corley}, {Cornish}, {Corre}, {Corsi}, {Cortese}, {Costa},
  {Cotesta}, {Coughlin}, {Coughlin}, {Coulon}, {Countryman}, {Couvares},
  {Covas}, {Coward}, {Cowart}, {Coyne}, {Coyne}, {Creighton}, {Creighton},
  {Cripe}, {Croquette}, {Crowder}, {Cudell}, {Cullen}, {Cumming}, {Cummings},
  {Cunningham}, {Cuoco}, {Curylo}, {Canton}, {D{\'a}lya}, {Dana},
  {Daneshgaran-Bajastani}, {D'Angelo}, {Danilishin}, {D'Antonio}, {Danzmann},
  {Darsow-Fromm}, {Dasgupta}, {Datrier}, {Dattilo}, {Dave}, {Davier}, {Davies},
  {Davis}, {Daw}, {DeBra}, {Deenadayalan}, {Degallaix}, {De Laurentis},
  {Del{\'e}glise}, {Delfavero}, {De Lillo}, {Del Pozzo}, {DeMarchi},
  {D'Emilio}, {Demos}, {Dent}, {De Pietri}, {De Rosa}, {De Rossi}, {DeSalvo},
  {de Varona}, {Dhurandhar}, {D{\'\i}az}, {Diaz-Ortiz}, {Dietrich}, {Di Fiore},
  {Di Fronzo}, {Di Giorgio}, {Di Giovanni}, {Di Giovanni}, {Di Girolamo}, {Di
  Lieto}, {Ding}, {Di Pace}, {Di Palma}, {Di Renzo}, {Divakarla}, {Dmitriev},
  {Doctor}, {Donovan}, {Dooley}, {Doravari}, {Dorrington}, {Downes}, {Drago},
  {Driggers}, {Du}, {Ducoin}, {Dupej}, {Durante}, {D'Urso}, {Dwyer}, {Easter},
  {Eddolls}, {Edelman}, {Edo}, {Edy}, {Effler}, {Ehrens}, {Eichholz},
  {Eikenberry}, {Eisenmann}, {Eisenstein}, {Ejlli}, {Errico}, {Essick},
  {Estelles}, {Estevez}, {Etienne}, {Etzel}, {Evans}, {Evans}, {Ewing},
  {Fafone}, {Fairhurst}, {Fan}, {Farinon}, {Farr}, {Farr}, {Fauchon-Jones},
  {Favata}, {Fays}, {Fazio}, {Feicht}, {Fejer}, {Feng}, {Fenyvesi}, {Ferguson},
  {Fernandez-Galiana}, {Ferrante}, {Ferreira}, {Ferreira}, {Fidecaro}, {Fiori},
  {Fiorucci}, {Fishbach}, {Fisher}, {Fittipaldi}, {Fitz-Axen}, {Fiumara},
  {Flaminio}, {Floden}, {Flynn}, {Fong}, {Font}, {Forsyth}, {Fournier},
  {Frasca}, {Frasconi}, {Frei}, {Freise}, {Frey}, {Frey}, {Fritschel},
  {Frolov}, {Fronz{\`e}}, {Fulda}, {Fyffe}, {Gabbard}, {Gadre}, {Gaebel},
  {Gair}, {Galaudage}, {Ganapathy}, {Gaonkar}, {Garc{\'\i}a-Quir{\'o}s},
  {Garufi}, {Gateley}, {Gaudio}, {Gayathri}, {Gemme}, {Genin}, {Gennai},
  {George}, {George}, {Gergely}, {Ghonge}, {Ghosh}, {Ghosh}, {Ghosh},
  {Giacomazzo}, {Giaime}, {Giardina}, {Gibson}, {Gier}, {Gill}, {Glanzer},
  {Gniesmer}, {Godwin}, {Goetz}, {Goetz}, {Gohlke}, {Goncharov},
  {Gonz{\'a}lez}, {Gopakumar}, {Gossan}, {Gosselin}, {Gouaty}, {Grace},
  {Grado}, {Granata}, {Grant}, {Gras}, {Grassia}, {Gray}, {Gray}, {Greco},
  {Green}, {Green}, {Gretarsson}, {Griggs}, {Grignani}, {Grimaldi}, {Grimm},
  {Grote}, {Grunewald}, {Gruning}, {Guidi}, {Guimaraes}, {Guix{\'e}}, {Gulati},
  {Guo}, {Gupta}, {Gupta}, {Gupta}, {Gustafson}, {Gustafson}, {Haegel},
  {Halim}, {Hall}, {Hamilton}, {Hammond}, {Haney}, {Hanke}, {Hanks}, {Hanna},
  {Hannam}, {Hannuksela}, {Hansen}, {Hanson}, {Harder}, {Hardwick}, {Haris},
  {Harms}, {Harry}, {Harry}, {Hasskew}, {Haster}, {Haughian}, {Hayes}, {Healy},
  {Heidmann}, {Heintze}, {Heinze}, {Heitmann}, {Hellman}, {Hello}, {Hemming},
  {Hendry}, {Heng}, {Hennes}, {Hennig}, {Heurs}, {Hild}, {Hinderer}, {Hoback},
  {Hochheim}, {Hofgard}, {Hofman}, {Holgado}, {Holland}, {Holt}, {Holz},
  {Hopkins}, {Horst}, {Hough}, {Howell}, {Hoy}, {Huang}, {H{\"u}bner},
  {Huerta}, {Huet}, {Hughey}, {Hui}, {Husa}, {Huttner}, {Huxford},
  {Huynh-Dinh}, {Idzkowski}, {Iess}, {Inchauspe}, {Ingram}, {Intini}, {Isac},
  {Isi}, {Iyer}, {Jacqmin}, {Jadhav}, {Jadhav}, {James}, {Jani}, {Janthalur},
  {Jaranowski}, {Jariwala}, {Jaume}, {Jenkins}, {Jiang}, {Johns},
  {Johnson-McDaniel}, {Jones}, {Jones}, {Jones}, {Jones}, {Jones}, {Jonker},
  {Ju}, {Junker}, {Kalaghatgi}, {Kalogera}, {Kamai}, {Kandhasamy}, {Kang},
  {Kanner}, {Kapadia}, {Karki}, {Kashyap}, {Kasprzack}, {Kastaun},
  {Katsanevas}, {Katsavounidis}, {Katzman}, {Kaufer}, {Kawabe},
  {K{\'e}f{\'e}lian}, {Keitel}, {Keivani}, {Kennedy}, {Key}, {Khadka},
  {Khalili}, {Khan}, {Khan}, {Khan}, {Khazanov}, {Khetan}, {Khursheed},
  {Kijbunchoo}, {Kim}, {Kim}, {Kim}, {Kim}, {Kim}, {Kim}, {Kim}, {Kimball},
  {King}, {Kinley-Hanlon}, {Kirchhoff}, {Kissel}, {Kleybolte}, {Klimenko},
  {Knowles}, {Knyazev}, {Koch}, {Koehlenbeck}, {Koekoek}, {Koley},
  {Kondrashov}, {Kontos}, {Koper}, {Korobko}, {Korth}, {Kovalam}, {Kozak},
  {Kringel}, {Krishnendu}, {Kr{\'o}lak}, {Krupinski}, {Kuehn}, {Kumar},
  {Kumar}, {Kumar}, {Kumar}, {Kumar}, {Kuo}, {Kutynia}, {Lackey}, {Laghi},
  {Lalande}, {Lam}, {Lamberts}, {Landry}, {Lane}, {Lang}, {Lange}, {Lantz},
  {Lanza}, {La Rosa}, {Lartaux-Vollard}, {Lasky}, {Laxen}, {Lazzarini},
  {Lazzaro}, {Leaci}, {Leavey}, {Lecoeuche}, {Lee}, {Lee}, {Lee}, {Lee}, {Lee},
  {Lehmann}, {Leroy}, {Letendre}, {Levin}, {Li}, {Li}, {li}, {Li}, {Li},
  {Linde}, {Linker}, {Linley}, {Littenberg}, {Liu}, {Liu},
  {Llorens-Monteagudo}, {Lo}, {Lockwood}, {London}, {Longo}, {Lorenzini},
  {Loriette}, {Lormand}, {Losurdo}, {Lough}, {Lousto}, {Lovelace}, {L{\"u}ck},
  {Lumaca}, {Lundgren}, {Ma}, {Macas}, {Macfoy}, {MacInnis}, {Macleod},
  {MacMillan}, {Macquet}, {Hernandez}, {Maga{\~n}a-Sandoval}, {Magee},
  {Majorana}, {Maksimovic}, {Malik}, {Man}, {Mandic}, {Mangano}, {Mansell},
  {Manske}, {Mantovani}, {Mapelli}, {Marchesoni}, {Marion}, {M{\'a}rka},
  {M{\'a}rka}, {Markakis}, {Markosyan}, {Markowitz}, {Maros}, {Marquina},
  {Marsat}, {Martelli}, {Martin}, {Martin}, {Martinez}, {Martynov},
  {Masalehdan}, {Mason}, {Massera}, {Masserot}, {Massinger}, {Masso-Reid},
  {Mastrogiovanni}, {Matas}, {Matichard}, {Mavalvala}, {Maynard}, {McCann},
  {McCarthy}, {McClelland}, {McCormick}, {McCuller}, {McGuire}, {McIsaac},
  {McIver}, {McManus}, {McRae}, {McWilliams}, {Meacher}, {Meadors}, {Mehmet},
  {Mehta}, {Villa}, {Melatos}, {Mendell}, {Mercer}, {Mereni}, {Merfeld},
  {Merilh}, {Merritt}, {Merzougui}, {Meshkov}, {Messenger}, {Messick},
  {Metzdorff}, {Meyers}, {Meylahn}, {Mhaske}, {Miani}, {Miao}, {Michaloliakos},
  {Michel}, {Middleton}, {Milano}, {Miller}, {Millhouse}, {Mills}, {Milotti},
  {Milovich-Goff}, {Minazzoli}, {Minenkov}, {Mishkin}, {Mishra}, {Mistry},
  {Mitra}, {Mitrofanov}, {Mitselmakher}, {Mittleman}, {Mo}, {Mogushi},
  {Mohapatra}, {Mohite}, {Molina-Ruiz}, {Mondin}, {Montani}, {Moore}, {Moraru},
  {Morawski}, {Moreno}, {Morisaki}, {Mours}, {Mow-Lowry}, {Mozzon},
  {Muciaccia}, {Mukherjee}, {Mukherjee}, {Mukherjee}, {Mukherjee}, {Mukund},
  {Mullavey}, {Munch}, {Mu{\~n}iz}, {Murray}, {Nagar}, {Nardecchia},
  {Naticchioni}, {Nayak}, {Neil}, {Neilson}, {Nelemans}, {Nelson}, {Nery},
  {Neunzert}, {Ng}, {Ng}, {Nguyen}, {Nguyen}, {Nichols}, {Nichols}, {Nissanke},
  {Nocera}, {Noh}, {North}, {Nothard}, {Nuttall}, {Oberling}, {O'Brien},
  {Oganesyan}, {Ogin}, {Oh}, {Oh}, {Ohme}, {Ohta}, {Okada}, {Oliver},
  {Olivetto}, {Oppermann}, {Oram}, {O'Reilly}, {Ormiston}, {Ortega},
  {O'Shaughnessy}, {Ossokine}, {Osthelder}, {Ottaway}, {Overmier}, {Owen},
  {Pace}, {Pagano}, {Page}, {Pagliaroli}, {Pai}, {Pai}, {Palamos}, {Palashov},
  {Palomba}, {Pan}, {Panda}, {Pang}, {Pankow}, {Pannarale}, {Pant}, {Paoletti},
  {Paoli}, {Parida}, {Parker}, {Pascucci}, {Pasqualetti}, {Passaquieti},
  {Passuello}, {Patricelli}, {Payne}, {Pearlstone}, {Pechsiri}, {Pedersen},
  {Pedraza}, {Pele}, {Penn}, {Perego}, {Perez}, {P{\'e}rigois}, {Perreca},
  {Perri{\`e}s}, {Petermann}, {Pfeiffer}, {Phelps}, {Phukon}, {Piccinni},
  {Pichot}, {Piendibene}, {Piergiovanni}, {Pierro}, {Pillant}, {Pinard},
  {Pinto}, {Piotrzkowski}, {Pirello}, {Pitkin}, {Plastino}, {Poggiani}, {Pong},
  {Ponrathnam}, {Popolizio}, {Porter}, {Powell}, {Prajapati}, {Prasai},
  {Prasanna}, {Pratten}, {Prestegard}, {Principe}, {Prodi}, {Prokhorov},
  {Punturo}, {Puppo}, {P{\"u}rrer}, {Qi}, {Quetschke}, {Quinonez}, {Raab},
  {Raaijmakers}, {Radkins}, {Radulesco}, {Raffai}, {Rafferty}, {Raja}, {Rajan},
  {Rajbhandari}, {Rakhmanov}, {Ramirez}, {Ramos-Buades}, {Rana}, {Rao},
  {Rapagnani}, {Raymond}, {Razzano}, {Read}, {Regimbau}, {Rei}, {Reid},
  {Reitze}, {Rettegno}, {Ricci}, {Richardson}, {Richardson}, {Ricker},
  {Riemenschneider}, {Riles}, {Rizzo}, {Robertson}, {Robinet}, {Rocchi},
  {Rodriguez-Soto}, {Rolland}, {Rollins}, {Roma}, {Romanelli}, {Romano},
  {Romel}, {Romero-Shaw}, {Romie}, {Rose}, {Rose}, {Rose}, {Rosi{\'n}ska},
  {Rosofsky}, {Ross}, {Rowan}, {Rowlinson}, {Roy}, {Roy}, {Roy}, {Ruggi},
  {Rutins}, {Ryan}, {Sachdev}, {Sadecki}, {Sakellariadou}, {Salafia},
  {Salconi}, {Saleem}, {Samajdar}, {Sanchez}, {Sanchez}, {Sanchis-Gual},
  {Sanders}, {Santiago}, {Santos}, {Sarin}, {Sassolas}, {Sathyaprakash},
  {Sauter}, {Savage}, {Savant}, {Sawant}, {Sayah}, {Schaetzl}, {Schale},
  {Scheel}, {Scheuer}, {Schmidt}, {Schnabel}, {Schofield}, {Sch{\"o}nbeck},
  {Schreiber}, {Schulte}, {Schutz}, {Schwarm}, {Schwartz}, {Scott}, {Scott},
  {Seidel}, {Sellers}, {Sengupta}, {Sennett}, {Sentenac}, {Sequino}, {Sergeev},
  {Setyawati}, {Shaddock}, {Shaffer}, {Shahriar}, {Sharifi}, {Sharma},
  {Sharma}, {Shawhan}, {Shen}, {Shikauchi}, {Shink}, {Shoemaker}, {Shoemaker},
  {Shukla}, {ShyamSundar}, {Siellez}, {Sieniawska}, {Sigg}, {Singer}, {Singh},
  {Singh}, {Singha}, {Singhal}, {Sintes}, {Sipala}, {Skliris}, {Slagmolen},
  {Slaven-Blair}, {Smetana}, {Smith}, {Smith}, {Somala}, {Son}, {Soni},
  {Sorazu}, {Sordini}, {Sorrentino}, {Souradeep}, {Sowell}, {Spencer}, {Spera},
  {Srivastava}, {Srivastava}, {Staats}, {Stachie}, {Standke}, {Steer},
  {Steinke}, {Steinlechner}, {Steinlechner}, {Steinmeyer}, {Stevenson},
  {Stocks}, {Stops}, {Stover}, {Strain}, {Stratta}, {Strunk}, {Sturani},
  {Stuver}, {Sudhagar}, {Sudhir}, {Summerscales}, {Sun}, {Sunil}, {Sur},
  {Suresh}, {Sutton}, {Swinkels}, {Szczepa{\'n}czyk}, {Tacca}, {Tait},
  {Talbot}, {Tanasijczuk}, {Tanner}, {Tao}, {T{\'a}pai}, {Tapia}, {San Martin},
  {Tasson}, {Taylor}, {Tenorio}, {Terkowski}, {Thirugnanasambandam}, {Thomas},
  {Thomas}, {Thompson}, {Thondapu}, {Thorne}, {Thrane}, {Tinsman}, {Saravanan},
  {Tiwari}, {Tiwari}, {Tiwari}, {Toland}, {Tonelli}, {Tornasi},
  {Torres-Forn{\'e}}, {Torrie}, {Tosta e Melo}, {T{\"o}yr{\"a}}, {Trail},
  {Travasso}, {Traylor}, {Tringali}, {Tripathee}, {Trovato}, {Trudeau},
  {Tsang}, {Tse}, {Tso}, {Tsukada}, {Tsuna}, {Tsutsui}, {Turconi}, {Ubhi},
  {Udall}, {Ueno}, {Ugolini}, {Unnikrishnan}, {Urban}, {Usman}, {Utina},
  {Vahlbruch}, {Vajente}, {Valdes}, {Valentini}, {van Bakel}, {van Beuzekom},
  {van den Brand}, {Van Den Broeck}, {Vander-Hyde}, {van der Schaaf}, {Van
  Heijningen}, {van Veggel}, {Vardaro}, {Varma}, {Vass}, {Vas{\'u}th},
  {Vecchio}, {Vedovato}, {Veitch}, {Veitch}, {Venkateswara}, {Venugopalan},
  {Verkindt}, {Veske}, {Vetrano}, {Vicer{\'e}}, {Viets}, {Vinciguerra}, {Vine},
  {Vinet}, {Vitale}, {Vivanco}, {Vo}, {Vocca}, {Vorvick}, {Vyatchanin}, {Wade},
  {Wade}, {Wade}, {Walet}, {Walker}, {Wallace}, {Wallace}, {Walsh}, {Wang},
  {Wang}, {Wang}, {Ward}, {Warden}, {Warner}, {Was}, {Watchi}, {Weaver}, {Wei},
  {Weinert}, {Weinstein}, {Weiss}, {Wellmann}, {Wen}, {We{\ss}els},
  {Westhouse}, {Wette}, {Whelan}, {Whiting}, {Whittle}, {Wilken}, {Williams},
  {Willis}, {Willke}, {Winkler}, {Wipf}, {Wittel}, {Woan}, {Woehler},
  {Wofford}, {Wong}, {Wright}, {Wu}, {Wysocki}, {Xiao}, {Yamamoto}, {Yang},
  {Yang}, {Yang}, {Yap}, {Yazback}, {Yeeles}, {Yu}, {Yu}, {Yuen},
  {Zadro{\.z}ny}, {Zadro{\.z}ny}, {Zanolin}, {Zelenova}, {Zendri}, {Zevin},
  {Zhang}, {Zhang}, {Zhang}, {Zhao}, {Zhao}, {Zhou}, {Zhou}, {Zhu},
  {Zimmerman}, {Zlochower}, {Zucker}, {Zweizig}, {LIGO Scientific
  Collaboration}, \& {Virgo Collaboration}}]{2020ApJ...900L..13A}
---. 2020{\natexlab{b}}, \apjl, 900, L13, \dodoi{10.3847/2041-8213/aba493}

\bibitem[{{Abbott} {et~al.}(2021){Abbott}, {Abbott}, {Abraham}, {Acernese},
  {Ackley}, {Adams}, {Adams}, {Adhikari}, {Adya}, {Affeldt}, {Agathos},
  {Agatsuma}, {Aggarwal}, {Aguiar}, {Aiello}, {Ain}, {Ajith}, {Akcay}, {Allen},
  {Allocca}, {Altin}, {Amato}, {Anand}, {Ananyeva}, {Anderson}, {Anderson},
  {Angelova}, {Ansoldi}, {Antelis}, {Antier}, {Appert}, {Arai}, {Araya},
  {Areeda}, {Ar{\`e}ne}, {Arnaud}, {Aronson}, {Arun}, {Asali}, {Ascenzi},
  {Ashton}, {Aston}, {Astone}, {Aubin}, {Aufmuth}, {AultONeal}, {Austin},
  {Avendano}, {Babak}, {Badaracco}, {Bader}, {Bae}, {Baer}, {Bagnasco},
  {Baird}, {Ball}, {Ballardin}, {Ballmer}, {Bals}, {Balsamo}, {Baltus},
  {Banagiri}, {Bankar}, {Bankar}, {Barayoga}, {Barbieri}, {Barish}, {Barker},
  {Barneo}, {Barnum}, {Barone}, {Barr}, {Barsotti}, {Barsuglia}, {Barta},
  {Bartlett}, {Bartos}, {Bassiri}, {Basti}, {Bawaj}, {Bayley}, {Bazzan},
  {Becher}, {B{\'e}csy}, {Bedakihale}, {Bejger}, {Belahcene}, {Beniwal},
  {Benjamin}, {Bennett}, {Bentley}, {Bergamin}, {Berger}, {Bergmann},
  {Bernuzzi}, {Berry}, {Bersanetti}, {Bertolini}, {Betzwieser}, {Bhandare},
  {Bhandari}, {Bhattacharjee}, {Bidler}, {Bilenko}, {Billingsley}, {Birney},
  {Birnholtz}, {Biscans}, {Bischi}, {Biscoveanu}, {Bisht}, {Bitossi},
  {Bizouard}, {Blackburn}, {Blackman}, {Blair}, {Blair}, {Blair}, {Blanch},
  {Bobba}, {Bode}, {Boer}, {Boetzel}, {Bogaert}, {Boldrini}, {Bondu},
  {Bonilla}, {Bonnand}, {Booker}, {Boom}, {Bork}, {Boschi}, {Bose},
  {Bossilkov}, {Boudart}, {Bouffanais}, {Bozzi}, {Bradaschia}, {Brady},
  {Bramley}, {Branchesi}, {Brau}, {Breschi}, {Briant}, {Briggs}, {Brighenti},
  {Brillet}, {Brinkmann}, {Brockill}, {Brooks}, {Brooks}, {Brown}, {Brunett},
  {Bruno}, {Bruntz}, {Buikema}, {Bulik}, {Bulten}, {Buonanno}, {Buscicchio},
  {Buskulic}, {Byer}, {Cabero}, {Cadonati}, {Caesar}, {Cagnoli}, {Cahillane},
  {Calder{\'o}n Bustillo}, {Callaghan}, {Callister}, {Calloni}, {Camp},
  {Canepa}, {Cannon}, {Cao}, {Cao}, {Carapella}, {Carbognani}, {Carney},
  {Carpinelli}, {Carullo}, {Carver}, {Casanueva Diaz}, {Casentini}, {Caudill},
  {Cavagli{\`a}}, {Cavalier}, {Cavalieri}, {Cella}, {Cerd{\'a}-Dur{\'a}n},
  {Cesarini}, {Chaibi}, {Chakravarti}, {Chan}, {Chan}, {Chandra}, {Chanial},
  {Chao}, {Charlton}, {Chase}, {Chassande-Mottin}, {Chatterjee},
  {Chattopadhyay}, {Chaturvedi}, {Chatziioannou}, {Chen}, {Chen}, {Chen},
  {Chen}, {Cheng}, {Cheong}, {Chia}, {Chiadini}, {Chierici}, {Chincarini},
  {Chiummo}, {Cho}, {Cho}, {Cho}, {Choate}, {Christensen}, {Chu}, {Chua},
  {Chung}, {Chung}, {Ciani}, {Ciecielag}, {Cie{\'s}lar}, {Cifaldi}, {Ciobanu},
  {Ciolfi}, {Cipriano}, {Cirone}, {Clara}, {Clark}, {Clark}, {Clarke},
  {Clearwater}, {Clesse}, {Cleva}, {Coccia}, {Cohadon}, {Cohen}, {Colleoni},
  {Collette}, {Collins}, {Colpi}, {Constancio}, {Conti}, {Cooper}, {Corban},
  {Corbitt}, {Cordero-Carri{\'o}n}, {Corezzi}, {Corley}, {Cornish}, {Corre},
  {Corsi}, {Cortese}, {Costa}, {Cotesta}, {Coughlin}, {Coughlin}, {Coulon},
  {Countryman}, {Cousins}, {Couvares}, {Covas}, {Coward}, {Cowart}, {Coyne},
  {Coyne}, {Creighton}, {Creighton}, {Croquette}, {Crowder}, {Cudell},
  {Cullen}, {Cumming}, {Cummings}, {Cunningham}, {Cuoco}, {Cury{\l}o},
  {Canton}, {D{\'a}lya}, {Dana}, {DaneshgaranBajastani}, {D'Angelo}, {Danila},
  {Danilishin}, {D'Antonio}, {Danzmann}, {Darsow-Fromm}, {Dasgupta}, {Datrier},
  {Dattilo}, {Dave}, {Davier}, {Davies}, {Davis}, {Daw}, {Dean}, {DeBra},
  {Deenadayalan}, {Degallaix}, {De Laurentis}, {Del{\'e}glise}, {Del Favero},
  {De Lillo}, {De Lillo}, {Del Pozzo}, {DeMarchi}, {De Matteis}, {D'Emilio},
  {Demos}, {Denker}, {Dent}, {Depasse}, {De Pietri}, {De Rosa}, {De Rossi},
  {DeSalvo}, {de Varona}, {Dhurandhar}, {D{\'\i}az}, {Diaz-Ortiz}, {Didio},
  {Dietrich}, {Di Fiore}, {DiFronzo}, {Di Giorgio}, {Di Giovanni}, {Di
  Giovanni}, {Di Girolamo}, {Di Lieto}, {Ding}, {Di Pace}, {Di Palma}, {Di
  Renzo}, {Divakarla}, {Dmitriev}, {Doctor}, {D'Onofrio}, {Donovan}, {Dooley},
  {Doravari}, {Dorrington}, {Downes}, {Drago}, {Driggers}, {Du}, {Ducoin},
  {Dupej}, {Durante}, {D'Urso}, {Duverne}, {Dwyer}, {Easter}, {Eddolls},
  {Edelman}, {Edo}, {Edy}, {Effler}, {Eichholz}, {Eikenberry}, {Eisenmann},
  {Eisenstein}, {Ejlli}, {Errico}, {Essick}, {Estell{\'e}s}, {Estevez},
  {Etienne}, {Etzel}, {Evans}, {Evans}, {Ewing}, {Fafone}, {Fair}, {Fairhurst},
  {Fan}, {Farah}, {Farinon}, {Farr}, {Farr}, {Fauchon-Jones}, {Favata}, {Fays},
  {Fazio}, {Feicht}, {Fejer}, {Feng}, {Fenyvesi}, {Ferguson},
  {Fernandez-Galiana}, {Ferrante}, {Ferreira}, {Fidecaro}, {Figura}, {Fiori},
  {Fiorucci}, {Fishbach}, {Fisher}, {Fishner}, {Fittipaldi}, {Fitz-Axen},
  {Fiumara}, {Flaminio}, {Floden}, {Flynn}, {Fong}, {Font}, {Forsyth},
  {Fournier}, {Frasca}, {Frasconi}, {Frei}, {Freise}, {Frey}, {Frey},
  {Fritschel}, {Frolov}, {Fronz{\'e}}, {Fulda}, {Fyffe}, {Gabbard}, {Gadre},
  {Gaebel}, {Gair}, {Gais}, {Galaudage}, {Gamba}, {Ganapathy}, {Ganguly},
  {Gaonkar}, {Garaventa}, {Garc{\'\i}a-Quir{\'o}s}, {Garufi}, {Gateley},
  {Gaudio}, {Gayathri}, {Gemme}, {Gennai}, {George}, {George}, {George},
  {Gergely}, {Ghonge}, {Ghosh}, {Ghosh}, {Ghosh}, {Giacomazzo}, {Giacoppo},
  {Giaime}, {Giardina}, {Gibson}, {Gier}, {Gill}, {Giri}, {Glanzer}, {Gleckl},
  {Godwin}, {Goetz}, {Goetz}, {Gohlke}, {Goncharov}, {Gonz{\'a}lez},
  {Gopakumar}, {Gossan}, {Gosselin}, {Gouaty}, {Grace}, {Grado}, {Granata},
  {Granata}, {Grant}, {Gras}, {Grassia}, {Gray}, {Gray}, {Greco}, {Green},
  {Green}, {Gretarsson}, {Griggs}, {Grignani}, {Grimaldi}, {Grimes}, {Grimm},
  {Grote}, {Grunewald}, {Gruning}, {Guerrero}, {Guidi}, {Guimaraes},
  {Guix{\'e}}, {Gulati}, {Guo}, {Gupta}, {Gupta}, {Gupta}, {Gustafson},
  {Gustafson}, {Guzman}, {Haegel}, {Halim}, {Hall}, {Hamilton}, {Hammond},
  {Haney}, {Hanke}, {Hanks}, {Hanna}, {Hannam}, {Hannuksela}, {Hannuksela},
  {Hansen}, {Hansen}, {Hanson}, {Harder}, {Hardwick}, {Haris}, {Harms},
  {Harry}, {Harry}, {Hartwig}, {Hasskew}, {Haster}, {Haughian}, {Hayes},
  {Healy}, {Heidmann}, {Heintze}, {Heinze}, {Heinzel}, {Heitmann}, {Hellman},
  {Hello}, {Helmling-Cornell}, {Hemming}, {Hendry}, {Heng}, {Hennes}, {Hennig},
  {Hennig}, {Hernandez Vivanco}, {Heurs}, {Hild}, {Hill}, {Hines}, {Hochheim},
  {Hofgard}, {Hofman}, {Hohmann}, {Holgado}, {Holland}, {Hollows}, {Holmes},
  {Holt}, {Holz}, {Hopkins}, {Horst}, {Hough}, {Howell}, {Hoy}, {Hoyland},
  {Huang}, {H{\"u}bner}, {Huddart}, {Huerta}, {Hughey}, {Hui}, {Husa},
  {Huttner}, {Hutzler}, {Huxford}, {Huynh-Dinh}, {Idzkowski}, {Iess},
  {Imperato}, {Inchauspe}, {Ingram}, {Intini}, {Isi}, {Iyer},
  {JaberianHamedan}, {Jacqmin}, {Jadhav}, {Jadhav}, {James}, {Jani},
  {Janssens}, {Janthalur}, {Jaranowski}, {Jariwala}, {Jaume}, {Jenkins},
  {Jeunon}, {Jiang}, {Johns}, {Johnson-McDaniel}, {Jones}, {Jones}, {Jones},
  {Jones}, {Jones}, {Jonker}, {Ju}, {Junker}, {Kalaghatgi}, {Kalogera},
  {Kamai}, {Kandhasamy}, {Kang}, {Kanner}, {Kapadia}, {Kapasi}, {Karathanasis},
  {Karki}, {Kashyap}, {Kasprzack}, {Kastaun}, {Katsanevas}, {Katsavounidis},
  {Katzman}, {Kawabe}, {K{\'e}f{\'e}lian}, {Keitel}, {Key}, {Khadka},
  {Khalili}, {Khan}, {Khan}, {Khazanov}, {Khetan}, {Khursheed}, {Kijbunchoo},
  {Kim}, {Kim}, {Kim}, {Kim}, {Kim}, {Kim}, {Kimball}, {King}, {Kinley-Hanlon},
  {Kirchhoff}, {Kissel}, {Kleybolte}, {Klimenko}, {Knowles}, {Knyazev}, {Koch},
  {Koehlenbeck}, {Koekoek}, {Koley}, {Kolstein}, {Komori}, {Kondrashov},
  {Kontos}, {Koper}, {Korobko}, {Korth}, {Kovalam}, {Kozak}, {Kr{\"a}mer},
  {Kringel}, {Krishnendu}, {Kr{\'o}lak}, {Kuehn}, {Kumar}, {Kumar}, {Kumar},
  {Kumar}, {Kuns}, {Kwang}, {Lackey}, {Laghi}, {Lalande}, {Lam}, {Lamberts},
  {Landry}, {Lane}, {Lang}, {Lange}, {Lantz}, {Lanza}, {La Rosa},
  {Lartaux-Vollard}, {Lasky}, {Laxen}, {Lazzarini}, {Lazzaro}, {Leaci},
  {Leavey}, {Lecoeuche}, {Lee}, {Lee}, {Lee}, {Lee}, {Lehmann}, {Leon},
  {Leroy}, {Letendre}, {Levin}, {Li}, {Li}, {Li}, {Li}, {Li}, {Linde},
  {Linker}, {Linley}, {Littenberg}, {Liu}, {Liu}, {Llorens-Monteagudo}, {Lo},
  {Lockwood}, {London}, {Longo}, {Lorenzini}, {Loriette}, {Lormand}, {Losurdo},
  {Lough}, {Lousto}, {Lovelace}, {L{\"u}ck}, {Lumaca}, {Lundgren}, {Ma},
  {Macas}, {MacInnis}, {Macleod}, {MacMillan}, {Macquet}, {Maga{\~n}a
  Hernandez}, {Maga{\~n}a-Sandoval}, {Magazz{\~A}{\textonesuperior}}, {Magee},
  {Majorana}, {Maksimovic}, {Maliakal}, {Malik}, {Man}, {Mandic}, {Mangano},
  {Mansell}, {Manske}, {Mantovani}, {Mapelli}, {Marchesoni}, {Marion},
  {M{\'a}rka}, {M{\'a}rka}, {Markakis}, {Markosyan}, {Markowitz}, {Maros},
  {Marquina}, {Marsat}, {Martelli}, {Martin}, {Martin}, {Martinez}, {Martinez},
  {Martynov}, {Masalehdan}, {Mason}, {Massera}, {Masserot}, {Massinger},
  {Masso-Reid}, {Mastrogiovanni}, {Matas}, {Mateu-Lucena}, {Matichard},
  {Matiushechkina}, {Mavalvala}, {Maynard}, {McCann}, {McCarthy}, {McClelland},
  {McCormick}, {McCuller}, {McGuire}, {McIsaac}, {McIver}, {McManus}, {McRae},
  {McWilliams}, {Meacher}, {Meadors}, {Mehmet}, {Mehta}, {Melatos}, {Melchor},
  {Mendell}, {Menendez-Vazquez}, {Mercer}, {Mereni}, {Merfeld}, {Merilh},
  {Merritt}, {Merzougui}, {Meshkov}, {Messenger}, {Messick}, {Metzdorff},
  {Meyers}, {Meylahn}, {Mhaske}, {Miani}, {Miao}, {Michaloliakos}, {Michel},
  {Middleton}, {Milano}, {Miller}, {Millhouse}, {Mills}, {Milotti},
  {Milovich-Goff}, {Minazzoli}, {Minenkov}, {Mir}, {Mishkin}, {Mishra},
  {Mistry}, {Mitra}, {Mitrofanov}, {Mitselmakher}, {Mittleman}, {Mo},
  {Mogushi}, {Mohapatra}, {Mohite}, {Molina}, {Molina-Ruiz}, {Mondin},
  {Montani}, {Moore}, {Moraru}, {Morawski}, {Moreno}, {Morisaki}, {Mours},
  {Mow-Lowry}, {Mozzon}, {Muciaccia}, {Mukherjee}, {Mukherjee}, {Mukherjee},
  {Mukherjee}, {Mukund}, {Mullavey}, {Munch}, {Mu{\~n}iz}, {Murray}, {Nadji},
  {Nagar}, {Nardecchia}, {Naticchioni}, {Nayak}, {Neil}, {Neilson}, {Nelemans},
  {Nelson}, {Nery}, {Neunzert}, {Nitz}, {Ng}, {Ng}, {Nguyen}, {Nguyen},
  {Nguyen}, {Nichols}, {Nissanke}, {Nocera}, {Noh}, {North}, {Nothard},
  {Nuttall}, {Oberling}, {O'Brien}, {O'Dell}, {Oganesyan}, {Ogin}, {Oh}, {Oh},
  {Ohme}, {Ohta}, {Okada}, {Olivetto}, {Oppermann}, {Oram}, {O'Reilly},
  {Ormiston}, {Ortega}, {O'Shaughnessy}, {Ossokine}, {Osthelder}, {Ottaway},
  {Overmier}, {Owen}, {Pace}, {Pagano}, {Page}, {Pagliaroli}, {Pai}, {Pai},
  {Palamos}, {Palashov}, {Palomba}, {Pan}, {Panda}, {Pang}, {Pankow},
  {Pannarale}, {Pant}, {Paoletti}, {Paoli}, {Paolone}, {Parker}, {Pascucci},
  {Pasqualetti}, {Passaquieti}, {Passuello}, {Patel}, {Patricelli}, {Payne},
  {Pechsiri}, {Pedraza}, {Pegoraro}, {Pele}, {Penn}, {Perego}, {Perez},
  {P{\'e}rigois}, {Perreca}, {Perri{\`e}s}, {Petermann}, {Petterson},
  {Pfeiffer}, {Pham}, {Phukon}, {Piccinni}, {Pichot}, {Piendibene},
  {Piergiovanni}, {Pierini}, {Pierro}, {Pillant}, {Pilo}, {Pinard}, {Pinto},
  {Piotrzkowski}, {Pirello}, {Pitkin}, {Placidi}, {Plastino}, {Pluchar},
  {Poggiani}, {Polini}, {Pong}, {Ponrathnam}, {Popolizio}, {Porter},
  {Poverman}, {Powell}, {Pracchia}, {Prajapati}, {Prasai}, {Prasanna},
  {Pratten}, {Prestegard}, {Principe}, {Prodi}, {Prokhorov}, {Prosposito},
  {Prudenzi}, {Puecher}, {Punturo}, {Puosi}, {Puppo}, {P{\"u}rrer}, {Qi},
  {Quetschke}, {Quinonez}, {Quitzow-James}, {Raab}, {Raaijmakers}, {Radkins},
  {Radulesco}, {Raffai}, {Rafferty}, {Rail}, {Raja}, {Rajan}, {Rajbhandari},
  {Rakhmanov}, {Ramirez}, {Ramirez}, {Ramos-Buades}, {Rana}, {Rao},
  {Rapagnani}, {Rapol}, {Ratto}, {Raymond}, {Razzano}, {Read}, {Regimbau},
  {Rei}, {Reid}, {Reitze}, {Rettegno}, {Ricci}, {Richardson}, {Richardson},
  {Richardson}, {Ricker}, {Riemenschneider}, {Riles}, {Rizzo}, {Robertson},
  {Robinet}, {Rocchi}, {Rocha}, {Rodriguez}, {Rodriguez-Soto}, {Rolland},
  {Rollins}, {Roma}, {Romanelli}, {Romano}, {Romel}, {Romero}, {Romero-Shaw},
  {Romie}, {Ronchini}, {Rose}, {Rose}, {Rose}, {Rosell}, {Rosi{\'n}ska},
  {Rosofsky}, {Ross}, {Rowan}, {Rowlinson}, {Roy}, {Roy}, {Ruggi}, {Ryan},
  {Sachdev}, {Sadecki}, {Sadiq}, {Sakellariadou}, {Salafia}, {Salconi},
  {Saleem}, {Samajdar}, {Sanchez}, {Sanchez}, {Sanchez}, {Sanchis-Gual},
  {Sanders}, {Sandles}, {Santiago}, {Santos}, {Saravanan}, {Sarin}, {Sassolas},
  {Sathyaprakash}, {Sauter}, {Savage}, {Savant}, {Sawant}, {Sayah}, {Schaetzl},
  {Schale}, {Scheel}, {Scheuer}, {Schindler-Tyka}, {Schmidt}, {Schnabel},
  {Schofield}, {Sch{\"o}nbeck}, {Schreiber}, {Schulte}, {Schutz}, {Schwarm},
  {Schwartz}, {Scott}, {Scott}, {Seglar-Arroyo}, {Seidel}, {Sellers},
  {Sengupta}, {Sennett}, {Sentenac}, {Sequino}, {Sergeev}, {Setyawati},
  {Shaffer}, {Shahriar}, {Sharifi}, {Sharma}, {Sharma}, {Shawhan}, {Shen},
  {Shikauchi}, {Shink}, {Shoemaker}, {Shoemaker}, {Shukla}, {ShyamSundar},
  {Sieniawska}, {Sigg}, {Singer}, {Singh}, {Singh}, {Singha}, {Singhal},
  {Sintes}, {Sipala}, {Skliris}, {Slagmolen}, {Slaven-Blair}, {Smetana},
  {Smith}, {Smith}, {Somala}, {Son}, {Soni}, {Soni}, {Sorazu}, {Sordini},
  {Sorrentino}, {Sorrentino}, {Soulard}, {Souradeep}, {Sowell}, {Spencer},
  {Spera}, {Srivastava}, {Srivastava}, {Staats}, {Stachie}, {Steer},
  {Steinhoff}, {Steinke}, {Steinlechner}, {Steinlechner}, {Steinmeyer},
  {Stevenson}, {Stolle-McAllister}, {Stops}, {Stover}, {Strain}, {Stratta},
  {Strunk}, {Sturani}, {Stuver}, {S{\"u}dbeck}, {Sudhagar}, {Sudhir}, {Suh},
  {Summerscales}, {Sun}, {Sun}, {Sunil}, {Sur}, {Suresh}, {Sutton}, {Swinkels},
  {Szczepa{\'n}czyk}, {Tacca}, {Tait}, {Talbot}, {Tanasijczuk}, {Tanner},
  {Tao}, {Tapia}, {Tapia San Martin}, {Tasson}, {Taylor}, {Tenorio},
  {Terkowski}, {Thirugnanasambandam}, {Thomas}, {Thomas}, {Thomas}, {Thompson},
  {Thondapu}, {Thorne}, {Thrane}, {Tiwari}, {Tiwari}, {Tiwari}, {Toland},
  {Tolley}, {Tonelli}, {Tornasi}, {Torres-Forn{\'e}}, {Torrie}, {e Melo},
  {T{\"o}yr{\"a}}, {Tran}, {Trapananti}, {Travasso}, {Traylor}, {Tringali},
  {Tripathee}, {Trovato}, {Trudeau}, {Tsai}, {Tsang}, {Tse}, {Tso}, {Tsukada},
  {Tsuna}, {Tsutsui}, {Turconi}, {Ubhi}, {Udall}, {Ueno}, {Ugolini},
  {Unnikrishnan}, {Urban}, {Usman}, {Utina}, {Vahlbruch}, {Vajente}, {Vajpeyi},
  {Valdes}, {Valentini}, {Valsan}, {van Bakel}, {van Beuzekom}, {van den
  Brand}, {Van Den Broeck}, {Vander-Hyde}, {van der Schaaf}, {van Heijningen},
  {Vardaro}, {Vargas}, {Varma}, {Vass}, {Vas{\'u}th}, {Vecchio}, {Vedovato},
  {Veitch}, {Veitch}, {Venkateswara}, {Venneberg}, {Venugopalan}, {Verkindt},
  {Verma}, {Veske}, {Vetrano}, {Vicer{\'e}}, {Viets}, {Vijaykumar},
  {Villa-Ortega}, {Vinet}, {Vitale}, {Vo}, {Vocca}, {Vorvick}, {Vyatchanin},
  {Wade}, {Wade}, {Wade}, {Walet}, {Walker}, {Wallace}, {Wallace}, {Walsh},
  {Wang}, {Wang}, {Wang}, {Wang}, {Ward}, {Warner}, {Was}, {Washington},
  {Watchi}, {Weaver}, {Wei}, {Weinert}, {Weinstein}, {Weiss}, {Wellmann},
  {Wen}, {We{\ss}els}, {Westhouse}, {Wette}, {Whelan}, {White}, {White},
  {Whiting}, {Whittle}, {Wilken}, {Williams}, {Williams}, {Williamson},
  {Willis}, {Willke}, {Wilson}, {Wimmer}, {Winkler}, {Wipf}, {Woan}, {Woehler},
  {Wofford}, {Wong}, {Wrangel}, {Wright}, {Wu}, {Wysocki}, {Xiao}, {Yamamoto},
  {Yang}, {Yang}, {Yang}, {Yap}, {Yeeles}, {Yoon}, {Yu}, {Yu}, {Yuen},
  {Zadro{\.Z}ny}, {Zanolin}, {Zelenova}, {Zendri}, {Zevin}, {Zhang}, {Zhang},
  {Zhang}, {Zhang}, {Zhao}, {Zhao}, {Zheng}, {Zhou}, {Zhou}, {Zhu},
  {Zimmerman}, {Zlochower}, {Zucker}, {Zweizig}, {LIGO Scientific
  Collaboration}, \& {Virgo Collaboration}}]{2021PhRvX..11b1053A}
---. 2021, Physical Review X, 11, 021053, \dodoi{10.1103/PhysRevX.11.021053}

\bibitem[{{Abbott} {et~al.}(2022){Abbott}, {Abbott}, {Acernese}, {Ackley},
  {Adams}, {Adhikari}, {Adhikari}, {Adya}, {Affeldt}, {Agarwal}, {Agathos},
  {Agatsuma}, {Aggarwal}, {Aguiar}, {Aiello}, {Ain}, {Ajith}, {Akutsu},
  {Albanesi}, {Allocca}, {Altin}, {Amato}, {Anand}, {Anand}, {Ananyeva},
  {Anderson}, {Anderson}, {Ando}, {Andrade}, {Andres}, {Andri{\'c}},
  {Angelova}, {Ansoldi}, {Antelis}, {Antier}, {Appert}, {Arai}, {Arai}, {Arai},
  {Araki}, {Araya}, {Araya}, {Areeda}, {Ar{\`e}ne}, {Aritomi}, {Arnaud},
  {Aronson}, {Arun}, {Asada}, {Asali}, {Ashton}, {Aso}, {Assiduo}, {Aston},
  {Astone}, {Aubin}, {Austin}, {Babak}, {Badaracco}, {Bader}, {Badger}, {Bae},
  {Bae}, {Baer}, {Bagnasco}, {Bai}, {Baiotti}, {Baird}, {Bajpai}, {Ball},
  {Ballardin}, {Ballmer}, {Balsamo}, {Baltus}, {Banagiri}, {Bankar},
  {Barayoga}, {Barbieri}, {Barish}, {Barker}, {Barneo}, {Barone}, {Barr},
  {Barsotti}, {Barsuglia}, {Barta}, {Bartlett}, {Barton}, {Bartos}, {Bassiri},
  {Basti}, {Bawaj}, {Bayley}, {Baylor}, {Bazzan}, {B{\'e}csy}, {Bedakihale},
  {Bejger}, {Belahcene}, {Benedetto}, {Beniwal}, {Bennett}, {Bentley},
  {BenYaala}, {Bergamin}, {Berger}, {Bernuzzi}, {Berry}, {Bersanetti},
  {Bertolini}, {Betzwieser}, {Beveridge}, {Bhandare}, {Bhardwaj},
  {Bhattacharjee}, {Bhaumik}, {Bilenko}, {Billingsley}, {Bini}, {Birney},
  {Birnholtz}, {Biscans}, {Bischi}, {Biscoveanu}, {Bisht}, {Biswas}, {Bitossi},
  {Bizouard}, {Blackburn}, {Blair}, {Blair}, {Blair}, {Bobba}, {Bode}, {Boer},
  {Bogaert}, {Boldrini}, {Bonavena}, {Bondu}, {Bonilla}, {Bonnand}, {Booker},
  {Boom}, {Bork}, {Boschi}, {Bose}, {Bose}, {Bossilkov}, {Boudart},
  {Bouffanais}, {Bozzi}, {Bradaschia}, {Brady}, {Bramley}, {Branch},
  {Branchesi}, {Brau}, {Breschi}, {Briant}, {Briggs}, {Brillet}, {Brinkmann},
  {Brockill}, {Brooks}, {Brooks}, {Brown}, {Brunett}, {Bruno}, {Bruntz},
  {Bryant}, {Bulik}, {Bulten}, {Buonanno}, {Buscicchio}, {Buskulic}, {Buy},
  {Byer}, {Cadonati}, {Cagnoli}, {Cahillane}, {Calder{\'o}n Bustillo},
  {Callaghan}, {Callister}, {Calloni}, {Cameron}, {Camp}, {Canepa},
  {Canevarolo}, {Cannavacciuolo}, {Cannon}, {Cao}, {Cao}, {Capocasa}, {Capote},
  {Carapella}, {Carbognani}, {Carlin}, {Carney}, {Carpinelli}, {Carrillo},
  {Carullo}, {Carver}, {Casanueva Diaz}, {Casentini}, {Castaldi}, {Caudill},
  {Cavagli{\`a}}, {Cavalier}, {Cavalieri}, {Ceasar}, {Cella},
  {Cerd{\'a}-Dur{\'a}n}, {Cesarini}, {Chaibi}, {Chakravarti}, {Chalathadka
  Subrahmanya}, {Champion}, {Chan}, {Chan}, {Chan}, {Chan}, {Chan}, {Chandra},
  {Chanial}, {Chao}, {Charlton}, {Chase}, {Chassande-Mottin}, {Chatterjee},
  {Chatterjee}, {Chatterjee}, {Chaturvedi}, {Chaty}, {Chatziioannou}, {Chen},
  {Chen}, {Chen}, {Chen}, {Chen}, {Chen}, {Chen}, {Chen}, {Cheng}, {Cheong},
  {Cheung}, {Chia}, {Chiadini}, {Chiang}, {Chiarini}, {Chierici}, {Chincarini},
  {Chiofalo}, {Chiummo}, {Cho}, {Cho}, {Choudhary}, {Choudhary}, {Christensen},
  {Chu}, {Chu}, {Chu}, {Chua}, {Chung}, {Ciani}, {Ciecielag}, {Cie{\'s}lar},
  {Cifaldi}, {Ciobanu}, {Ciolfi}, {Cipriano}, {Cirone}, {Clara}, {Clark},
  {Clark}, {Clarke}, {Clearwater}, {Clesse}, {Cleva}, {Coccia}, {Codazzo},
  {Cohadon}, {Cohen}, {Cohen}, {Colleoni}, {Collette}, {Colombo}, {Colpi},
  {Compton}, {Constancio}, {Conti}, {Cooper}, {Corban}, {Corbitt},
  {Cordero-Carri{\'o}n}, {Corezzi}, {Corley}, {Cornish}, {Corre}, {Corsi},
  {Cortese}, {Costa}, {Cotesta}, {Coughlin}, {Coulon}, {Countryman}, {Cousins},
  {Couvares}, {Coward}, {Cowart}, {Coyne}, {Coyne}, {Creighton}, {Creighton},
  {Criswell}, {Croquette}, {Crowder}, {Cudell}, {Cullen}, {Cumming},
  {Cummings}, {Cunningham}, {Cuoco}, {Cury{\l}o}, {Dabadie}, {Dal Canton},
  {Dall'Osso}, {D{\'a}lya}, {Dana}, {Daneshgaran Bajastani}, {D'Angelo},
  {Danilishin}, {D'Antonio}, {Danzmann}, {Darsow-Fromm}, {Dasgupta}, {Datrier},
  {Datta}, {Dattilo}, {Dave}, {Davier}, {Davies}, {Davis}, {Davis}, {Daw},
  {Dean}, {DeBra}, {Deenadayalan}, {Degallaix}, {De Laurentis},
  {Del{\'e}glise}, {Del Favero}, {De Lillo}, {De Lillo}, {Del Pozzo}, {De
  Marchi}, {De Matteis}, {D'Emilio}, {Demos}, {Dent}, {Depasse}, {De Pietri},
  {De Rosa}, {De Rossi}, {De Salvo}, {De Simone}, {Dhurandhar}, {D{\'\i}az},
  {Diaz-Ortiz}, {Didio}, {Dietrich}, {Di Fiore}, {Di Fronzo}, {Di Giorgio}, {Di
  Giovanni}, {Di Giovanni}, {Di Girolamo}, {Di Lieto}, {Ding}, {Di Pace}, {Di
  Palma}, {Di Renzo}, {Divakarla}, {Dmitriev}, {Doctor}, {D'Onofrio},
  {Donovan}, {Dooley}, {Doravari}, {Dorrington}, {Drago}, {Driggers}, {Drori},
  {Ducoin}, {Dupej}, {Durante}, {D'Urso}, {Duverne}, {Dwyer}, {Eassa},
  {Easter}, {Ebersold}, {Eckhardt}, {Eddolls}, {Edelman}, {Edo}, {Edy},
  {Effler}, {Eguchi}, {Eichholz}, {Eikenberry}, {Eisenmann}, {Eisenstein},
  {Ejlli}, {Engelby}, {Enomoto}, {Errico}, {Essick}, {Estell{\'e}s}, {Estevez},
  {Etienne}, {Etzel}, {Evans}, {Evans}, {Ewing}, {Fafone}, {Fair}, {Fairhurst},
  {Farah}, {Farinon}, {Farr}, {Farr}, {Farrow}, {Fauchon-Jones}, {Favaro},
  {Favata}, {Fays}, {Fazio}, {Feicht}, {Fejer}, {Fenyvesi}, {Ferguson},
  {Fernandez-Galiana}, {Ferrante}, {Ferreira}, {Fidecaro}, {Figura}, {Fiori},
  {Fishbach}, {Fisher}, {Fittipaldi}, {Fiumara}, {Flaminio}, {Floden}, {Fong},
  {Font}, {Fornal}, {Forsyth}, {Franke}, {Frasca}, {Frasconi}, {Frederick},
  {Freed}, {Frei}, {Freise}, {Frey}, {Fritschel}, {Frolov}, {Fronz{\'e}},
  {Fujii}, {Fujikawa}, {Fukunaga}, {Fukushima}, {Fulda}, {Fyffe}, {Gabbard},
  {Gadre}, {Gair}, {Gais}, {Galaudage}, {Gamba}, {Ganapathy}, {Ganguly}, {Gao},
  {Gaonkar}, {Garaventa}, {Garc{\'\i}a-N{\'u}{\~n}ez},
  {Garc{\'\i}a-Quir{\'o}s}, {Garufi}, {Gateley}, {Gaudio}, {Gayathri}, {Ge},
  {Gemme}, {Gennai}, {George}, {Gerberding}, {Gergely}, {Gewecke}, {Ghonge},
  {Ghosh}, {Ghosh}, {Ghosh}, {Ghosh}, {Giacomazzo}, {Giacoppo}, {Giaime},
  {Giardina}, {Gibson}, {Gier}, {Giesler}, {Giri}, {Gissi}, {Glanzer},
  {Gleckl}, {Godwin}, {Goetz}, {Goetz}, {Gohlke}, {Goncharov}, {Gonz{\'a}lez},
  {Gopakumar}, {Gosselin}, {Gouaty}, {Gould}, {Grace}, {Grado}, {Granata},
  {Granata}, {Grant}, {Gras}, {Grassia}, {Gray}, {Gray}, {Greco}, {Green},
  {Green}, {Gretarsson}, {Gretarsson}, {Griffith}, {Griffiths}, {Griggs},
  {Grignani}, {Grimaldi}, {Grimm}, {Grote}, {Grunewald}, {Gruning}, {Guerra},
  {Guidi}, {Guimaraes}, {Guix{\'e}}, {Gulati}, {Guo}, {Guo}, {Gupta}, {Gupta},
  {Gupta}, {Gustafson}, {Gustafson}, {Guzman}, {Ha}, {Haegel}, {Hagiwara},
  {Haino}, {Halim}, {Hall}, {Hamilton}, {Hammond}, {Han}, {Haney}, {Hanks},
  {Hanna}, {Hannam}, {Hannuksela}, {Hansen}, {Hansen}, {Hanson}, {Harder},
  {Hardwick}, {Haris}, {Harms}, {Harry}, {Harry}, {Hartwig}, {Hasegawa},
  {Haskell}, {Hasskew}, {Haster}, {Hattori}, {Haughian}, {Hayakawa}, {Hayama},
  {Hayes}, {Healy}, {Heidmann}, {Heidt}, {Heintze}, {Heinze}, {Heinzel},
  {Heitmann}, {Hellman}, {Hello}, {Helmling-Cornell}, {Hemming}, {Hendry},
  {Heng}, {Hennes}, {Hennig}, {Hennig}, {Hernandez}, {Hernandez Vivanco},
  {Heurs}, {Hild}, {Hill}, {Himemoto}, {Hines}, {Hiranuma}, {Hirata}, {Hirose},
  {Hochheim}, {Hofman}, {Hohmann}, {Holcomb}, {Holland}, {Hollows}, {Holmes},
  {Holt}, {Holz}, {Hong}, {Hopkins}, {Hough}, {Hourihane}, {Howell}, {Hoy},
  {Hoyland}, {Hreibi}, {Hsieh}, {Hsu}, {Huang}, {Huang}, {Huang}, {Huang},
  {Huang}, {Huang}, {H{\"u}bner}, {Huddart}, {Hughey}, {Hui}, {Hui}, {Husa},
  {Huttner}, {Huxford}, {Huynh-Dinh}, {Ide}, {Idzkowski}, {Iess}, {Ikenoue},
  {Imam}, {Inayoshi}, {Ingram}, {Inoue}, {Ioka}, {Isi}, {Isleif}, {Ito},
  {Itoh}, {Iyer}, {Izumi}, {Jaberian Hamedan}, {Jacqmin}, {Jadhav}, {Jadhav},
  {James}, {Jan}, {Jani}, {Janquart}, {Janssens}, {Janthalur}, {Jaranowski},
  {Jariwala}, {Jaume}, {Jenkins}, {Jenner}, {Jeon}, {Jeunon}, {Jia}, {Jin},
  {Johns}, {Jones}, {Jones}, {Jones}, {Jones}, {Jones}, {Jonker}, {Ju}, {Jung},
  {Jung}, {Junker}, {Juste}, {Kaihotsu}, {Kajita}, {Kakizaki}, {Kalaghatgi},
  {Kalogera}, {Kamai}, {Kamiizumi}, {Kanda}, {Kandhasamy}, {Kang}, {Kanner},
  {Kao}, {Kapadia}, {Kapasi}, {Karat}, {Karathanasis}, {Karki}, {Kashyap},
  {Kasprzack}, {Kastaun}, {Katsanevas}, {Katsavounidis}, {Katzman}, {Kaur},
  {Kawabe}, {Kawaguchi}, {Kawai}, {Kawasaki}, {K{\'e}f{\'e}lian}, {Keitel},
  {Key}, {Khadka}, {Khalili}, {Khan}, {Khazanov}, {Khetan}, {Khursheed},
  {Kijbunchoo}, {Kim}, {Kim}, {Kim}, {Kim}, {Kim}, {Kim}, {Kimball}, {Kimura},
  {Kinley-Hanlon}, {Kirchhoff}, {Kissel}, {Kita}, {Kitazawa}, {Kleybolte},
  {Klimenko}, {Knee}, {Knowles}, {Knyazev}, {Koch}, {Koekoek}, {Kojima},
  {Kokeyama}, {Koley}, {Kolitsidou}, {Kolstein}, {Komori}, {Kondrashov},
  {Kong}, {Kontos}, {Koper}, {Korobko}, {Kotake}, {Kovalam}, {Kozak},
  {Kozakai}, {Kozu}, {Kringel}, {Krishnendu}, {Kr{\'o}lak}, {Kuehn}, {Kuei},
  {Kuijer}, {Kumar}, {Kumar}, {Kumar}, {Kumar}, {Kume}, {Kuns}, {Kuo}, {Kuo},
  {Kuromiya}, {Kuroyanagi}, {Kusayanagi}, {Kuwahara}, {Kwak}, {Lagabbe},
  {Laghi}, {Lalande}, {Lam}, {Lamberts}, {Landry}, {Lane}, {Lang}, {Lange},
  {Lantz}, {La Rosa}, {Lartaux-Vollard}, {Lasky}, {Laxen}, {Lazzarini},
  {Lazzaro}, {Leaci}, {Leavey}, {Lecoeuche}, {Lee}, {Lee}, {Lee}, {Lee}, {Lee},
  {Lee}, {Lehmann}, {Lema{\^\i}tre}, {Leonardi}, {Leroy}, {Letendre},
  {Levesque}, {Levin}, {Leviton}, {Leyde}, {Li}, {Li}, {Li}, {Li}, {Li}, {Li},
  {Lin}, {Lin}, {Lin}, {Lin}, {Lin}, {Linde}, {Linker}, {Linley}, {Littenberg},
  {Liu}, {Liu}, {Liu}, {Liu}, {Llamas}, {Llorens-Monteagudo}, {Lo}, {Lockwood},
  {London}, {Longo}, {Lopez}, {Lopez Portilla}, {Lorenzini}, {Loriette},
  {Lormand}, {Losurdo}, {Lott}, {Lough}, {Lousto}, {Lovelace}, {Lucaccioni},
  {L{\"u}ck}, {Lumaca}, {Lundgren}, {Luo}, {Lynam}, {Macas}, {MacInnis},
  {Macleod}, {MacMillan}, {Macquet}, {Maga{\~n}a Hernandez}, {Magazz{\`u}},
  {Magee}, {Maggiore}, {Magnozzi}, {Mahesh}, {Majorana}, {Makarem},
  {Maksimovic}, {Maliakal}, {Malik}, {Man}, {Mandic}, {Mangano}, {Mango},
  {Mansell}, {Manske}, {Mantovani}, {Mapelli}, {Marchesoni}, {Marchio},
  {Marion}, {Mark}, {M{\'a}rka}, {M{\'a}rka}, {Markakis}, {Markosyan},
  {Markowitz}, {Maros}, {Marquina}, {Marsat}, {Martelli}, {Martin}, {Martin},
  {Martinez}, {Martinez}, {Martinez}, {Martinovic}, {Martynov}, {Marx},
  {Masalehdan}, {Mason}, {Massera}, {Masserot}, {Massinger}, {Masso-Reid},
  {Mastrogiovanni}, {Matas}, {Mateu-Lucena}, {Matichard}, {Matiushechkina},
  {Mavalvala}, {McCann}, {McCarthy}, {McClelland}, {McClincy}, {McCormick},
  {McCuller}, {McGhee}, {McGuire}, {McIsaac}, {McIver}, {McRae}, {McWilliams},
  {Meacher}, {Mehmet}, {Mehta}, {Meijer}, {Melatos}, {Melchor}, {Mendell},
  {Menendez-Vazquez}, {Menoni}, {Mercer}, {Mereni}, {Merfeld}, {Merilh},
  {Merritt}, {Merzougui}, {Meshkov}, {Messenger}, {Messick}, {Meyers},
  {Meylahn}, {Mhaske}, {Miani}, {Miao}, {Michaloliakos}, {Michel}, {Michimura},
  {Middleton}, {Milano}, {Miller}, {Miller}, {Miller}, {Millhouse}, {Mills},
  {Milotti}, {Minazzoli}, {Minenkov}, {Mio}, {Mir}, {Miravet-Ten{\'e}s},
  {Mishra}, {Mishra}, {Mistry}, {Mitra}, {Mitrofanov}, {Mitselmakher},
  {Mittleman}, {Miyakawa}, {Miyamoto}, {Miyazaki}, {Miyo}, {Miyoki}, {Mo},
  {Moguel}, {Mogushi}, {Mohapatra}, {Mohite}, {Molina}, {Molina-Ruiz},
  {Mondin}, {Montani}, {Moore}, {Moraru}, {Morawski}, {More}, {Moreno},
  {Moreno}, {Mori}, {Morisaki}, {Moriwaki}, {Mours}, {Mow-Lowry}, {Mozzon},
  {Muciaccia}, {Mukherjee}, {Mukherjee}, {Mukherjee}, {Mukherjee}, {Mukherjee},
  {Mukund}, {Mullavey}, {Munch}, {Mu{\~n}iz}, {Murray}, {Musenich}, {Muusse},
  {Nadji}, {Nagano}, {Nagano}, {Nagar}, {Nakamura}, {Nakano}, {Nakano},
  {Nakashima}, {Nakayama}, {Napolano}, {Nardecchia}, {Narikawa}, {Naticchioni},
  {Nayak}, {Nayak}, {Negishi}, {Neil}, {Neilson}, {Nelemans}, {Nelson}, {Nery},
  {Neubauer}, {Neunzert}, {Ng}, {Ng}, {Nguyen}, {Nguyen}, {Nguyen}, {Nguyen
  Quynh}, {Ni}, {Nichols}, {Nishizawa}, {Nissanke}, {Nitoglia}, {Nocera},
  {Norman}, {North}, {Nozaki}, {Nuttall}, {Oberling}, {O'Brien}, {Obuchi},
  {O'Dell}, {Oelker}, {Ogaki}, {Oganesyan}, {Oh}, {Oh}, {Oh}, {Ohashi},
  {Ohishi}, {Ohkawa}, {Ohme}, {Ohta}, {Okada}, {Okutani}, {Okutomi},
  {Olivetto}, {Oohara}, {Ooi}, {Oram}, {O'Reilly}, {Ormiston}, {Ormsby},
  {Ortega}, {O'Shaughnessy}, {O'Shea}, {Oshino}, {Ossokine}, {Osthelder},
  {Otabe}, {Ottaway}, {Overmier}, {Pace}, {Pagano}, {Page}, {Pagliaroli},
  {Pai}, {Pai}, {Palamos}, {Palashov}, {Palomba}, {Pan}, {Pan}, {Panda},
  {Pang}, {Pang}, {Pankow}, {Pannarale}, {Pant}, {Panther}, {Paoletti},
  {Paoli}, {Paolone}, {Parisi}, {Park}, {Park}, {Parker}, {Pascucci},
  {Pasqualetti}, {Passaquieti}, {Passuello}, {Patel}, {Pathak}, {Patricelli},
  {Patron}, {Patrone}, {Paul}, {Payne}, {Pedraza}, {Pegoraro}, {Pele},
  {Pe{\~n}a Arellano}, {Penn}, {Perego}, {Pereira}, {Pereira}, {Perez},
  {P{\'e}rigois}, {Perkins}, {Perreca}, {Perri{\`e}s}, {Petermann},
  {Petterson}, {Pfeiffer}, {Pham}, {Phukon}, {Piccinni}, {Pichot},
  {Piendibene}, {Piergiovanni}, {Pierini}, {Pierro}, {Pillant}, {Pillas},
  {Pilo}, {Pinard}, {Pinto}, {Pinto}, {Piotrzkowski}, {Pirello}, {Pitkin},
  {Placidi}, {Planas}, {Plastino}, {Pluchar}, {Poggiani}, {Polini}, {Pong},
  {Ponrathnam}, {Popolizio}, {Porter}, {Poulton}, {Powell}, {Pracchia},
  {Pradier}, {Prajapati}, {Prasai}, {Prasanna}, {Pratten}, {Principe}, {Prodi},
  {Prokhorov}, {Prosposito}, {Prudenzi}, {Puecher}, {Punturo}, {Puosi},
  {Puppo}, {P{\"u}rrer}, {Qi}, {Quetschke}, {Quitzow-James}, {Raab},
  {Raaijmakers}, {Radkins}, {Radulesco}, {Raffai}, {Rail}, {Raja}, {Rajan},
  {Ramirez}, {Ramirez}, {Ramos-Buades}, {Rana}, {Rapagnani}, {Rapol}, {Ray},
  {Raymond}, {Raza}, {Razzano}, {Read}, {Rees}, {Regimbau}, {Rei}, {Reid},
  {Reid}, {Reitze}, {Relton}, {Renzini}, {Rettegno}, {Rezac}, {Ricci},
  {Richards}, {Richardson}, {Richardson}, {Riemenschneider}, {Riles},
  {Rinaldi}, {Rink}, {Rizzo}, {Robertson}, {Robie}, {Robinet}, {Rocchi},
  {Rodriguez}, {Rolland}, {Rollins}, {Romanelli}, {Romano}, {Romel},
  {Romero-Rodr{\'\i}guez}, {Romero-Shaw}, {Romie}, {Ronchini}, {Rosa}, {Rose},
  {Rosi{\'n}ska}, {Ross}, {Rowan}, {Rowlinson}, {Roy}, {Roy}, {Roy}, {Rozza},
  {Ruggi}, {Ryan}, {Sachdev}, {Sadecki}, {Sadiq}, {Sago}, {Saito}, {Saito},
  {Sakai}, {Sakai}, {Sakellariadou}, {Sakuno}, {Salafia}, {Salconi}, {Saleem},
  {Salemi}, {Samajdar}, {Sanchez}, {Sanchez}, {Sanchez}, {Sanchis-Gual},
  {Sanders}, {Sanuy}, {Saravanan}, {Sarin}, {Sassolas}, {Satari},
  {Sathyaprakash}, {Sato}, {Sato}, {Sauter}, {Savage}, {Sawada}, {Sawant},
  {Sawant}, {Sayah}, {Schaetzl}, {Scheel}, {Scheuer}, {Schiworski}, {Schmidt},
  {Schmidt}, {Schnabel}, {Schneewind}, {Schofield}, {Sch{\"o}nbeck}, {Schulte},
  {Schutz}, {Schwartz}, {Scott}, {Scott}, {Seglar-Arroyo}, {Sekiguchi},
  {Sekiguchi}, {Sellers}, {Sengupta}, {Sentenac}, {Seo}, {Sequino}, {Sergeev},
  {Setyawati}, {Shaffer}, {Shahriar}, {Shams}, {Shao}, {Sharma}, {Sharma},
  {Shawhan}, {Shcheblanov}, {Shibagaki}, {Shikauchi}, {Shimizu}, {Shimoda},
  {Shimode}, {Shinkai}, {Shishido}, {Shoda}, {Shoemaker}, {Shoemaker},
  {ShyamSundar}, {Sieniawska}, {Sigg}, {Singer}, {Singh}, {Singh}, {Singha},
  {Sintes}, {Sipala}, {Skliris}, {Slagmolen}, {Slaven-Blair}, {Smetana},
  {Smith}, {Smith}, {Soldateschi}, {Somala}, {Somiya}, {Son}, {Soni}, {Soni},
  {Sordini}, {Sorrentino}, {Sorrentino}, {Sotani}, {Soulard}, {Souradeep},
  {Sowell}, {Spagnuolo}, {Spencer}, {Spera}, {Srinivasan}, {Srivastava},
  {Srivastava}, {Staats}, {Stachie}, {Steer}, {Steinlechner}, {Steinlechner},
  {Stops}, {Stover}, {Strain}, {Strang}, {Stratta}, {Strunk}, {Sturani},
  {Stuver}, {Sudhagar}, {Sudhir}, {Sugimoto}, {Suh}, {Summerscales}, {Sun},
  {Sun}, {Sunil}, {Sur}, {Suresh}, {Sutton}, {Suzuki}, {Suzuki}, {Swinkels},
  {Szczepa{\'n}czyk}, {Szewczyk}, {Tacca}, {vTagoshi}, {Tait}, {Takahashi},
  {Takahashi}, {Takamori}, {Takano}, {Takeda}, {Takeda}, {Talbot}, {Talbot},
  {Tanaka}, {Tanaka}, {Tanaka}, {Tanaka}, {Tanaka}, {Tanasijczuk}, {Tanioka},
  {Tanner}, {Tao}, {Tao}, {Tapia San Martin}, {Tapia San Mart{\'\i}n},
  {Taranto}, {Tasson}, {Telada}, {Tenorio}, {Terhune}, {Terkowski},
  {Thirugnanasambandam}, {Thomas}, {Thomas}, {Thompson}, {Thondapu}, {Thorne},
  {Thrane}, {Tiwari}, {Tiwari}, {Tiwari}, {Toivonen}, {Toland}, {Tolley},
  {Tomaru}, {Tomigami}, {Tomura}, {Tonelli}, {Torres-Forn{\'e}}, {Torrie},
  {Tosta e Melo}, {T{\"o}yr{\"a}}, {Trapananti}, {Travasso}, {Traylor},
  {Trevor}, {Tringali}, {Tripathee}, {Troiano}, {Trovato}, {Trozzo}, {Trudeau},
  {Tsai}, {Tsai}, {Tsang}, {Tsang}, {Tsao}, {Tse}, {Tso}, {Tsubono},
  {Tsuchida}, {Tsukada}, {Tsuna}, {Tsutsui}, {Tsuzuki}, {Turbang}, {Turconi},
  {Tuyenbayev}, {Ubhi}, {Uchikata}, {Uchiyama}, {Udall}, {Ueda}, {Uehara},
  {Ueno}, {Ueshima}, {Unnikrishnan}, {Uraguchi}, {Urban}, {Ushiba}, {Utina},
  {Vahlbruch}, {Vajente}, {Vajpeyi}, {Valdes}, {Valentini}, {Valsan}, {van
  Bakel}, {van Beuzekom}, {van den Brand}, {Van Den Broeck}, {Vander-Hyde},
  {van der Schaaf}, {van Heijningen}, {Vanosky}, {van Putten}, {van Remortel},
  {Vardaro}, {Vargas}, {Varma}, {Vas{\'u}th}, {Vecchio}, {Vedovato}, {Veitch},
  {Veitch}, {Venneberg}, {Venugopalan}, {Verkindt}, {Verma}, {Verma}, {Veske},
  {Vetrano}, {Vicer{\'e}}, {Vidyant}, {Viets}, {Vijaykumar}, {Villa-Ortega},
  {Vinet}, {Virtuoso}, {Vitale}, {Vo}, {Vocca}, {von Reis}, {von Wrangel},
  {Vorvick}, {Vyatchanin}, {Wade}, {Wade}, {Wagner}, {Walet}, {Walker},
  {Wallace}, {Wallace}, {Walsh}, {Wang}, {Wang}, {Wang}, {Ward}, {Warner},
  {Was}, {Washimi}, {Washington}, {Watchi}, {Weaver}, {Webster}, {Weinert},
  {Weinstein}, {Weiss}, {Weller}, {Wellmann}, {Wen}, {We{\ss}els}, {Wette},
  {Whelan}, {White}, {Whiting}, {Whittle}, {Wilken}, {Williams}, {Williams},
  {Williamson}, {Willis}, {Willke}, {Wilson}, {Winkler}, {Wipf}, {Wlodarczyk},
  {Woan}, {Woehler}, {Wofford}, {Wong}, {Wu}, {Wu}, {Wu}, {Wu}, {Wysocki},
  {Xiao}, {Xu}, {Yamada}, {Yamamoto}, {Yamamoto}, {Yamamoto}, {Yamamoto},
  {Yamashita}, {Yamazaki}, {Yang}, {Yang}, {Yang}, {Yang}, {Yang}, {Yap},
  {Yeeles}, {Yelikar}, {Ying}, {Yokogawa}, {Yokoyama}, {Yokozawa}, {Yoo},
  {Yoshioka}, {Yu}, {Yu}, {Yuzurihara}, {Zadro{\.z}ny}, {Zanolin}, {Zeidler},
  {Zelenova}, {Zendri}, {Zevin}, {Zhan}, {Zhang}, {Zhang}, {Zhang}, {Zhang},
  {Zhang}, {Zhao}, {Zhao}, {Zhao}, {Zhao}, {Zhou}, {Zhou}, {Zhu}, {Zhu},
  {Zimmerman}, {Zucker}, \& {Zweizig}}]{2022A&A...659A..84A}
{Abbott}, R., {Abbott}, T.~D., {Acernese}, F., {et~al.} 2022, \aap, 659, A84,
  \dodoi{10.1051/0004-6361/202141452}

\bibitem[{{Abbott} {et~al.}(2023{\natexlab{a}}){Abbott}, {Abbott}, {Acernese},
  {Ackley}, {Adams}, {Adhikari}, {Adhikari}, {Adya}, {Affeldt}, {Agarwal},
  {Agathos}, {Agatsuma}, {Aggarwal}, {Aguiar}, {Aiello}, {Ain}, {Ajith},
  {Akcay}, {Akutsu}, {Albanesi}, {Allocca}, {Altin}, {Amato}, {Anand}, {Anand},
  {Ananyeva}, {Anderson}, {Anderson}, {Ando}, {Andrade}, {Andres},
  {Andri{\'c}}, {Angelova}, {Ansoldi}, {Antelis}, {Antier}, {Appert}, {Arai},
  {Arai}, {Arai}, {Araki}, {Araya}, {Araya}, {Areeda}, {Ar{\`e}ne}, {Aritomi},
  {Arnaud}, {et~al.}}]{2023PhRvX..13d1039A}
---. 2023{\natexlab{a}}, Physical Review X, 13, 041039,
  \dodoi{10.1103/PhysRevX.13.041039}

\bibitem[{{Abbott} {et~al.}(2023{\natexlab{b}}){Abbott}, {Abbott}, {Acernese},
  {Ackley}, {Adams}, {Adhikari}, {Adhikari}, {Adya}, {Affeldt}, {Agarwal},
  {Agathos}, {Agatsuma}, {Aggarwal}, {Aguiar}, {Aiello}, {Ain}, {Ajith},
  {Akutsu}, {de Alarc{\'o}n}, {Akcay}, {Albanesi}, {Allocca}, {Altin}, {Amato},
  {Anand}, {Anand}, {Ananyeva}, {Anderson}, {Anderson}, {Ando}, {Andrade},
  {Andres}, {Andri{\'c}}, {Angelova}, {Ansoldi}, {Antelis}, {Antier},
  {Antonini}, {Appert}, {Arai}, {Arai}, {Arai}, {Araki}, {Araya}, {Araya},
  {Areeda}, {Ar{\`e}ne}, {Aritomi}, {Arnaud}, {Arogeti}, {Aronson}, {Arun},
  {Asada}, {Asali}, {Ashton}, {Aso}, {Assiduo}, {Aston}, {Astone}, {Aubin},
  {Austin}, {Babak}, {Badaracco}, {Bader}, {Badger}, {Bae}, {Bae}, {Baer},
  {Bagnasco}, {Bai}, {Baiotti}, {Baird}, {Bajpai}, {Ball}, {Ballardin},
  {Ballmer}, {Balsamo}, {Baltus}, {Banagiri}, {Bankar}, {Barayoga}, {Barbieri},
  {Barish}, {Barker}, {Barneo}, {Barone}, {Barr}, {Barsotti}, {Barsuglia},
  {Barta}, {Bartlett}, {Barton}, {Bartos}, {Bassiri}, {Basti}, {Bawaj},
  {Bayley}, {Baylor}, {Bazzan}, {B{\'e}csy}, {Bedakihale}, {Bejger},
  {Belahcene}, {Benedetto}, {Beniwal}, {Bennett}, {Bentley}, {Benyaala},
  {Bergamin}, {Berger}, {Bernuzzi}, {Berry}, {Bersanetti}, {Bertolini},
  {Betzwieser}, {Beveridge}, {Bhandare}, {Bhardwaj}, {Bhattacharjee},
  {Bhaumik}, {Bilenko}, {Billingsley}, {Bini}, {Birney}, {Birnholtz},
  {Biscans}, {Bischi}, {Biscoveanu}, {Bisht}, {Biswas}, {Bitossi}, {Bizouard},
  {Blackburn}, {Blair}, {Blair}, {Blair}, {Bobba}, {Bode}, {Boer}, {Bogaert},
  {Boldrini}, {Bonavena}, {Bondu}, {Bonilla}, {Bonnand}, {Booker}, {Boom},
  {Bork}, {Boschi}, {Bose}, {Bose}, {Bossilkov}, {Boudart}, {Bouffanais},
  {Bozzi}, {Bradaschia}, {Brady}, {Bramley}, {Branch}, {Branchesi}, {Brandt},
  {Brau}, {Breschi}, {Briant}, {Briggs}, {Brillet}, {Brinkmann}, {Brockill},
  {Brooks}, {Brooks}, {Brown}, {Brunett}, {Bruno}, {Bruntz}, {Bryant}, {Bulik},
  {Bulten}, {Buonanno}, {Buscicchio}, {Buskulic}, {Buy}, {Byer}, {Cadonati},
  {Cagnoli}, {Cahillane}, {Bustillo}, {Callaghan}, {Callister}, {Calloni},
  {Cameron}, {Camp}, {Canepa}, {Canevarolo}, {Cannavacciuolo}, {Cannon}, {Cao},
  {Cao}, {Capocasa}, {Capote}, {Carapella}, {Carbognani}, {Carlin}, {Carney},
  {Carpinelli}, {Carrillo}, {Carullo}, {Carver}, {Diaz}, {Casentini},
  {Castaldi}, {Caudill}, {Cavagli{\`a}}, {Cavalier}, {Cavalieri}, {Ceasar},
  {Cella}, {Cerd{\'a}-Dur{\'a}n}, {Cesarini}, {Chaibi}, {Chakravarti},
  {Subrahmanya}, {Champion}, {Chan}, {Chan}, {Chan}, {Chan}, {Chan}, {Chandra},
  {Chanial}, {Chao}, {Chapman-Bird}, {Charlton}, {Chase}, {Chassande-Mottin},
  {Chatterjee}, {Chatterjee}, {Chatterjee}, {Chaturvedi}, {Chaty},
  {Chatziioannou}, {Chen}, {Chen}, {Chen}, {Chen}, {Chen}, {Chen}, {Chen},
  {Chen}, {Cheng}, {Cheong}, {Cheung}, {Chia}, {Chiadini}, {Chiang},
  {Chiarini}, {Chierici}, {Chincarini}, {Chiofalo}, {Chiummo}, {Cho}, {Cho},
  {Choudhary}, {Choudhary}, {Christensen}, {Chu}, {Chu}, {Chu}, {Chua},
  {Chung}, {Ciani}, {Ciecielag}, {Cie{\'s}lar}, {Cifaldi}, {Ciobanu}, {Ciolfi},
  {Cipriano}, {Cirone}, {Clara}, {Clark}, {Clark}, {Clarke}, {Clearwater},
  {Clesse}, {Cleva}, {Coccia}, {Codazzo}, {Cohadon}, {Cohen}, {Cohen},
  {Colleoni}, {Collette}, {Colombo}, {Colpi}, {Compton}, {Constancio}, {Conti},
  {Cooper}, {Corban}, {Corbitt}, {Cordero-Carri{\'o}n}, {Corezzi}, {Corley},
  {Cornish}, {Corre}, {Corsi}, {Cortese}, {Costa}, {Cotesta}, {Coughlin},
  {Coulon}, {Countryman}, {Cousins}, {Couvares}, {Coward}, {Cowart}, {Coyne},
  {Coyne}, {Creighton}, {Creighton}, {Criswell}, {Croquette}, {Crowder},
  {Cudell}, {Cullen}, {Cumming}, {Cummings}, {Cunningham}, {Cuoco},
  {Cury{\l}o}, {Dabadie}, {Canton}, {Dall'Osso}, {D{\'a}lya}, {Dana},
  {Daneshgaranbajastani}, {D'Angelo}, {Danila}, {Danilishin}, {D'Antonio},
  {Danzmann}, {Darsow-Fromm}, {Dasgupta}, {Datrier}, {Datta}, {Dattilo},
  {Dave}, {Davier}, {Davies}, {Davis}, {Davis}, {Daw}, {Dean}, {Debra},
  {Deenadayalan}, {Degallaix}, {de Laurentis}, {Del{\'e}glise}, {Del Favero},
  {de Lillo}, {de Lillo}, {Del Pozzo}, {Demarchi}, {de Matteis}, {D'Emilio},
  {Demos}, {Dent}, {Depasse}, {de Pietri}, {De Rosa}, {de Rossi}, {Desalvo},
  {de Simone}, {Dhurandhar}, {D{\'\i}az}, {Diaz-Ortiz}, {Didio}, {Dietrich},
  {di Fiore}, {di Fronzo}, {di Giorgio}, {di Giovanni}, {di Giovanni}, {di
  Girolamo}, {di Lieto}, {Ding}, {di Pace}, {di Palma}, {di Renzo},
  {Divakarla}, {Dmitriev}, {Doctor}, {D'Onofrio}, {Donovan}, {Dooley},
  {Doravari}, {Dorrington}, {Drago}, {Driggers}, {Drori}, {Ducoin}, {Dupej},
  {Durante}, {D'Urso}, {Duverne}, {Dwyer}, {Eassa}, {Easter}, {Ebersold},
  {Eckhardt}, {Eddolls}, {Edelman}, {Edo}, {Edy}, {Effler}, {Eguchi},
  {Eichholz}, {Eikenberry}, {Eisenmann}, {Eisenstein}, {Ejlli}, {Engelby},
  {Enomoto}, {Errico}, {Essick}, {Estell{\'e}s}, {Estevez}, {Etienne}, {Etzel},
  {Evans}, {Evans}, {Ewing}, {Fafone}, {Fair}, {Fairhurst}, {Farah}, {Farinon},
  {Farr}, {Farr}, {Farrow}, {Fauchon-Jones}, {Favaro}, {Favata}, {Fays},
  {Fazio}, {Feicht}, {Fejer}, {Fenyvesi}, {Ferguson}, {Fernandez-Galiana},
  {Ferrante}, {Ferreira}, {Fidecaro}, {Figura}, {Fiori}, {Fishbach}, {Fisher},
  {Fittipaldi}, {Fiumara}, {Flaminio}, {Floden}, {Fong}, {Font}, {Fornal},
  {Forsyth}, {Franke}, {Frasca}, {Frasconi}, {Frederick}, {Freed}, {Frei},
  {Freise}, {Frey}, {Fritschel}, {Frolov}, {Fronz{\'e}}, {Fujii}, {Fujikawa},
  {Fukunaga}, {Fukushima}, {Fulda}, {Fyffe}, {Gabbard}, {Gadre}, {Gair},
  {Gais}, {Galaudage}, {Gamba}, {Ganapathy}, {Ganguly}, {Gao}, {Gaonkar},
  {Garaventa}, {Garc{\'\i}a}, {Garc{\'\i}a-N{\'u}{\~n}ez},
  {Garc{\'\i}a-Quir{\'o}s}, {Garufi}, {Gateley}, {Gaudio}, {Gayathri}, {Ge},
  {Gemme}, {Gennai}, {George}, {George}, {Gerberding}, {Gergely}, {Gewecke},
  {Ghonge}, {Ghosh}, {Ghosh}, {Ghosh}, {Ghosh}, {Giacomazzo}, {Giacoppo},
  {Giaime}, {Giardina}, {Gibson}, {Gier}, {Giesler}, {Giri}, {Gissi},
  {Glanzer}, {Gleckl}, {Godwin}, {Golomb}, {Goetz}, {Goetz}, {Gohlke},
  {Goncharov}, {Gonz{\'a}lez}, {Gopakumar}, {Gosselin}, {Gouaty}, {Gould},
  {Grace}, {Grado}, {Granata}, {Granata}, {Grant}, {Gras}, {Grassia}, {Gray},
  {Gray}, {Greco}, {Green}, {Green}, {Gretarsson}, {Gretarsson}, {Griffith},
  {Griffiths}, {Griggs}, {Grignani}, {Grimaldi}, {Grimm}, {Grote}, {Grunewald},
  {Gruning}, {Guerra}, {Guidi}, {Guimaraes}, {Guix{\'e}}, {Gulati}, {Guo},
  {Guo}, {Gupta}, {Gupta}, {Gupta}, {Gustafson}, {Gustafson}, {Guzman}, {Ha},
  {Haegel}, {Hagiwara}, {Haino}, {Halim}, {Hall}, {Hamilton}, {Hammond}, {Han},
  {Haney}, {Hanks}, {Hanna}, {Hannam}, {Hannuksela}, {Hansen}, {Hansen},
  {Hanson}, {Harder}, {Hardwick}, {Haris}, {Harms}, {Harry}, {Harry},
  {Hartwig}, {Hasegawa}, {Haskell}, {Hasskew}, {Haster}, {Hattori}, {Haughian},
  {Hayakawa}, {Hayama}, {Hayes}, {Healy}, {Heidmann}, {Heidt}, {Heintze},
  {Heinze}, {Heinzel}, {Heitmann}, {Hellman}, {Hello}, {Helmling-Cornell},
  {Hemming}, {Hendry}, {Heng}, {Hennes}, {Hennig}, {Hennig}, {Hernandez},
  {Vivanco}, {Heurs}, {Hild}, {Hill}, {Himemoto}, {Hines}, {Hiranuma},
  {Hirata}, {Hirose}, {Hochheim}, {Hofman}, {Hohmann}, {Holcomb}, {Holland},
  {Hollows}, {Holmes}, {Holt}, {Holz}, {Hong}, {Hopkins}, {Hough}, {Hourihane},
  {Howell}, {Hoy}, {Hoyland}, {Hreibi}, {Hsieh}, {Hsu}, {Huang}, {Huang},
  {Huang}, {Huang}, {Huang}, {Huang}, {H{\"u}bner}, {Huddart}, {Hughey}, {Hui},
  {Hui}, {Husa}, {Huttner}, {Huxford}, {Huynh-Dinh}, {Ide}, {Idzkowski},
  {Iess}, {Ikenoue}, {Imam}, {Inayoshi}, {Ingram}, {Inoue}, {Ioka}, {Isi},
  {Isleif}, {Ito}, {Itoh}, {Iyer}, {Izumi}, {Jaberianhamedan}, {Jacqmin},
  {Jadhav}, {Jadhav}, {James}, {Jan}, {Jani}, {Janquart}, {Janssens},
  {Janthalur}, {Jaranowski}, {Jariwala}, {Jaume}, {Jenkins}, {Jenner}, {Jeon},
  {Jeunon}, {Jia}, {Jin}, {Johns}, {Jones}, {Jones}, {Jones}, {Jones}, {Jones},
  {Jonker}, {Ju}, {Jung}, {Jung}, {Junker}, {Juste}, {Kaihotsu}, {Kajita},
  {Kakizaki}, {Kalaghatgi}, {Kalogera}, {Kamai}, {Kamiizumi}, {Kanda},
  {Kandhasamy}, {Kang}, {Kanner}, {Kao}, {Kapadia}, {Kapasi}, {Karat},
  {Karathanasis}, {Karki}, {Kashyap}, {Kasprzack}, {Kastaun}, {Katsanevas},
  {Katsavounidis}, {Katzman}, {Kaur}, {Kawabe}, {Kawaguchi}, {Kawai},
  {Kawasaki}, {K{\'e}f{\'e}lian}, {Keitel}, {Key}, {Khadka}, {Khalili}, {Khan},
  {Khazanov}, {Khetan}, {Khursheed}, {Kijbunchoo}, {Kim}, {Kim}, {Kim}, {Kim},
  {Kim}, {Kim}, {Kimball}, {Kimura}, {Kinley-Hanlon}, {Kirchhoff}, {Kissel},
  {Kita}, {Kitazawa}, {Kleybolte}, {Klimenko}, {Knee}, {Knowles}, {Knyazev},
  {Koch}, {Koekoek}, {Kojima}, {Kokeyama}, {Koley}, {Kolitsidou}, {Kolstein},
  {Komori}, {Kondrashov}, {Kong}, {Kontos}, {Koper}, {Korobko}, {Kotake},
  {Kovalam}, {Kozak}, {Kozakai}, {Kozu}, {Kringel}, {Krishnendu}, {Kr{\'o}lak},
  {Kuehn}, {Kuei}, {Kuijer}, {Kulkarni}, {Kumar}, {Kumar}, {Kumar}, {Kumar},
  {Kume}, {Kuns}, {Kuo}, {Kuo}, {Kuromiya}, {Kuroyanagi}, {Kusayanagi},
  {Kuwahara}, {Kwak}, {Lagabbe}, {Laghi}, {Lalande}, {Lam}, {Lamberts},
  {Landry}, {Landry}, {Lane}, {Lang}, {Lange}, {Lantz}, {La Rosa},
  {Lartaux-Vollard}, {Lasky}, {Laxen}, {Lazzarini}, {Lazzaro}, {Leaci},
  {Leavey}, {Lecoeuche}, {Lee}, {Lee}, {Lee}, {Lee}, {Lee}, {Lee}, {Lehmann},
  {Lema{\^\i}tre}, {Leonardi}, {Leroy}, {Letendre}, {Levesque}, {Levin},
  {Leviton}, {Leyde}, {Li}, {Li}, {Li}, {Li}, {Li}, {Li}, {Lin}, {Lin}, {Lin},
  {Lin}, {Lin}, {Linde}, {Linker}, {Linley}, {Littenberg}, {Liu}, {Liu}, {Liu},
  {Liu}, {Llamas}, {Llorens-Monteagudo}, {Lo}, {Lockwood}, {Loh}, {London},
  {Longo}, {Lopez}, {Portilla}, {Lorenzini}, {Loriette}, {Lormand}, {Losurdo},
  {Lott}, {Lough}, {Lousto}, {Lovelace}, {Lucaccioni}, {L{\"u}ck}, {Lumaca},
  {Lundgren}, {Luo}, {Lynam}, {Macas}, {Macinnis}, {MacLeod}, {MacMillan},
  {Macquet}, {Hernandez}, {Magazz{\`u}}, {Magee}, {Maggiore}, {Magnozzi},
  {Mahesh}, {Majorana}, {Makarem}, {Maksimovic}, {Maliakal}, {Malik}, {Man},
  {Mandic}, {Mangano}, {Mango}, {Mansell}, {Manske}, {Mantovani}, {Mapelli},
  {Marchesoni}, {Marchio}, {Marion}, {Mark}, {M{\'a}rka}, {M{\'a}rka},
  {Markakis}, {Markosyan}, {Markowitz}, {Maros}, {Marquina}, {Marsat},
  {Martelli}, {Martin}, {Martin}, {Martinez}, {Martinez}, {Martinez},
  {Martinovic}, {Martynov}, {Marx}, {Masalehdan}, {Mason}, {Massera},
  {Masserot}, {Massinger}, {Masso-Reid}, {Mastrogiovanni}, {Matas},
  {Mateu-Lucena}, {Matichard}, {Matiushechkina}, {Mavalvala}, {McCann},
  {McCarthy}, {McClelland}, {McClincy}, {McCormick}, {McCuller}, {McGhee},
  {McGuire}, {McIsaac}, {McIver}, {McRae}, {McWilliams}, {Meacher}, {Mehmet},
  {Mehta}, {Meijer}, {Melatos}, {Melchor}, {Mendell}, {Menendez-Vazquez},
  {Menoni}, {Mercer}, {Mereni}, {Merfeld}, {Merilh}, {Merritt}, {Merzougui},
  {Meshkov}, {Messenger}, {Messick}, {Meyers}, {Meylahn}, {Mhaske}, {Miani},
  {Miao}, {Michaloliakos}, {Michel}, {Michimura}, {Middleton}, {Milano},
  {Miller}, {Miller}, {Miller}, {Miller}, {Millhouse}, {Mills}, {Milotti},
  {Minazzoli}, {Minenkov}, {Mio}, {Mir}, {Miravet-Ten{\'e}s}, {Mishra},
  {Mishra}, {Mistry}, {Mitra}, {Mitrofanov}, {Mitselmakher}, {Mittleman},
  {Miyakawa}, {Miyamoto}, {Miyazaki}, {Miyo}, {Miyoki}, {Mo}, {Modafferi},
  {Moguel}, {Mogushi}, {Mohapatra}, {Mohite}, {Molina}, {Molina-Ruiz},
  {Mondin}, {Montani}, {Moore}, {Moraru}, {Morawski}, {More}, {Moreno},
  {Moreno}, {Mori}, {Morisaki}, {Moriwaki}, {Morr{\'a}s}, {Mours}, {Mow-Lowry},
  {Mozzon}, {Muciaccia}, {Mukherjee}, {Mukherjee}, {Mukherjee}, {Mukherjee},
  {Mukherjee}, {Mukund}, {Mullavey}, {Munch}, {Mu{\~n}iz}, {Murray},
  {Musenich}, {Muusse}, {Nadji}, {Nagano}, {Nagano}, {Nagar}, {Nakamura},
  {Nakano}, {Nakano}, {Nakashima}, {Nakayama}, {Napolano}, {Nardecchia},
  {Narikawa}, {Naticchioni}, {Nayak}, {Nayak}, {Negishi}, {Neil}, {Neilson},
  {Nelemans}, {Nelson}, {Nery}, {Neubauer}, {Neunzert}, {Ng}, {Ng}, {Nguyen},
  {Nguyen}, {Nguyen}, {Quynh}, {Ni}, {Nichols}, {Nishizawa}, {Nissanke},
  {Nitoglia}, {Nocera}, {Norman}, {North}, {Nozaki}, {Siles}, {Nuttall},
  {Oberling}, {O'Brien}, {Obuchi}, {O'Dell}, {Oelker}, {Ogaki}, {Oganesyan},
  {Oh}, {Oh}, {Oh}, {Ohashi}, {Ohishi}, {Ohkawa}, {Ohme}, {Ohta}, {Okada},
  {Okutani}, {Okutomi}, {Olivetto}, {Oohara}, {Ooi}, {Oram}, {O'Reilly},
  {Ormiston}, {Ormsby}, {Ortega}, {O'Shaughnessy}, {O'Shea}, {Oshino},
  {Ossokine}, {Osthelder}, {Otabe}, {Ottaway}, {Overmier}, {Pace}, {Pagano},
  {Page}, {Pagliaroli}, {Pai}, {Pai}, {Palamos}, {Palashov}, {Palomba}, {Pan},
  {Pan}, {Panda}, {Pang}, {Pang}, {Pankow}, {Pannarale}, {Pant}, {Panther},
  {Paoletti}, {Paoli}, {Paolone}, {Parisi}, {Park}, {Park}, {Parker},
  {Pascucci}, {Pasqualetti}, {Passaquieti}, {Passuello}, {Patel}, {Pathak},
  {Patricelli}, {Patron}, {Paul}, {Payne}, {Pedraza}, {Pegoraro}, {Pele},
  {Arellano}, {Penn}, {Perego}, {Pereira}, {Pereira}, {Perez}, {P{\'e}rigois},
  {Perkins}, {Perreca}, {Perri{\`e}s}, {Petermann}, {Petterson}, {Pfeiffer},
  {Pham}, {Phukon}, {Piccinni}, {Pichot}, {Piendibene}, {Piergiovanni},
  {Pierini}, {Pierro}, {Pillant}, {Pillas}, {Pilo}, {Pinard}, {Pinto}, {Pinto},
  {Piotrzkowski}, {Piotrzkowski}, {Pirello}, {Pitkin}, {Placidi}, {Planas},
  {Plastino}, {Pluchar}, {Poggiani}, {Polini}, {Pong}, {Ponrathnam},
  {Popolizio}, {Porter}, {Poulton}, {Powell}, {Pracchia}, {Pradier},
  {Prajapati}, {Prasai}, {Prasanna}, {Pratten}, {Principe}, {Prodi},
  {Prokhorov}, {Prosposito}, {Prudenzi}, {Puecher}, {Punturo}, {Puosi},
  {Puppo}, {P{\"u}rrer}, {Qi}, {Quetschke}, {Quitzow-James}, {Raab},
  {Raaijmakers}, {Radkins}, {Radulesco}, {Raffai}, {Rail}, {Raja}, {Rajan},
  {Ramirez}, {Ramirez}, {Ramos-Buades}, {Rana}, {Rapagnani}, {Rapol}, {Ray},
  {Raymond}, {Raza}, {Razzano}, {Read}, {Rees}, {Regimbau}, {Rei}, {Reid},
  {Reid}, {Reitze}, {Relton}, {Renzini}, {Rettegno}, {Reza}, {Rezac}, {Ricci},
  {Richards}, {Richardson}, {Richardson}, {Riemenschneider}, {Riles},
  {Rinaldi}, {Rink}, {Rizzo}, {Robertson}, {Robie}, {Robinet}, {Rocchi},
  {Rodriguez}, {Rolland}, {Rollins}, {Romanelli}, {Romano}, {Romel},
  {Romero-Rodr{\'\i}guez}, {Romero-Shaw}, {Romie}, {Ronchini}, {Rosa}, {Rose},
  {Rosi{\'n}ska}, {Ross}, {Rowan}, {Rowlinson}, {Roy}, {Roy}, {Roy}, {Rozza},
  {Ruggi}, {Ryan}, {Sachdev}, {Sadecki}, {Sadiq}, {Sago}, {Saito}, {Saito},
  {Sakai}, {Sakai}, {Sakellariadou}, {Sakuno}, {Salafia}, {Salconi}, {Saleem},
  {Salemi}, {Samajdar}, {Sanchez}, {Sanchez}, {Sanchez}, {Sanchis-Gual},
  {Sanders}, {Sanuy}, {Saravanan}, {Sarin}, {Sassolas}, {Satari},
  {Sathyaprakash}, {Sato}, {Sato}, {Sauter}, {Savage}, {Sawada}, {Sawant},
  {Sawant}, {Sayah}, {Schaetzl}, {Scheel}, {Scheuer}, {Schiworski}, {Schmidt},
  {Schmidt}, {Schnabel}, {Schneewind}, {Schofield}, {Sch{\"o}nbeck}, {Schulte},
  {Schutz}, {Schwartz}, {Scott}, {Scott}, {Seglar-Arroyo}, {Sekiguchi},
  {Sekiguchi}, {Sellers}, {Sengupta}, {Sentenac}, {Seo}, {Sequino}, {Sergeev},
  {Setyawati}, {Shaffer}, {Shahriar}, {Shams}, {Shao}, {Sharma}, {Sharma},
  {Shawhan}, {Shcheblanov}, {Shibagaki}, {Shikauchi}, {Shimizu}, {Shimoda},
  {Shimode}, {Shinkai}, {Shishido}, {Shoda}, {Shoemaker}, {Shoemaker},
  {Shyamsundar}, {Sieniawska}, {Sigg}, {Singer}, {Singh}, {Singh}, {Singha},
  {Sintes}, {Sipala}, {Skliris}, {Slagmolen}, {Slaven-Blair}, {Smetana},
  {Smith}, {Smith}, {Soldateschi}, {Somala}, {Somiya}, {Son}, {Soni}, {Soni},
  {Sordini}, {Sorrentino}, {Sorrentino}, {Sotani}, {Soulard}, {Souradeep},
  {Sowell}, {Spagnuolo}, {Spencer}, {Spera}, {Srinivasan}, {Srivastava},
  {Srivastava}, {Staats}, {Stachie}, {Steer}, {Steinhoff}, {Steinlechner},
  {Steinlechner}, {Stevenson}, {Stops}, {Stover}, {Strain}, {Strang},
  {Stratta}, {Strunk}, {Sturani}, {Stuver}, {Sudhagar}, {Sudhir}, {Sugimoto},
  {Suh}, {Sullivan}, {Summerscales}, {Sun}, {Sun}, {Sunil}, {Sur}, {Suresh},
  {Sutton}, {Suzuki}, {Suzuki}, {Swinkels}, {Szczepa{\'n}czyk}, {Szewczyk},
  {Tacca}, {Tagoshi}, {Tait}, {Takahashi}, {Takahashi}, {Takamori}, {Takano},
  {Takeda}, {Takeda}, {Talbot}, {Talbot}, {Tanaka}, {Tanaka}, {Tanaka},
  {Tanaka}, {Tanaka}, {Tanasijczuk}, {Tanioka}, {Tanner}, {Tao}, {Tao},
  {Mart{\'\i}n}, {Taranto}, {Tasson}, {Telada}, {Tenorio}, {Terhune},
  {Terkowski}, {Thirugnanasambandam}, {Thomas}, {Thomas}, {Thomas}, {Thompson},
  {Thondapu}, {Thorne}, {Thrane}, {Tiwari}, {Tiwari}, {Tiwari}, {Toivonen},
  {Toland}, {Tolley}, {Tomaru}, {Tomigami}, {Tomura}, {Tonelli},
  {Torres-Forn{\'e}}, {Torrie}, {E Melo}, {T{\"o}yr{\"a}}, {Trapananti},
  {Travasso}, {Traylor}, {Trevor}, {Tringali}, {Tripathee}, {Troiano},
  {Trovato}, {Trozzo}, {Trudeau}, {Tsai}, {Tsai}, {Tsang}, {Tsang}, {Tsao},
  {Tse}, {Tso}, {Tsubono}, {Tsuchida}, {Tsukada}, {Tsuna}, {Tsutsui},
  {Tsuzuki}, {Turbang}, {Turconi}, {Tuyenbayev}, {Ubhi}, {Uchikata},
  {Uchiyama}, {Udall}, {Ueda}, {Uehara}, {Ueno}, {Ueshima}, {Unnikrishnan},
  {Uraguchi}, {Urban}, {Ushiba}, {Utina}, {Vahlbruch}, {Vajente}, {Vajpeyi},
  {Valdes}, {Valentini}, {Valsan}, {van Bakel}, {van Beuzekom}, {van den
  Brand}, {van den Broeck}, {Vander-Hyde}, {van der Schaaf}, {van Heijningen},
  {Vanosky}, {van Putten}, {van Remortel}, {Vardaro}, {Vargas}, {Varma},
  {Vas{\'u}th}, {Vecchio}, {Vedovato}, {Veitch}, {Veitch}, {Venneberg},
  {Venugopalan}, {Verkindt}, {Verma}, {Verma}, {Veske}, {Vetrano},
  {Vicer{\'e}}, {Vidyant}, {Viets}, {Vijaykumar}, {Villa-Ortega}, {Vinet},
  {Virtuoso}, {Vitale}, {Vo}, {Vocca}, {von Reis}, {von Wrangel}, {Vorvick},
  {Vyatchanin}, {Wade}, {Wade}, {Wagner}, {Walet}, {Walker}, {Wallace},
  {Wallace}, {Walsh}, {Wang}, {Wang}, {Wang}, {Ward}, {Warner}, {Was},
  {Washimi}, {Washington}, {Watchi}, {Weaver}, {Webster}, {Weinert},
  {Weinstein}, {Weiss}, {Weller}, {Wellmann}, {Wen}, {We{\ss}els}, {Wette},
  {Whelan}, {White}, {Whiting}, {Whittle}, {Wilken}, {Williams}, {Williams},
  {Williamson}, {Willis}, {Willke}, {Wilson}, {Winkler}, {Wipf}, {Wlodarczyk},
  {Woan}, {Woehler}, {Wofford}, {Wong}, {Wu}, {Wu}, {Wu}, {Wu}, {Wysocki},
  {Xiao}, {Xu}, {Yamada}, {Yamamoto}, {Yamamoto}, {Yamamoto}, {Yamamoto},
  {Yamashita}, {Yamazaki}, {Yang}, {Yang}, {Yang}, {Yang}, {Yang}, {Yap},
  {Yeeles}, {Yelikar}, {Ying}, {Yokogawa}, {Yokoyama}, {Yokozawa}, {Yoo},
  {Yoshioka}, {Yu}, {Yu}, {Yuzurihara}, {Zadro{\.z}ny}, {Zanolin}, {Zeidler},
  {Zelenova}, {Zendri}, {Zevin}, {Zhan}, {Zhang}, {Zhang}, {Zhang}, {Zhang},
  {Zhang}, {Zhao}, {Zhao}, {Zhao}, {Zhao}, {Zheng}, {Zhou}, {Zhou}, {Zhu},
  {Zhu}, {Zimmerman}, {Zlochower}, {Zucker}, {Zweizig}, {LIGO Scientific
  Collaboration}, {VIRGO Collaboration}, \& {KAGRA
  Collaboration}}]{2023PhRvX..13a1048A}
---. 2023{\natexlab{b}}, Physical Review X, 13, 011048,
  \dodoi{10.1103/PhysRevX.13.011048}

\bibitem[{{Acernese} {et~al.}(2015){Acernese}, {Agathos}, {Agatsuma}, {Aisa},
  {Allemandou}, {Allocca}, {Amarni}, {Astone}, {Balestri}, {Ballardin},
  {Barone}, {Baronick}, {Barsuglia}, {Basti}, {Basti}, {Bauer}, {Bavigadda},
  {Bejger}, {Beker}, {Belczynski}, {Bersanetti}, {Bertolini}, {Bitossi},
  {Bizouard}, {Bloemen}, {Blom}, {Boer}, {Bogaert}, {Bondi}, {Bondu},
  {Bonelli}, {Bonnand}, {Boschi}, {Bosi}, {Bouedo}, {Bradaschia}, {Branchesi},
  {Briant}, {Brillet}, {Brisson}, {Bulik}, {Bulten}, {Buskulic}, {Buy},
  {Cagnoli}, {Calloni}, {Campeggi}, {Canuel}, {Carbognani}, {Cavalier},
  {Cavalieri}, {Cella}, {Cesarini}, {Mottin}, {Chincarini}, {Chiummo}, {Chua},
  {Cleva}, {Coccia}, {Cohadon}, {Colla}, {Colombini}, {Conte}, {Coulon},
  {Cuoco}, {Dalmaz}, {D'Antonio}, {Dattilo}, {Davier}, {Day}, {Debreczeni},
  {Degallaix}, {Del{\'e}glise}, {Pozzo}, {Dereli}, {Rosa}, {Fiore}, {Lieto},
  {Virgilio}, {Doets}, {Dolique}, {Drago}, {Ducrot}, {Endr{\H{o}}czi},
  {Fafone}, {Farinon}, {Ferrante}, {Ferrini}, {Fidecaro}, {Fiori}, {Flaminio},
  {Fournier}, {Franco}, {Frasca}, {Frasconi}, {Gammaitoni}, {Garufi},
  {Gaspard}, {Gatto}, {Gemme}, {Gendre}, {Genin}, {Gennai}, {Ghosh},
  {Giacobone}, {Giazotto}, {Gouaty}, {Granata}, {Greco}, {Groot}, {Guidi},
  {Harms}, {Heidmann}, {Heitmann}, {Hello}, {Hemming}, {Hennes}, {Hofman},
  {Jaranowski}, {Jonker}, {Kasprzack}, {K{\'e}f{\'e}lian}, {Kowalska}, {Kraan},
  {Kr{\'o}lak}, {Kutynia}, {Lazzaro}, {Leonardi}, {Leroy}, {Letendre}, {Li},
  {Lieunard}, {Lorenzini}, {Loriette}, {Losurdo}, {Magazz{\`u}}, {Majorana},
  {Maksimovic}, {Malvezzi}, {Man}, {Mangano}, {Mantovani}, {Marchesoni},
  {Marion}, {Marque}, {Martelli}, {Martellini}, {Masserot}, {Meacher},
  {Meidam}, {Mezzani}, {Michel}, {Milano}, {Minenkov}, {Moggi}, {Mohan},
  {Montani}, {Morgado}, {Mours}, {Mul}, {Nagy}, {Nardecchia}, {Naticchioni},
  {Nelemans}, {Neri}, {Neri}, {Nocera}, {Pacaud}, {Palomba}, {Paoletti},
  {Paoli}, {Pasqualetti}, {Passaquieti}, {Passuello}, {Perciballi}, {Petit},
  {Pichot}, {Piergiovanni}, {Pillant}, {Piluso}, {Pinard}, {Poggiani},
  {Prijatelj}, {Prodi}, {Punturo}, {Puppo}, {Rabeling}, {R{\'a}cz},
  {Rapagnani}, {Razzano}, {Re}, {Regimbau}, {Ricci}, {Robinet}, {Rocchi},
  {Rolland}, {Romano}, {Rosi{\'n}ska}, {Ruggi}, {Saracco}, {Sassolas},
  {Schimmel}, {Sentenac}, {Sequino}, {Shah}, {Siellez}, {Straniero},
  {Swinkels}, {Tacca}, {Tonelli}, {Travasso}, {Turconi}, {Vajente}, {van
  Bakel}, {van Beuzekom}, {van den Brand}, {Van Den Broeck}, {van der Sluys},
  {van Heijningen}, {Vas{\'u}th}, {Vedovato}, {Veitch}, {Verkindt}, {Vetrano},
  {Vicer{\'e}}, {Vinet}, {Visser}, {Vocca}, {Ward}, {Was}, {Wei}, {Yvert},
  {{\.z}ny}, \& {Zendri}}]{2015CQGra..32b4001A}
{Acernese}, F., {Agathos}, M., {Agatsuma}, K., {et~al.} 2015, Classical and
  Quantum Gravity, 32, 024001, \dodoi{10.1088/0264-9381/32/2/024001}

\bibitem[{{Adamcewicz} {et~al.}(2024){Adamcewicz}, {Galaudage}, {Lasky}, \&
  {Thrane}}]{2024ApJ...964L...6A}
{Adamcewicz}, C., {Galaudage}, S., {Lasky}, P.~D., \& {Thrane}, E. 2024, \apjl,
  964, L6, \dodoi{10.3847/2041-8213/ad2df2}

\bibitem[{{Afroz} \& {Mukherjee}(2025)}]{2025arXiv250909123A}
{Afroz}, S., \& {Mukherjee}, S. 2025, arXiv e-prints, arXiv:2509.09123,
  \dodoi{10.48550/arXiv.2509.09123}

\bibitem[{{Akutsu} {et~al.}(2018){Akutsu}, {Ando}, {Araki}, {Araya}, {Arima},
  {Aritomi}, {Asada}, {Aso}, {Atsuta}, {Awai}, {Baiotti}, {Barton}, {Chen},
  {Cho}, {Craig}, {DeSalvo}, {Doi}, {Eda}, {Enomoto}, {Flaminio},
  {Fujibayashi}, {Fujii}, {Fujimoto}, {Fukushima}, {Furuhata}, {Hagiwara},
  {Haino}, {Harita}, {Hasegawa}, {Hasegawa}, {Hashino}, {Hayama}, {Hirata},
  {Hirose}, {Ikenoue}, {Inoue}, {Ioka}, {Ishizaki}, {Itoh}, {Jia}, {Kagawa},
  {Kaji}, {Kajita}, {Kakizaki}, {Kakuhata}, {Kamiizumi}, {Kanbara}, {Kanda},
  {Kanemura}, {Kaneyama}, {Kasuya}, {Kataoka}, {Kawaguchi}, {Kawai},
  {Kawamura}, {Kawazoe}, {Kim}, {Kim}, {Kim}, {Kim}, {Kimura}, {Kitaoka},
  {Kobayashi}, {Kojima}, {Kokeyama}, {Komori}, {Kotake}, {Kubo}, {Kumar},
  {Kume}, {Kuroda}, {Kuwahara}, {Lee}, {Lee}, {Lin}, {Liu}, {Majorana}, {Mano},
  {Marchio}, {Matsui}, {Matsumoto}, {Matsushima}, {Michimura}, {Mio},
  {Miyakawa}, {Miyake}, {Miyamoto}, {Miyamoto}, {Miyo}, {Miyoki}, {Morii},
  {Morisaki}, {Moriwaki}, {Muraki}, {Murakoshi}, {Musha}, {Nagano}, {Nagano},
  {Nakamura}, {Nakamura}, {Nakano}, {Nakano}, {Nakano}, {Nakao}, {Nakao},
  {Narikawa}, {Ni}, {Nonomura}, {Obuchi}, {Oh}, {Oh}, {Ohashi}, {Ohishi},
  {Ohkawa}, {Ohmae}, {Okino}, {Okutomi}, {Ono}, {Ono}, {Oohara}, {Ota}, {Park},
  {Pe{\~n}a Arellano}, {Pinto}, {Principe}, {Sago}, {Saijo}, {Saito}, {Saito},
  {Saitou}, {Sakai}, {Sakakibara}, {Sasaki}, {Sato}, {Sato}, {Sato},
  {Sekiguchi}, {Sekiguchi}, {Shibata}, {Shiga}, {Shikano}, {Shimoda},
  {Shinkai}, {Shoda}, {Someya}, {Somiya}, {Son}, {Starecki}, {Suemasa},
  {Sugimoto}, {Susa}, {Suwabe}, {Suzuki}, {Tachibana}, {Tagoshi}, {Takada},
  {Takahashi}, {Takahashi}, {Takamori}, {Takeda}, {Tanaka}, {Tanaka}, {Tanaka},
  {Tatsumi}, {Telada}, {Tomaru}, {Tsubono}, {Tsuchida}, {Tsukada}, {Tsuzuki},
  {Uchikata}, {Uchiyama}, {Uehara}, {Ueki}, {Ueno}, {Uraguchi}, {Ushiba}, {van
  Putten}, {Wada}, {Wakamatsu}, {Yaginuma}, {Yamamoto}, {Yamamoto}, {Yamamoto},
  {Yano}, {Yokoyama}, {Yokozawa}, {Yoon}, {Yuzurihara}, {Zeidler}, {Zhao},
  {Zheng}, {Agatsuma}, {Akiyama}, {Arai}, {Asano}, {Bertolini}, {Fujisawa},
  {Goetz}, {Guscott}, {Hashimoto}, {Hayashida}, {Hennes}, {Hirai}, {Hirayama},
  {Ishitsuka}, {Kato}, {Khalaidovski}, {Koike}, {Kumeta}, {Miener}, {Morioka},
  {Mueller}, {Narita}, {Oda}, {Ogawa}, {Sekiguchi}, {Tamura}, {Tanner},
  {Tokoku}, {Toritani}, {Utsuki}, {Uyeshima}, {van den Brand}, {van
  Heijningen}, {Yamaguchi}, \& {Yanagida}}]{2018PTEP.2018a3F01A}
{Akutsu}, T., {Ando}, M., {Araki}, S., {et~al.} 2018, Progress of Theoretical
  and Experimental Physics, 2018, 013F01, \dodoi{10.1093/ptep/ptx180}

\bibitem[{{Alvarez-Lopez} {et~al.}(2025){Alvarez-Lopez}, {Heinzel}, {Mould}, \&
  {Vitale}}]{2025arXiv250620731A}
{Alvarez-Lopez}, S., {Heinzel}, J., {Mould}, M., \& {Vitale}, S. 2025, arXiv
  e-prints, arXiv:2506.20731, \dodoi{10.48550/arXiv.2506.20731}

\bibitem[{{Amaro-Seoane} {et~al.}(2017){Amaro-Seoane}, {Audley}, {Babak},
  {Baker}, {Barausse}, {Bender}, {Berti}, {Binetruy}, {Born}, {Bortoluzzi},
  {Camp}, {Caprini}, {Cardoso}, {Colpi}, {Conklin}, {Cornish}, {Cutler},
  {Danzmann}, {Dolesi}, {Ferraioli}, {Ferroni}, {Fitzsimons}, {Gair}, {Gesa
  Bote}, {Giardini}, {Gibert}, {Grimani}, {Halloin}, {Heinzel}, {Hertog},
  {Hewitson}, {Holley-Bockelmann}, {Hollington}, {Hueller}, {Inchauspe},
  {Jetzer}, {Karnesis}, {Killow}, {Klein}, {Klipstein}, {Korsakova}, {Larson},
  {Livas}, {Lloro}, {Man}, {Mance}, {Martino}, {Mateos}, {McKenzie},
  {McWilliams}, {Miller}, {Mueller}, {Nardini}, {Nelemans}, {Nofrarias},
  {Petiteau}, {Pivato}, {Plagnol}, {Porter}, {Reiche}, {Robertson},
  {Robertson}, {Rossi}, {Russano}, {Schutz}, {Sesana}, {Shoemaker}, {Slutsky},
  {Sopuerta}, {Sumner}, {Tamanini}, {Thorpe}, {Troebs}, {Vallisneri},
  {Vecchio}, {Vetrugno}, {Vitale}, {Volonteri}, {Wanner}, {Ward}, {Wass},
  {Weber}, {Ziemer}, \& {Zweifel}}]{2017arXiv170200786A}
{Amaro-Seoane}, P., {Audley}, H., {Babak}, S., {et~al.} 2017, arXiv e-prints,
  arXiv:1702.00786, \dodoi{10.48550/arXiv.1702.00786}

\bibitem[{{Antonini} {et~al.}(2025{\natexlab{a}}){Antonini}, {Callister},
  {Dosopoulou}, {Romero-Shaw}, \& {Chattopadhyay}}]{2025PhRvD.112f3040A}
{Antonini}, F., {Callister}, T., {Dosopoulou}, F., {Romero-Shaw}, I.~M., \&
  {Chattopadhyay}, D. 2025{\natexlab{a}}, \prd, 112, 063040,
  \dodoi{10.1103/nxnr-pdyx}

\bibitem[{{Antonini} \& {Gieles}(2020)}]{2020PhRvD.102l3016A}
{Antonini}, F., \& {Gieles}, M. 2020, \prd, 102, 123016,
  \dodoi{10.1103/PhysRevD.102.123016}

\bibitem[{{Antonini} {et~al.}(2023){Antonini}, {Gieles}, {Dosopoulou}, \&
  {Chattopadhyay}}]{2023MNRAS.522..466A}
{Antonini}, F., {Gieles}, M., {Dosopoulou}, F., \& {Chattopadhyay}, D. 2023,
  \mnras, 522, 466, \dodoi{10.1093/mnras/stad972}

\bibitem[{{Antonini} {et~al.}(2025{\natexlab{b}}){Antonini}, {Romero-Shaw},
  {Callister}, {Dosopoulou}, {Chattopadhyay}, {Gieles}, \&
  {Mapelli}}]{2025arXiv250904637A}
{Antonini}, F., {Romero-Shaw}, I., {Callister}, T., {et~al.}
  2025{\natexlab{b}}, arXiv e-prints, arXiv:2509.04637,
  \dodoi{10.48550/arXiv.2509.04637}

\bibitem[{{Antonini} {et~al.}(2025{\natexlab{c}}){Antonini}, {Romero-Shaw}, \&
  {Callister}}]{2025PhRvL.134a1401A}
{Antonini}, F., {Romero-Shaw}, I.~M., \& {Callister}, T. 2025{\natexlab{c}},
  \prl, 134, 011401, \dodoi{10.1103/PhysRevLett.134.011401}

\bibitem[{{Ashton} {et~al.}(2019){Ashton}, {H{\"u}bner}, {Lasky}, {Talbot},
  {Ackley}, {et~al.}}]{2019ascl.soft01011A}
{Ashton}, G., {H{\"u}bner}, M., {Lasky}, P.~D., {et~al.} 2019, {Bilby: Bayesian
  inference library}, Astrophysics Source Code Library, record ascl:1901.011.
\newblock \doeprint{1901.011}

\bibitem[{{Banagiri} {et~al.}(2025){Banagiri}, {Thrane}, \&
  {Lasky}}]{2025arXiv250915646B}
{Banagiri}, S., {Thrane}, E., \& {Lasky}, P.~D. 2025, arXiv e-prints,
  arXiv:2509.15646, \dodoi{10.48550/arXiv.2509.15646}

\bibitem[{{Barkat} {et~al.}(1967){Barkat}, {Rakavy}, \&
  {Sack}}]{1967PhRvL..18..379B}
{Barkat}, Z., {Rakavy}, G., \& {Sack}, N. 1967, \prl, 18, 379,
  \dodoi{10.1103/PhysRevLett.18.379}

\bibitem[{{Bartos} \& {Haiman}(2025)}]{2025arXiv250808558B}
{Bartos}, I., \& {Haiman}, Z. 2025, arXiv e-prints, arXiv:2508.08558,
  \dodoi{10.48550/arXiv.2508.08558}

\bibitem[{{Baumgarte} \& {Shapiro}(2025)}]{2025arXiv250904574B}
{Baumgarte}, T.~W., \& {Shapiro}, S.~L. 2025, arXiv e-prints, arXiv:2509.04574,
  \dodoi{10.48550/arXiv.2509.04574}

\bibitem[{{Bavera} {et~al.}(2020){Bavera}, {Fragos}, {Qin}, {Zapartas},
  {Neijssel}, {Mandel}, {Batta}, {Gaebel}, {Kimball}, \&
  {Stevenson}}]{2020A&A...635A..97B}
{Bavera}, S.~S., {Fragos}, T., {Qin}, Y., {et~al.} 2020, \aap, 635, A97,
  \dodoi{10.1051/0004-6361/201936204}

\bibitem[{{Belczynski} {et~al.}(2020){Belczynski}, {Hirschi}, {Kaiser}, {Liu},
  {Casares}, {Lu}, {O'Shaughnessy}, {Heger}, {Justham}, \&
  {Soria}}]{2020ApJ...890..113B}
{Belczynski}, K., {Hirschi}, R., {Kaiser}, E.~A., {et~al.} 2020, \apj, 890,
  113, \dodoi{10.3847/1538-4357/ab6d77}

\bibitem[{{Bini} {et~al.}(2026){Bini}, {Kr{\'o}l}, {Chatziioannou}, \&
  {Isi}}]{2026arXiv260109678B}
{Bini}, S., {Kr{\'o}l}, K., {Chatziioannou}, K., \& {Isi}, M. 2026, arXiv
  e-prints, arXiv:2601.09678, \dodoi{10.48550/arXiv.2601.09678}

\bibitem[{{Buchner}(2016)}]{2016ascl.soft06005B}
{Buchner}, J. 2016, {PyMultiNest: Python interface for MultiNest}, Astrophysics
  Source Code Library, record ascl:1606.005.
\newblock \doeprint{1606.005}

\bibitem[{{Cai} {et~al.}(2018){Cai}, {Tong}, {Wang}, \&
  {Yan}}]{2018PhRvL.121h1306C}
{Cai}, Y.-F., {Tong}, X., {Wang}, D.-G., \& {Yan}, S.-F. 2018, \prl, 121,
  081306, \dodoi{10.1103/PhysRevLett.121.081306}

\bibitem[{Callister(2021)}]{Callister:2021rew}
Callister, T. 2021, Reweighting Single Event Posteriors with Hyperparameter
  Marginalization, Tech. Rep. LIGO DCC T2100301, LIGO Scientific Collaboration.
\newblock
  \url{https://dcc.ligo.org/public/0177/T2100301/003/Reweighting_Single_Event_Posteriors.pdf}

\bibitem[{{Callister} {et~al.}(2022){Callister}, {Miller}, {Chatziioannou}, \&
  {Farr}}]{2022ApJ...937L..13C}
{Callister}, T.~A., {Miller}, S.~J., {Chatziioannou}, K., \& {Farr}, W.~M.
  2022, \apjl, 937, L13, \dodoi{10.3847/2041-8213/ac847e}

\bibitem[{{Caputo} {et~al.}(2025){Caputo}, {Franciolini}, \&
  {Witte}}]{2025arXiv250721788C}
{Caputo}, A., {Franciolini}, G., \& {Witte}, S.~J. 2025, arXiv e-prints,
  arXiv:2507.21788, \dodoi{10.48550/arXiv.2507.21788}

\bibitem[{{Carr} \& {Hawking}(1974)}]{1974MNRAS.168..399C}
{Carr}, B.~J., \& {Hawking}, S.~W. 1974, \mnras, 168, 399,
  \dodoi{10.1093/mnras/168.2.399}

\bibitem[{{Chakraborty} \& {Mukherjee}(2025)}]{2025arXiv251219077C}
{Chakraborty}, A., \& {Mukherjee}, S. 2025, arXiv e-prints, arXiv:2512.19077,
  \dodoi{10.48550/arXiv.2512.19077}

\bibitem[{{Chiba} \& {Yokoyama}(2017)}]{2017PTEP.2017h3E01C}
{Chiba}, T., \& {Yokoyama}, S. 2017, Progress of Theoretical and Experimental
  Physics, 2017, 083E01, \dodoi{10.1093/ptep/ptx087}

\bibitem[{{Croker} {et~al.}(2021){Croker}, {Zevin}, {Farrah}, {Nishimura}, \&
  {Tarl{\'e}}}]{2021ApJ...921L..22C}
{Croker}, K.~S., {Zevin}, M., {Farrah}, D., {Nishimura}, K.~A., \& {Tarl{\'e}},
  G. 2021, \apjl, 921, L22, \dodoi{10.3847/2041-8213/ac2fad}

\bibitem[{{Croon} {et~al.}(2025){Croon}, {Sakstein}, \&
  {Gerosa}}]{2025arXiv250810088C}
{Croon}, D., {Sakstein}, J., \& {Gerosa}, D. 2025, arXiv e-prints,
  arXiv:2508.10088, \dodoi{10.48550/arXiv.2508.10088}

\bibitem[{{De Luca} {et~al.}(2019){De Luca}, {Desjacques}, {Franciolini},
  {Malhotra}, \& {Riotto}}]{2019JCAP...05..018D}
{De Luca}, V., {Desjacques}, V., {Franciolini}, G., {Malhotra}, A., \&
  {Riotto}, A. 2019, \jcap, 2019, 018, \dodoi{10.1088/1475-7516/2019/05/018}

\bibitem[{{De Luca} {et~al.}(2020){De Luca}, {Franciolini}, {Pani}, \&
  {Riotto}}]{2020JCAP...04..052D}
{De Luca}, V., {Franciolini}, G., {Pani}, P., \& {Riotto}, A. 2020, \jcap,
  2020, 052, \dodoi{10.1088/1475-7516/2020/04/052}

\bibitem[{{De Luca} {et~al.}(2025){De Luca}, {Franciolini}, \&
  {Riotto}}]{2025arXiv250809965D}
{De Luca}, V., {Franciolini}, G., \& {Riotto}, A. 2025, arXiv e-prints,
  arXiv:2508.09965, \dodoi{10.48550/arXiv.2508.09965}

\bibitem[{{Edelman} {et~al.}(2023){Edelman}, {Farr}, \&
  {Doctor}}]{2023ApJ...946...16E}
{Edelman}, B., {Farr}, B., \& {Doctor}, Z. 2023, \apj, 946, 16,
  \dodoi{10.3847/1538-4357/acb5ed}

\bibitem[{{Escriv{\`a}} {et~al.}(2024){Escriv{\`a}}, {K{\"u}hnel}, \&
  {Tada}}]{2024bheg.book..261E}
{Escriv{\`a}}, A., {K{\"u}hnel}, F., \& {Tada}, Y. 2024, in Black Holes in the
  Era of Gravitational-Wave Astronomy, ed. M.~{Arca Sedda}, E.~{Bortolas}, \&
  M.~{Spera}, 261--377, \dodoi{10.1016/B978-0-32-395636-9.00012-8}

\bibitem[{{Farmer} {et~al.}(2020){Farmer}, {Renzo}, {de Mink}, {Fishbach}, \&
  {Justham}}]{2020ApJ...902L..36F}
{Farmer}, R., {Renzo}, M., {de Mink}, S.~E., {Fishbach}, M., \& {Justham}, S.
  2020, \apjl, 902, L36, \dodoi{10.3847/2041-8213/abbadd}

\bibitem[{{Farmer} {et~al.}(2019){Farmer}, {Renzo}, {de Mink}, {Marchant}, \&
  {Justham}}]{2019ApJ...887...53F}
{Farmer}, R., {Renzo}, M., {de Mink}, S.~E., {Marchant}, P., \& {Justham}, S.
  2019, \apj, 887, 53, \dodoi{10.3847/1538-4357/ab518b}

\bibitem[{{Fishbach} {et~al.}(2020){Fishbach}, {Farr}, \&
  {Holz}}]{2020ApJ...891L..31F}
{Fishbach}, M., {Farr}, W.~M., \& {Holz}, D.~E. 2020, \apjl, 891, L31,
  \dodoi{10.3847/2041-8213/ab77c9}

\bibitem[{{Fishbach} \& {Holz}(2020)}]{2020ApJ...904L..26F}
{Fishbach}, M., \& {Holz}, D.~E. 2020, \apjl, 904, L26,
  \dodoi{10.3847/2041-8213/abc827}

\bibitem[{{Fishbach} {et~al.}(2017){Fishbach}, {Holz}, \&
  {Farr}}]{2017ApJ...840L..24F}
{Fishbach}, M., {Holz}, D.~E., \& {Farr}, B. 2017, \apjl, 840, L24,
  \dodoi{10.3847/2041-8213/aa7045}

\bibitem[{{Ford} \& {McKernan}(2025)}]{2025arXiv250608801F}
{Ford}, K.~E.~S., \& {McKernan}, B. 2025, arXiv e-prints, arXiv:2506.08801,
  \dodoi{10.48550/arXiv.2506.08801}

\bibitem[{{Fowler} \& {Hoyle}(1964)}]{1964ApJS....9..201F}
{Fowler}, W.~A., \& {Hoyle}, F. 1964, \apjs, 9, 201, \dodoi{10.1086/190103}

\bibitem[{{Franciolini} \& {Pani}(2022)}]{2022PhRvD.105l3024F}
{Franciolini}, G., \& {Pani}, P. 2022, \prd, 105, 123024,
  \dodoi{10.1103/PhysRevD.105.123024}

\bibitem[{{Franciolini} {et~al.}(2022){Franciolini}, {Baibhav}, {De Luca},
  {Ng}, {Wong}, {Berti}, {Pani}, {Riotto}, \& {Vitale}}]{2022PhRvD.105h3526F}
{Franciolini}, G., {Baibhav}, V., {De Luca}, V., {et~al.} 2022, \prd, 105,
  083526, \dodoi{10.1103/PhysRevD.105.083526}

\bibitem[{{Fuller} \& {Ma}(2019)}]{2019ApJ...881L...1F}
{Fuller}, J., \& {Ma}, L. 2019, \apjl, 881, L1,
  \dodoi{10.3847/2041-8213/ab339b}

\bibitem[{{Galaudage} {et~al.}(2021){Galaudage}, {Talbot}, {Nagar}, {Jain},
  {Thrane}, \& {Mandel}}]{2021ApJ...921L..15G}
{Galaudage}, S., {Talbot}, C., {Nagar}, T., {et~al.} 2021, \apjl, 921, L15,
  \dodoi{10.3847/2041-8213/ac2f3c}

\bibitem[{{Gerosa} \& {Berti}(2017)}]{2017PhRvD..95l4046G}
{Gerosa}, D., \& {Berti}, E. 2017, \prd, 95, 124046,
  \dodoi{10.1103/PhysRevD.95.124046}

\bibitem[{{Gerosa} \& {Fishbach}(2021)}]{2021NatAs...5..749G}
{Gerosa}, D., \& {Fishbach}, M. 2021, Nature Astronomy, 5, 749,
  \dodoi{10.1038/s41550-021-01398-w}

\bibitem[{{Gerosa} \& {Kesden}(2016)}]{2016PhRvD..93l4066G}
{Gerosa}, D., \& {Kesden}, M. 2016, \prd, 93, 124066,
  \dodoi{10.1103/PhysRevD.93.124066}

\bibitem[{{Giacobbo} {et~al.}(2018){Giacobbo}, {Mapelli}, \&
  {Spera}}]{2018MNRAS.474.2959G}
{Giacobbo}, N., {Mapelli}, M., \& {Spera}, M. 2018, \mnras, 474, 2959,
  \dodoi{10.1093/mnras/stx2933}

\bibitem[{{Godfrey} {et~al.}(2023){Godfrey}, {Edelman}, \&
  {Farr}}]{2023arXiv230401288G}
{Godfrey}, J., {Edelman}, B., \& {Farr}, B. 2023, arXiv e-prints,
  arXiv:2304.01288, \dodoi{10.48550/arXiv.2304.01288}

\bibitem[{{Golomb} \& {Talbot}(2023)}]{2023PhRvD.108j3009G}
{Golomb}, J., \& {Talbot}, C. 2023, \prd, 108, 103009,
  \dodoi{10.1103/PhysRevD.108.103009}

\bibitem[{{Green} \& {Kavanagh}(2021)}]{2021JPhG...48d3001G}
{Green}, A.~M., \& {Kavanagh}, B.~J. 2021, Journal of Physics G Nuclear
  Physics, 48, 043001, \dodoi{10.1088/1361-6471/abc534}

\bibitem[{{Guo} {et~al.}(2024){Guo}, {Li}, {Wang}, {Shao}, {Wu}, {Zhu}, \&
  {Fan}}]{2024ApJ...975...54G}
{Guo}, W.-H., {Li}, Y.-J., {Wang}, Y.-Z., {et~al.} 2024, \apj, 975, 54,
  \dodoi{10.3847/1538-4357/ad758a}

\bibitem[{{He} {et~al.}(2025){He}, {Zhu}, {Liu}, {Niu}, {Wei}, {Gao}, {Zhou},
  {Liang}, {Chen}, {Wang}, {Jiang}, {Cai}, {Jiang}, {Dai}, {Yuan}, {Li}, \&
  {Zhao}}]{2025arXiv251105144H}
{He}, L., {Zhu}, L.-G., {Liu}, Z.-Y., {et~al.} 2025, arXiv e-prints,
  arXiv:2511.05144, \dodoi{10.48550/arXiv.2511.05144}

\bibitem[{{Hoy} {et~al.}(2022){Hoy}, {Fairhurst}, {Hannam}, \&
  {Tiwari}}]{2022ApJ...928...75H}
{Hoy}, C., {Fairhurst}, S., {Hannam}, M., \& {Tiwari}, V. 2022, \apj, 928, 75,
  \dodoi{10.3847/1538-4357/ac54a3}

\bibitem[{{Hu} {et~al.}(2025){Hu}, {Narola}, {Heynen}, {Wright}, {Veitch},
  {Janquart}, \& {Van Den Broeck}}]{2025arXiv251217550H}
{Hu}, Q., {Narola}, H., {Heynen}, J., {et~al.} 2025, arXiv e-prints,
  arXiv:2512.17550, \dodoi{10.48550/arXiv.2512.17550}

\bibitem[{{Hu} \& {Wu}(2017)}]{2017NSRev...4..685H}
{Hu}, W.-R., \& {Wu}, Y.-L. 2017, National Science Review, 4, 685,
  \dodoi{10.1093/nsr/nwx116}

\bibitem[{{Hussain} {et~al.}(2024){Hussain}, {Isi}, \&
  {Zimmerman}}]{2024arXiv241102252H}
{Hussain}, A., {Isi}, M., \& {Zimmerman}, A. 2024, arXiv e-prints,
  arXiv:2411.02252, \dodoi{10.48550/arXiv.2411.02252}

\bibitem[{JackLee0214(2025)}]{jacklee0214_2025_17572869}
JackLee0214. 2025, JackLee0214/Exploring-field-evolution-and-
  dynamical-capture-coalescing-binary-black-holes- in-GWTC-3: Field VS
  Dynamical, v1.0.0,  Zenodo, \dodoi{10.5281/zenodo.17572869}

\bibitem[{{Khlopov}(2010)}]{2010RAA....10..495K}
{Khlopov}, M.~Y. 2010, Research in Astronomy and Astrophysics, 10, 495,
  \dodoi{10.1088/1674-4527/10/6/001}

\bibitem[{{Kimball} {et~al.}(2021){Kimball}, {Talbot}, {Berry}, {Zevin},
  {Thrane}, {Kalogera}, {Buscicchio}, {Carney}, {Dent}, {Middleton}, {Payne},
  {Veitch}, \& {Williams}}]{2021ApJ...915L..35K}
{Kimball}, C., {Talbot}, C., {Berry}, C. P.~L., {et~al.} 2021, \apjl, 915, L35,
  \dodoi{10.3847/2041-8213/ac0aef}

\bibitem[{{Lei} {et~al.}(2025{\natexlab{a}}){Lei}, {Wang}, {Yuan}, {Wang},
  {Groenewegen}, \& {Fan}}]{2025ApJ...980..249L}
{Lei}, L., {Wang}, Y.-Y., {Yuan}, G.-W., {et~al.} 2025{\natexlab{a}}, \apj,
  980, 249, \dodoi{10.3847/1538-4357/ada93b}

\bibitem[{{Lei} {et~al.}(2025{\natexlab{b}}){Lei}, {Wang}, {Wang}, {Feng}, \&
  {Fan}}]{2025arXiv250619589L}
{Lei}, L., {Wang}, Z.-F., {Wang}, Y.-Y., {Feng}, L., \& {Fan}, Y.-Z.
  2025{\natexlab{b}}, arXiv e-prints, arXiv:2506.19589,
  \dodoi{10.48550/arXiv.2506.19589}

\bibitem[{{Li} \& {Fan}(2025)}]{2025ApJ...984...63L}
{Li}, G.-P., \& {Fan}, X.-L. 2025, \apj, 984, 63,
  \dodoi{10.3847/1538-4357/adc7bb}

\bibitem[{{Li} {et~al.}(2024{\natexlab{a}}){Li}, {Tang}, {Gao}, {Wu}, \&
  {Wang}}]{2024ApJ...977...67L}
{Li}, Y.-J., {Tang}, S.-P., {Gao}, S.-J., {Wu}, D.-C., \& {Wang}, Y.-Z.
  2024{\natexlab{a}}, \apj, 977, 67, \dodoi{10.3847/1538-4357/ad83b5}

\bibitem[{{Li} {et~al.}(2021){Li}, {Wang}, {Han}, {Tang}, {Yuan}, {Fan}, \&
  {Wei}}]{2021ApJ...917...33L}
{Li}, Y.-J., {Wang}, Y.-Z., {Han}, M.-Z., {et~al.} 2021, \apj, 917, 33,
  \dodoi{10.3847/1538-4357/ac0971}

\bibitem[{{Li} {et~al.}(2025{\natexlab{a}}){Li}, {Wang}, {Tang}, {Chen}, \&
  {Fan}}]{2025ApJ...987...65L}
{Li}, Y.-J., {Wang}, Y.-Z., {Tang}, S.-P., {Chen}, T., \& {Fan}, Y.-Z.
  2025{\natexlab{a}}, \apj, 987, 65, \dodoi{10.3847/1538-4357/add535}

\bibitem[{{Li} {et~al.}(2024{\natexlab{b}}){Li}, {Wang}, {Tang}, \&
  {Fan}}]{2024PhRvL.133e1401L}
{Li}, Y.-J., {Wang}, Y.-Z., {Tang}, S.-P., \& {Fan}, Y.-Z. 2024{\natexlab{b}},
  \prl, 133, 051401, \dodoi{10.1103/PhysRevLett.133.051401}

\bibitem[{{Li} {et~al.}(2025{\natexlab{b}}){Li}, {Wang}, {Tang}, \&
  {Fan}}]{2025arXiv250923897L}
---. 2025{\natexlab{b}}, arXiv e-prints, arXiv:2509.23897,
  \dodoi{10.48550/arXiv.2509.23897}

\bibitem[{{Li} {et~al.}(2022){Li}, {Wang}, {Tang}, {Yuan}, {Fan}, \&
  {Wei}}]{2022ApJ...933L..14L}
{Li}, Y.-J., {Wang}, Y.-Z., {Tang}, S.-P., {et~al.} 2022, \apjl, 933, L14,
  \dodoi{10.3847/2041-8213/ac78dd}

\bibitem[{{Liang} {et~al.}(2017){Liang}, {Wang}, {Wang}, {Li}, {Hu}, {Jin},
  {Fan}, {Liang}, \& {Wei}}]{2017arXiv170501881L}
{Liang}, Y.-F., {Wang}, Y.-Z., {Wang}, H., {et~al.} 2017, arXiv e-prints,
  arXiv:1705.01881, \dodoi{10.48550/arXiv.1705.01881}

\bibitem[{{LIGO Scientific Collaboration} {et~al.}(2015){LIGO Scientific
  Collaboration}, {Aasi}, {Abbott}, {Abbott}, {Abbott}, {Abernathy}, {Ackley},
  {Adams}, {Adams}, {Addesso}, {Adhikari}, {Adya}, {Affeldt}, {Aggarwal},
  {Aguiar}, {Ain}, {Ajith}, {Alemic}, {Allen}, {Amariutei}, {Anderson},
  {Anderson}, {Arai}, {Araya}, {Arceneaux}, {Areeda}, {Ashton}, {Ast}, {Aston},
  {Aufmuth}, {Aulbert}, {Aylott}, {Babak}, {Baker}, {Ballmer}, {Barayoga},
  {Barbet}, {Barclay}, {Barish}, {Barker}, {Barr}, {Barsotti}, {Bartlett},
  {Barton}, {Bartos}, {Bassiri}, {Batch}, {Baune}, {Behnke}, {Bell}, {Bell},
  {Benacquista}, {Bergman}, {Bergmann}, {Berry}, {Betzwieser}, {Bhagwat},
  {Bhandare}, {Bilenko}, {Billingsley}, {Birch}, {Biscans}, {Biwer},
  {Blackburn}, {Blackburn}, {Blair}, {Blair}, {Bock}, {Bodiya}, {Bojtos},
  {Bond}, {Bork}, {Born}, {Bose}, {Brady}, {Braginsky}, {Brau}, {Bridges},
  {Brinkmann}, {Brooks}, {Brown}, {Brown}, {Brown}, {Buchman}, {Buikema},
  {Buonanno}, {Cadonati}, {Calder{\'o}n Bustillo}, {Camp}, {Cannon}, {Cao},
  {Capano}, {Caride}, {Caudill}, {Cavagli{\`a}}, {Cepeda}, {Chakraborty},
  {Chalermsongsak}, {Chamberlin}, {Chao}, {Charlton}, {Chen}, {Cho}, {Cho},
  {Chow}, {Christensen}, {Chu}, {Chung}, {Ciani}, {Clara}, {Clark}, {Collette},
  {Cominsky}, {Constancio}, {Cook}, {Corbitt}, {Cornish}, {Corsi}, {Costa},
  {Coughlin}, {Countryman}, {Couvares}, {Coward}, {Cowart}, {Coyne}, {Coyne},
  {Craig}, {Creighton}, {Creighton}, {Cripe}, {Crowder}, {Cumming},
  {Cunningham}, {Cutler}, {Dahl}, {Dal Canton}, {Damjanic}, {Danilishin},
  {Danzmann}, {Dartez}, {Dave}, {Daveloza}, {Davies}, {Daw}, {DeBra}, {Del
  Pozzo}, {Denker}, {Dent}, {Dergachev}, {DeRosa}, {DeSalvo}, {Dhurandhar},
  {D{\textasciiacute}{\i}az}, {Di Palma}, {Dojcinoski}, {Dominguez}, {Donovan},
  {Dooley}, {Doravari}, {Douglas}, {Downes}, {Driggers}, {Du}, {Dwyer},
  {Eberle}, {Edo}, {Edwards}, {Edwards}, {Effler}, {Eggenstein}, {Ehrens},
  {Eichholz}, {Eikenberry}, {Essick}, {Etzel}, {Evans}, {Evans},
  {Factourovich}, {Fairhurst}, {Fan}, {Fang}, {Farr}, {Farr}, {Favata}, {Fays},
  {Fehrmann}, {Fejer}, {Feldbaum}, {Ferreira}, {Fisher}, {Frei}, {Freise},
  {Frey}, {Fricke}, {Fritschel}, {Frolov}, {Fuentes-Tapia}, {Fulda}, {Fyffe},
  {Gair}, {Gaonkar}, {Gehrels}, {Gergely}, {Giaime}, {Giardina}, {Gleason},
  {Goetz}, {Goetz}, {Gondan}, {Gonz{\'a}lez}, {Gordon}, {Gorodetsky}, {Gossan},
  {Go{\ss}ler}, {Gr{\"a}f}, {Graff}, {Grant}, {Gras}, {Gray}, {Greenhalgh},
  {Gretarsson}, {Grote}, {Grunewald}, {Guido}, {Guo}, {Gushwa}, {Gustafson},
  {Gustafson}, {Hacker}, {Hall}, {Hammond}, {Hanke}, {Hanks}, {Hanna},
  {Hannam}, {Hanson}, {Hardwick}, {Harry}, {Harry}, {Hart}, {Hartman},
  {Haster}, {Haughian}, {Hee}, {Heintze}, {Heinzel}, {Hendry}, {Heng},
  {Heptonstall}, {Heurs}, {Hewitson}, {Hild}, {Hoak}, {Hodge}, {Hollitt},
  {Holt}, {Hopkins}, {Hosken}, {Hough}, {Houston}, {Howell}, {Hu}, {Huerta},
  {Hughey}, {Husa}, {Huttner}, {Huynh}, {Huynh-Dinh}, {Idrisy}, {Indik},
  {Ingram}, {Inta}, {Islas}, {Isler}, {Isogai}, {Iyer}, {Izumi}, {Jacobson},
  {Jang}, {Jawahar}, {Ji}, {Jim{\'e}nez-Forteza}, {Johnson}, {Jones}, {Jones},
  {Ju}, {Haris}, {Kalogera}, {Kandhasamy}, {Kang}, {Kanner}, {Katsavounidis},
  {Katzman}, {Kaufer}, {Kaufer}, {Kaur}, {Kawabe}, {Kawazoe}, {Keiser},
  {Keitel}, {Kelley}, {Kells}, {Keppel}, {Key}, {Khalaidovski}, {Khalili},
  {Khazanov}, {Kim}, {Kim}, {Kim}, {Kim}, {Kim}, {King}, {King}, {Kinzel},
  {Kissel}, {Klimenko}, {Kline}, {Koehlenbeck}, {Kokeyama}, {Kondrashov},
  {Korobko}, {Korth}, {Kozak}, {Kringel}, {Krishnan}, {Krueger}, {Kuehn},
  {Kumar}, {Kumar}, {Kuo}, {Landry}, {Lantz}, {Larson}, {Lasky}, {Lazzarini},
  {Lazzaro}, {Le}, {Leaci}, {Leavey}, {Lebigot}, {Lee}, {Lee}, {Lee}, {Leong},
  {Levin}, {Levine}, {Lewis}, {Li}, {Libbrecht}, {Libson}, {Lin}, {Littenberg},
  {Lockerbie}, {Lockett}, {Logue}, {Lombardi}, {Lormand}, {Lough}, {Lubinski},
  {L{\"u}ck}, {Lundgren}, {Lynch}, {Ma}, {Macarthur}, {MacDonald},
  {Machenschalk}, {MacInnis}, {Macleod}, {Maga{\~n}a-Sandoval}, {Magee},
  {Mageswaran}, {Maglione}, {Mailand}, {Mandel}, {Mandic}, {Mangano},
  {Mansell}, {M{\'a}rka}, {M{\'a}rka}, {Markosyan}, {Maros}, {Martin},
  {Martin}, {Martynov}, {Marx}, {Mason}, {Massinger}, {Matichard}, {Matone},
  {Mavalvala}, {Mazumder}, {Mazzolo}, {McCarthy}, {McClelland}, {McCormick},
  {McGuire}, {McIntyre}, {McIver}, {McLin}, {McWilliams}, {Meadors},
  {Meinders}, {Melatos}, {Mendell}, {Mercer}, {Meshkov}, {Messenger}, {Meyers},
  {Miao}, {Middleton}, {Mikhailov}, {Miller}, {Miller}, {Millhouse}, {Ming},
  {Mirshekari}, {Mishra}, {Mitra}, {Mitrofanov}, {Mitselmakher}, {Mittleman},
  {Moe}, {Mohanty}, {Mohapatra}, {Moore}, {Moraru}, {Moreno}, {Morriss},
  {Mossavi}, {Mow-Lowry}, {Mueller}, {Mueller}, {Mukherjee}, {Mullavey},
  {Munch}, {Murphy}, {Murray}, {Mytidis}, {Nash}, {Nayak}, {Necula}, {Nedkova},
  {Newton}, {Nguyen}, {Nielsen}, {Nissanke}, {Nitz}, {Nolting}, {Normandin},
  {Nuttall}, {Ochsner}, {O'Dell}, {Oelker}, {Ogin}, {Oh}, {Oh}, {Ohme},
  {Oppermann}, {Oram}, {O'Reilly}, {Ortega}, {O'Shaughnessy}, {Osthelder},
  {Ott}, {Ottaway}, {Ottens}, {Overmier}, {Owen}, {Padilla}, {Pai}, {Pai},
  {Palashov}, {Pal-Singh}, {Pan}, {Pankow}, {Pannarale}, {Pant}, {Papa},
  {Paris}, {Patrick}, {Pedraza}, {Pekowsky}, {Pele}, {Penn}, {Perreca},
  {Phelps}, {Pierro}, {Pinto}, {Pitkin}, {Poeld}, {Post}, {Poteomkin},
  {Powell}, {Prasad}, {Predoi}, {Premachandra}, {Prestegard}, {Price},
  {Principe}, {Privitera}, {Prix}, {Prokhorov}, {Puncken}, {P{\"u}rrer}, {Qin},
  {Quetschke}, {Quintero}, {Quiroga}, {Quitzow-James}, {Raab}, {Rabeling},
  {Radkins}, {Raffai}, {Raja}, {Rajalakshmi}, {Rakhmanov}, {Ramirez},
  {Raymond}, {Reed}, {Reid}, {Reitze}, {Reula}, {Riles}, {Robertson}, {Robie},
  {Rollins}, {Roma}, {Romano}, {Romanov}, {Romie}, {Rowan}, {R{\"u}diger},
  {Ryan}, {Sachdev}, {Sadecki}, {Sadeghian}, {Saleem}, {Salemi}, {Sammut},
  {Sandberg}, {Sanders}, {Sannibale}, {Santiago-Prieto}, {Sathyaprakash},
  {Saulson}, {Savage}, {Sawadsky}, {Scheuer}, {Schilling}, {Schmidt},
  {Schnabel}, {Schofield}, {Schreiber}, {Schuette}, {Schutz}, {Scott}, {Scott},
  {Sellers}, {Sengupta}, {Sergeev}, {Serna}, {Sevigny}, {Shaddock}, {Shahriar},
  {Shaltev}, {Shao}, {Shapiro}, {Shawhan}, {Shoemaker}, {Sidery}, {Siemens},
  {Sigg}, {Silva}, {Simakov}, {Singer}, {Singer}, {Singh}, {Sintes},
  {Slagmolen}, {Smith}, {Smith}, {Smith}, {Smith-Lefebvre}, {Son}, {Sorazu},
  {Souradeep}, {Staley}, {Stebbins}, {Steinke}, {Steinlechner}, {Steinlechner},
  {Steinmeyer}, {Stephens}, {Steplewski}, {Stevenson}, {Stone}, {Strain},
  {Strigin}, {Sturani}, {Stuver}, {Summerscales}, {Sutton}, {Szczepanczyk},
  {Szeifert}, {Talukder}, {Tanner}, {T{\'a}pai}, {Tarabrin}, {Taracchini},
  {Taylor}, {Tellez}, {Theeg}, {Thirugnanasambandam}, {Thomas}, {Thomas},
  {Thorne}, {Thorne}, {Thrane}, {Tiwari}, {Tomlinson}, {Torres}, {Torrie},
  {Traylor}, {Tse}, {Tshilumba}, {Ugolini}, {Unnikrishnan}, {Urban}, {Usman},
  {Vahlbruch}, {Vajente}, {Valdes}, {Vallisneri}, {van Veggel}, {Vass},
  {Vaulin}, {Vecchio}, {Veitch}, {Veitch}, {Venkateswara}, {Vincent-Finley},
  {Vitale}, {Vo}, {Vorvick}, {Vousden}, {Vyatchanin}, {Wade}, {Wade}, {Wade},
  {Walker}, {Wallace}, {Walsh}, {Wang}, {Wang}, {Wang}, {Ward}, {Warner},
  {Was}, {Weaver}, {Weinert}, {Weinstein}, {Weiss}, {Welborn}, {Wen},
  {Wessels}, {Westphal}, {Wette}, {Whelan}, {Whitcomb}, {White}, {Whiting},
  {Wilkinson}, {Williams}, {Williams}, {Williamson}, {Willis}, {Willke},
  {Wimmer}, {Winkler}, {Wipf}, {Wittel}, {Woan}, {Worden}, {Xie}, {Yablon},
  {Yakushin}, {Yam}, {Yamamoto}, {Yancey}, {Yang}, {Zanolin}, {Zhang}, {Zhang},
  {Zhang}, {Zhang}, {Zhao}, {Zhou}, {Zhu}, {Zucker}, {Zuraw}, \&
  {Zweizig}}]{2015CQGra..32g4001L}
{LIGO Scientific Collaboration}, {Aasi}, J., {Abbott}, B.~P., {et~al.} 2015,
  Classical and Quantum Gravity, 32, 074001,
  \dodoi{10.1088/0264-9381/32/7/074001}

\bibitem[{{Liu} \& {Lai}(2025)}]{2025arXiv251113820L}
{Liu}, B., \& {Lai}, D. 2025, arXiv e-prints, arXiv:2511.13820,
  \dodoi{10.48550/arXiv.2511.13820}

\bibitem[{{Liu} {et~al.}(2025){Liu}, {Wang}, {Tanikawa}, {Wu}, \&
  {Fujii}}]{2025ApJ...993L..30L}
{Liu}, S., {Wang}, L., {Tanikawa}, A., {Wu}, W., \& {Fujii}, M.~S. 2025, \apjl,
  993, L30, \dodoi{10.3847/2041-8213/ae1024}

\bibitem[{{Luo} {et~al.}(2016){Luo}, {Chen}, {Duan}, {Gong}, {Hu}, {Ji}, {Liu},
  {Mei}, {Milyukov}, {Sazhin}, {Shao}, {Toth}, {Tu}, {Wang}, {Wang}, {Yeh},
  {Zhan}, {Zhang}, {Zharov}, \& {Zhou}}]{2016CQGra..33c5010L}
{Luo}, J., {Chen}, L.-S., {Duan}, H.-Z., {et~al.} 2016, Classical and Quantum
  Gravity, 33, 035010, \dodoi{10.1088/0264-9381/33/3/035010}

\bibitem[{{Mahapatra} {et~al.}(2024){Mahapatra}, {Chattopadhyay}, {Gupta},
  {Antonini}, {Favata}, {Sathyaprakash}, \& {Arun}}]{2024ApJ...975..117M}
{Mahapatra}, P., {Chattopadhyay}, D., {Gupta}, A., {et~al.} 2024, \apj, 975,
  117, \dodoi{10.3847/1538-4357/ad781b}

\bibitem[{{Mandel}(2010)}]{2010PhRvD..81h4029M}
{Mandel}, I. 2010, \prd, 81, 084029, \dodoi{10.1103/PhysRevD.81.084029}

\bibitem[{{Miller} \& {Colbert}(2004)}]{2004IJMPD..13....1M}
{Miller}, M.~C., \& {Colbert}, E.~J.~M. 2004, International Journal of Modern
  Physics D, 13, 1, \dodoi{10.1142/S0218271804004426}

\bibitem[{{Miller} {et~al.}(2020){Miller}, {Callister}, \&
  {Farr}}]{2020ApJ...895..128M}
{Miller}, S., {Callister}, T.~A., \& {Farr}, W.~M. 2020, \apj, 895, 128,
  \dodoi{10.3847/1538-4357/ab80c0}

\bibitem[{{Mould} {et~al.}(2022){Mould}, {Gerosa}, {Broekgaarden}, \&
  {Steinle}}]{2022MNRAS.517.2738M}
{Mould}, M., {Gerosa}, D., {Broekgaarden}, F.~S., \& {Steinle}, N. 2022,
  \mnras, 517, 2738, \dodoi{10.1093/mnras/stac2859}

\bibitem[{{Nitz} \& {Capano}(2021)}]{2021ApJ...907L...9N}
{Nitz}, A.~H., \& {Capano}, C.~D. 2021, \apjl, 907, L9,
  \dodoi{10.3847/2041-8213/abccc5}

\bibitem[{{Pierra} {et~al.}(2024{\natexlab{a}}){Pierra}, {Mastrogiovanni}, \&
  {Perri{\`e}s}}]{2024arXiv240601679P}
{Pierra}, G., {Mastrogiovanni}, S., \& {Perri{\`e}s}, S. 2024{\natexlab{a}},
  arXiv e-prints, arXiv:2406.01679, \dodoi{10.48550/arXiv.2406.01679}

\bibitem[{{Pierra} {et~al.}(2024{\natexlab{b}}){Pierra}, {Mastrogiovanni}, \&
  {Perri{\`e}s}}]{2024A&A...692A..80P}
---. 2024{\natexlab{b}}, \aap, 692, A80, \dodoi{10.1051/0004-6361/202452545}

\bibitem[{{Popa} \& {de Mink}(2025)}]{2025arXiv250900154P}
{Popa}, S.~A., \& {de Mink}, S.~E. 2025, arXiv e-prints, arXiv:2509.00154,
  \dodoi{10.48550/arXiv.2509.00154}

\bibitem[{{Punturo} {et~al.}(2010){Punturo}, {Abernathy}, {Acernese}, {Allen},
  {Andersson}, {Arun}, {Barone}, {Barr}, {Barsuglia}, {Beker}, {Beveridge},
  {Birindelli}, {Bose}, {Bosi}, {Braccini}, {Bradaschia}, {Bulik}, {Calloni},
  {Cella}, {Chassande Mottin}, {Chelkowski}, {Chincarini}, {Clark}, {Coccia},
  {Colacino}, {Colas}, {Cumming}, {Cunningham}, {Cuoco}, {Danilishin},
  {Danzmann}, {De Luca}, {De Salvo}, {Dent}, {De Rosa}, {Di Fiore}, {Di
  Virgilio}, {Doets}, {Fafone}, {Falferi}, {Flaminio}, {Franc}, {Frasconi},
  {Freise}, {Fulda}, {Gair}, {Gemme}, {Gennai}, {Giazotto}, {Glampedakis},
  {Granata}, {Grote}, {Guidi}, {Hammond}, {Hannam}, {Harms}, {Heinert},
  {Hendry}, {Heng}, {Hennes}, {Hild}, {Hough}, {Husa}, {Huttner}, {Jones},
  {Khalili}, {Kokeyama}, {Kokkotas}, {Krishnan}, {Lorenzini}, {L{\"u}ck},
  {Majorana}, {Mandel}, {Mandic}, {Martin}, {Michel}, {Minenkov}, {Morgado},
  {Mosca}, {Mours}, {M{\"u}ller{\textendash}Ebhardt}, {Murray}, {Nawrodt},
  {Nelson}, {Oshaughnessy}, {Ott}, {Palomba}, {Paoli}, {Parguez},
  {Pasqualetti}, {Passaquieti}, {Passuello}, {Pinard}, {Poggiani}, {Popolizio},
  {Prato}, {Puppo}, {Rabeling}, {Rapagnani}, {Read}, {Regimbau}, {Rehbein},
  {Reid}, {Rezzolla}, {Ricci}, {Richard}, {Rocchi}, {Rowan}, {R{\"u}diger},
  {Sassolas}, {Sathyaprakash}, {Schnabel}, {Schwarz}, {Seidel}, {Sintes},
  {Somiya}, {Speirits}, {Strain}, {Strigin}, {Sutton}, {Tarabrin},
  {Th{\"u}ring}, {van den Brand}, {van Leewen}, {van Veggel}, {van den Broeck},
  {Vecchio}, {Veitch}, {Vetrano}, {Vicere}, {Vyatchanin}, {Willke}, {Woan},
  {Wolfango}, \& {Yamamoto}}]{2010CQGra..27s4002P}
{Punturo}, M., {Abernathy}, M., {Acernese}, F., {et~al.} 2010, Classical and
  Quantum Gravity, 27, 194002, \dodoi{10.1088/0264-9381/27/19/194002}

\bibitem[{{P{\"u}rrer} {et~al.}(2016){P{\"u}rrer}, {Hannam}, \&
  {Ohme}}]{2016PhRvD..93h4042P}
{P{\"u}rrer}, M., {Hannam}, M., \& {Ohme}, F. 2016, \prd, 93, 084042,
  \dodoi{10.1103/PhysRevD.93.084042}

\bibitem[{{Qin} {et~al.}(2018){Qin}, {Fragos}, {Meynet}, {Andrews},
  {S{\o}rensen}, \& {Song}}]{2018A&A...616A..28Q}
{Qin}, Y., {Fragos}, T., {Meynet}, G., {et~al.} 2018, \aap, 616, A28,
  \dodoi{10.1051/0004-6361/201832839}

\bibitem[{{Ray} {et~al.}(2025){Ray}, {Banagiri}, {Thrane}, \&
  {Lasky}}]{2025arXiv251007228R}
{Ray}, A., {Banagiri}, S., {Thrane}, E., \& {Lasky}, P.~D. 2025, arXiv
  e-prints, arXiv:2510.07228, \dodoi{10.48550/arXiv.2510.07228}

\bibitem[{{Ray} {et~al.}(2024){Ray}, {Maga{\~n}a Hernandez}, {Breivik}, \&
  {Creighton}}]{2024arXiv240403166R}
{Ray}, A., {Maga{\~n}a Hernandez}, I., {Breivik}, K., \& {Creighton}, J. 2024,
  arXiv e-prints, arXiv:2404.03166, \dodoi{10.48550/arXiv.2404.03166}

\bibitem[{{Roulet} {et~al.}(2021){Roulet}, {Chia}, {Olsen}, {Dai},
  {Venumadhav}, {Zackay}, \& {Zaldarriaga}}]{2021PhRvD.104h3010R}
{Roulet}, J., {Chia}, H.~S., {Olsen}, S., {et~al.} 2021, \prd, 104, 083010,
  \dodoi{10.1103/PhysRevD.104.083010}

\bibitem[{{Shao} \& {Li}(2022)}]{2022ApJ...930...26S}
{Shao}, Y., \& {Li}, X.-D. 2022, \apj, 930, 26,
  \dodoi{10.3847/1538-4357/ac61da}

\bibitem[{{Speagle}(2020)}]{2020MNRAS.493.3132S}
{Speagle}, J.~S. 2020, \mnras, 493, 3132, \dodoi{10.1093/mnras/staa278}

\bibitem[{{Tagawa} {et~al.}(2020){Tagawa}, {Haiman}, \&
  {Kocsis}}]{2020ApJ...898...25T}
{Tagawa}, H., {Haiman}, Z., \& {Kocsis}, B. 2020, \apj, 898, 25,
  \dodoi{10.3847/1538-4357/ab9b8c}

\bibitem[{{Tang} {et~al.}(2026){Tang}, {Wang}, {Li}, \&
  {Fan}}]{2025arXiv250903480T}
{Tang}, S.-P., {Wang}, H.-T., {Li}, Y.-J., \& {Fan}, Y.-Z. 2026, Science
  Bulletin, 71, \dodoi{10.1016/j.scib.2025.11.002}

\bibitem[{{Tiwari} {et~al.}(2018){Tiwari}, {Fairhurst}, \&
  {Hannam}}]{2018ApJ...868..140T}
{Tiwari}, V., {Fairhurst}, S., \& {Hannam}, M. 2018, \apj, 868, 140,
  \dodoi{10.3847/1538-4357/aae8df}

\bibitem[{{Tong} {et~al.}(2022){Tong}, {Galaudage}, \&
  {Thrane}}]{2022PhRvD.106j3019T}
{Tong}, H., {Galaudage}, S., \& {Thrane}, E. 2022, \prd, 106, 103019,
  \dodoi{10.1103/PhysRevD.106.103019}

\bibitem[{{Tong} {et~al.}(2025){Tong}, {Fishbach}, {Thrane}, {Mould},
  {Callister}, {Farah}, {Guttman}, {Banagiri}, {Beltran-Martinez}, {Farr},
  {Galaudage}, {Godfrey}, {Heinzel}, {Kalomenopoulos}, {Miller}, \&
  {Vijaykumar}}]{2025arXiv250904151T}
{Tong}, H., {Fishbach}, M., {Thrane}, E., {et~al.} 2025, arXiv e-prints,
  arXiv:2509.04151, \dodoi{10.48550/arXiv.2509.04151}

\bibitem[{{Torniamenti} {et~al.}(2022){Torniamenti}, {Rastello}, {Mapelli}, {Di
  Carlo}, {Ballone}, \& {Pasquato}}]{2022MNRAS.517.2953T}
{Torniamenti}, S., {Rastello}, S., {Mapelli}, M., {et~al.} 2022, \mnras, 517,
  2953, \dodoi{10.1093/mnras/stac2841}

\bibitem[{{Vaccaro} {et~al.}(2024){Vaccaro}, {Mapelli}, {P{\'e}rigois},
  {Barone}, {Artale}, {Dall'Amico}, {Iorio}, \&
  {Torniamenti}}]{2024A&A...685A..51V}
{Vaccaro}, M.~P., {Mapelli}, M., {P{\'e}rigois}, C., {et~al.} 2024, \aap, 685,
  A51, \dodoi{10.1051/0004-6361/202348509}

\bibitem[{{Wang} {et~al.}(2025{\natexlab{a}}){Wang}, {Tang}, {Li}, \&
  {Fan}}]{2025arXiv250902047W}
{Wang}, H.-T., {Tang}, S.-P., {Li}, P.-C., \& {Fan}, Y.-Z. 2025{\natexlab{a}},
  arXiv e-prints, arXiv:2509.02047, \dodoi{10.48550/arXiv.2509.02047}

\bibitem[{{Wang} {et~al.}(2025{\natexlab{b}}){Wang}, {Li}, {Gao}, {Tang}, \&
  {Fan}}]{2025arXiv251022698W}
{Wang}, Y.-Z., {Li}, Y.-J., {Gao}, S.-J., {Tang}, S.-P., \& {Fan}, Y.-Z.
  2025{\natexlab{b}}, arXiv e-prints, arXiv:2510.22698,
  \dodoi{10.48550/arXiv.2510.22698}

\bibitem[{{Wang} {et~al.}(2022){Wang}, {Li}, {Vink}, {Fan}, {Tang}, {Qin}, \&
  {Wei}}]{2022ApJ...941L..39W}
{Wang}, Y.-Z., {Li}, Y.-J., {Vink}, J.~S., {et~al.} 2022, \apjl, 941, L39,
  \dodoi{10.3847/2041-8213/aca89f}

\bibitem[{{Wang} {et~al.}(2021){Wang}, {Tang}, {Liang}, {Han}, {Li}, {Jin},
  {Fan}, \& {Wei}}]{2021ApJ...913...42W}
{Wang}, Y.-Z., {Tang}, S.-P., {Liang}, Y.-F., {et~al.} 2021, \apj, 913, 42,
  \dodoi{10.3847/1538-4357/abf5df}

\bibitem[{{Woosley}(2017)}]{2017ApJ...836..244W}
{Woosley}, S.~E. 2017, \apj, 836, 244, \dodoi{10.3847/1538-4357/836/2/244}

\bibitem[{{Woosley}(2019)}]{2019ApJ...878...49W}
---. 2019, \apj, 878, 49, \dodoi{10.3847/1538-4357/ab1b41}

\bibitem[{{Woosley} \& {Heger}(2021)}]{2021ApJ...912L..31W}
{Woosley}, S.~E., \& {Heger}, A. 2021, \apjl, 912, L31,
  \dodoi{10.3847/2041-8213/abf2c4}

\bibitem[{{Wysocki} {et~al.}(2018){Wysocki}, {Gerosa}, {O'Shaughnessy},
  {Belczynski}, {Gladysz}, {Berti}, {Kesden}, \& {Holz}}]{2018PhRvD..97d3014W}
{Wysocki}, D., {Gerosa}, D., {O'Shaughnessy}, R., {et~al.} 2018, \prd, 97,
  043014, \dodoi{10.1103/PhysRevD.97.043014}

\bibitem[{{Xue} {et~al.}(2025){Xue}, {Tagawa}, {Haiman}, \&
  {Bartos}}]{2025arXiv250419570X}
{Xue}, L., {Tagawa}, H., {Haiman}, Z., \& {Bartos}, I. 2025, arXiv e-prints,
  arXiv:2504.19570, \dodoi{10.48550/arXiv.2504.19570}

\bibitem[{{Yang} {et~al.}(2025){Yang}, {You}, \& {Fan}}]{2025arXiv251220890Y}
{Yang}, Q., {You}, Z.-Q., \& {Fan}, X. 2025, arXiv e-prints, arXiv:2512.20890,
  \dodoi{10.48550/arXiv.2512.20890}

\bibitem[{{Yang} {et~al.}(2019){Yang}, {Bartos}, {Gayathri}, {Ford}, {Haiman},
  {Klimenko}, {Kocsis}, {M{\'a}rka}, {M{\'a}rka}, {McKernan}, \&
  {O'Shaughnessy}}]{2019PhRvL.123r1101Y}
{Yang}, Y., {Bartos}, I., {Gayathri}, V., {et~al.} 2019, \prl, 123, 181101,
  \dodoi{10.1103/PhysRevLett.123.181101}

\bibitem[{{Yuan} {et~al.}(2025){Yuan}, {Chen}, \& {Liu}}]{2025PhRvD.112h1306Y}
{Yuan}, C., {Chen}, Z.-C., \& {Liu}, L. 2025, \prd, 112, L081306,
  \dodoi{10.1103/2vfn-48kh}

\bibitem[{{Zevin} \& {Holz}(2022)}]{2022ApJ...935L..20Z}
{Zevin}, M., \& {Holz}, D.~E. 2022, \apjl, 935, L20,
  \dodoi{10.3847/2041-8213/ac853d}

\bibitem[{{Zevin} {et~al.}(2021){Zevin}, {Bavera}, {Berry}, {Kalogera},
  {Fragos}, {Marchant}, {Rodriguez}, {Antonini}, {Holz}, \&
  {Pankow}}]{2021ApJ...910..152Z}
{Zevin}, M., {Bavera}, S.~S., {Berry}, C. P.~L., {et~al.} 2021, \apj, 910, 152,
  \dodoi{10.3847/1538-4357/abe40e}

\end{thebibliography}
\bibliographystyle{aasjournal}

\end{CJK*}
\end{document}